\documentclass{article}

\usepackage{arxiv}

\usepackage[T1]{fontenc}    
\usepackage{booktabs}       
\usepackage{amsfonts}       
\usepackage{nicefrac}       
\usepackage{microtype}      
\usepackage{amsmath}
\usepackage{amssymb}
\usepackage{mathtools}
\usepackage{nicefrac}
\usepackage{graphicx}
\usepackage{natbib}
\usepackage{hyperref}       
\usepackage{url}            
\usepackage{cleveref}       

\title{The nonparametric Behrens-Fisher problem \\in small samples}

\date{}

\author{ Claus P. Nowak \\
	Faculty of Statistics\\
	TU Dortmund University\\
	Dortmund, Germany 
	\And
	Markus Pauly \\
	Faculty of Statistics\\
	TU Dortmund University\\
	Dortmund, Germany 
	\And
	Edgar Brunner \\
	Department of Medical Statistics\\
	University Medical Center G{\"o}ttingen\\
	G{\"o}ttingen, Germany 
}

\renewcommand{\headeright}{}
\renewcommand{\undertitle}{}
\renewcommand{\shorttitle}{The nonparametric Behrens-Fisher problem in small samples}

\hypersetup{
pdftitle={The nonparametric Behrens-Fisher problem in small samples},
pdfsubject={},
pdfauthor={Claus P.\ Nowak, Markus Pauly, Edgar Brunner},
pdfkeywords={Brunner-Munzel test, Wilcoxon-Mann-Whitney test},
}

\begin{document}
\maketitle

\begin{abstract}
While there appears to be a general consensus in the literature on the definition of the estimand and estimator associated with the Wilcoxon-Mann-Whitney test, it seems somewhat less clear as to how best to estimate the variance. In addition to the Wilcoxon-Mann-Whitney test, we review different proposals of variance estimators consistent under both the null hypothesis and the alternative. 
Moreover, in case of small sample sizes, an approximation of the distribution of the test statistic based on the $t$-distribution, a logit transformation and a permutation approach have been proposed.
Focussing as well on different estimators of the degrees of freedom as regards the $t$-approximation, we carried out simulations for a range of scenarios, with results indicating that the performance of different variance estimators in terms of controlling the type I error rate largely depends on the heteroskedasticity pattern and the sample size allocation ratio, not on the specific type of distributions employed. By and large, a particular $t$-approximation together with Perme and Manevski's variance estimator best maintains the nominal significance level
\end{abstract}


\section{Introduction}

In the biomedical context, nonparametric methods are frequently indicated by ordered categorical data such as pain or clinical severity scores. In order to nonparametrically test the null hypothesis of whether two unpaired samples produce similar outcomes, the Wilcoxon-Mann-Whitney test \citep{mw,wil1,wil2} is arguably the one most commonly used in practice. 

Usually, the estimand related to the Wilcoxon-Mann-Whitney test is defined as the probability
$$
p = \mathbb{P}(X_{1} < X_{2}) + \nicefrac{1}{2}\cdot\mathbb{P}(X_{1}=X_{2}),
$$
where $X_{1}\sim F_{1}$ and $X_{2}\sim F_{2}$ denote two independent random variables corresponding to the two samples. The quantity $p$ is referred to as nonparametric relative effect of $X_{2}$ with respect to $X_{1}$ \citep{bm,np}, probabilistic index \citep{jan} or Mann-Whitney parameter \citep{fay}. In the setting of a parallel two-arm clinical trial, one may regard the random variables $X_{1}$ and $X_{2}$ as responses from treatment arms 1 and 2 respectively. Assuming that lower values imply a more beneficial outcome, one may interpret $p$ as the probability that a patient on arm 1 will fare better than one on arm 2, including $\nicefrac{1}{2}$ times the probability of equal outcomes.

While the literature seems to agree that the most suitable estimator of $p$, which we will refer to as $\widehat{p}$, are the corresponding relative frequencies arising from all pairwise comparisons of the sample data, the question of how best to estimate the variance of $\widehat{p}$ does not appear to be quite that settled.

The variance estimator employed in the Wilcoxon-Mann-Whitney test is unbiased and consistent, but only under the assumption of equal distributions, i.e., $F_{1}=F_{2}$. Hence the Wilcoxon-Mann-Whitney test can neither be inverted to produce confidence intervals nor does it directly address the nonparametric Behrens-Fisher problem. In that regard, assume both distribution $F_{1}$ and $F_{2}$ are symmetric with the same centre of symmetry but heterskedastic such as two normal distributions with the same expectation but different variances, yielding a Mann-Whitney parameter of $p=\nicefrac{1}{2}$. To test the null hypothesis $H_{0}: p=\nicefrac{1}{2}$, \cite{shirahata} considers a number of variance estimators of $\widehat{p}$ under the assumption that both distributions $F_{1}$ and $F_{2}$ are continuous while \cite{bamber} proposed an unbiased variance estimator for general $F_{1}$ and $F_{2}$, be they continuous, discreet or neither. Moreover, \cite{delong}, \cite{bm} as well as \cite{pm} put forward variance estimators consistent for arbitrary $F_{1}$ and $F_{2}$ as well.

In small samples, \cite{bm} suggest the use of a $t$-approximation analogous to the Satterthwaite-Smith-Welch approach \citep{satter,smith,welch} as regards the parametric Behrens-Fisher problem. Using different degrees of freedom and different variance estimators allowing for ties, we carry out simulation studies for a range of scenarios to gauge the performance of the resulting tests in terms of the type I error rate and power. In addition, we consider a permutation test proposed by \cite{asen}.

This manuscript proceeds as follows. In Section \ref{s:model} we review nonparametric theory and give definitions of the test statistics, whose empirical behaviour we examine as regards type I error rates and power in Section \ref{s:sim} and close with a discussion of the results in Section \ref{s:discuss}.
All proofs and derivations as well as more and more detailed simulation results are given in the appendix.

\section{Nonparametric model}
\label{s:model}
We start with notation from nonparametric theory convenient for stating variance formulas and test statistics. Then we go over the variance associated with the Wilcoxon-Mann-Whitney test, as well as variance estimators consistent for arbitrary distributions $F_{1}$ and $F_{2}$. 
For sake of completeness, we will also briefly mention Shirahata's [\citeyear{shirahata}] formulas. With the asymptotic normality of the resulting test statistics already established \citep{np}, we will discuss different estimators for the degrees of freedom in a small sample $t$-approximation as well as as the permutation approach developed by \cite{asen}.

\subsection{Notation}
Let $X$ denote a univariate random variable defined on a probability space $(\Omega,\mathcal{A},\mathbb{P})$, which stands for real-valued or ordered categorical responses. As is commonly done, we call
\begin{align*}
F^{-}(x) &= \mathbb{P}(X< x) \quad \text{ the left-continuous},\\
F^{+}(x) &= \mathbb{P}(X\leq x) \quad \text{ the right-continuous},\\
F(x) &= \mathbb{P}(X < x) + \nicefrac{1}{2}\cdot\mathbb{P}(X=x) \quad \text{ the normalised}
\end{align*}
version of the cumulative distribution function of $X$ \citep{levy1925calcul, ruymgaart1980unified, bm}.

For a particular sample of observations $X_{1},\dots,X_{n} \overset{iid}{\sim} F$, we further denote by
\begin{align*}
\widehat{F}(x) &= \frac{1}{n}\sum_{j=1}^{n} c(x,X_{j}), \quad c(x,X_{j}) = 
\begin{cases}
 0 & \text{ if } x < X_{j} \\
\nicefrac{1}{2} & \text{ if } x = X_{j} \\
 1 & \text{ if } x > X_{j}
\end{cases},
\end{align*}
the normalised version of the empirical cumulative distribution function. Moreover, we call
\begin{align*}
R_{i} = \nicefrac{1}{2} + \sum_{j=1}^{n} c(X_{i},X_{j}), \quad i=1,\dots,n,
\end{align*}
the mid-rank of $X_{i}$ among the observations $X_{1},\dots,X_{n}$.

As for two independent random variables $X_{1} \sim F_{1}$ and $X_{2} \sim F_{2}$, the Mann-Whitney parameter
as given in the Introduction has the following integral representation, i.e.,
\begin{align*}
p = \mathbb{P}(X_{1}<X_{2}) + \nicefrac{1}{2}\cdot\mathbb{P}(X_{1}=X_{2}) = \int F_{1}dF_{2}.
\end{align*}

We say that 
\begin{itemize}
\item $X_{1}$ tends to smaller values than $X_{2}$ if $p > \nicefrac{1}{2}$,
\item $X_{1}$ tends to larger values than $X_{2}$ if $p < \nicefrac{1}{2}$,
\item $X_{1}$ and $X_{2}$ are stochastically comparable if $p = \nicefrac{1}{2}$.
\end{itemize}
For a more comprehensive treatment of nonparametric theory we refer to \cite{np}.

Throughout the remainder of this manuscript we will focus on a parallel two-arm clinical trial with responses 
\begin{align*}
X_{1i}&\overset{iid}{\sim} F_{1}, \quad i=1,\dots,n_{1}, \\ 
X_{2j} &\overset{iid}{\sim} F_{2}, \quad j = 1,\dots,n_{2},
\end{align*}
from treatment arms 1 and 2 respectively. 
With $N=n_{1}+n_{2}$, we can estimate the nonparametric relative effect $p$ by
\begin{align*}
\widehat{p}&=\int\widehat{F}_{1}d\widehat{F}_{2} = \frac{1}{n_{1}}\frac{1}{n_{2}}
\sum_{j=1}^{n_{2}}\sum_{i=1}^{n_{1}}c(X_{2j},X_{1i})
=\frac{1}{N}(\bar{R}_{2\bullet}-\bar{R}_{1\bullet})+ \nicefrac{1}{2},
\end{align*}
with $\bar{R}_{g\bullet}=\frac{1}{n_{g}}\sum_{i=1}^{n_{g}}R_{gi}$, $g=1,2$, where $R_{gi}$ is the mid-rank of $X_{gi}$ among all $N$ observations $X_{11},\dots,X_{1n_{1}},X_{21},\dots,X_{2n_{2}}$. 

To address the nonparametric Behrens-Fisher problem, we consider the null hypothesis $H_{0}: p = \nicefrac{1}{2}$ against $H_{1}: p \neq \nicefrac{1}{2}$. Unsurprisingly, all resulting test statistics are based on the deviation of the Mann-Whitney parameter estimate from $\nicefrac{1}{2}$, i.e.,
\begin{equation*}
\widehat{p}-\nicefrac{1}{2}.
\end{equation*}
For asymptotic results, we let both sample sizes tend to infinity such that neither vanishes, i.e., $n_{g}/N \rightarrow \gamma_{g} > 0$ for both $n_{1}\rightarrow \infty$ and $n_{2}\rightarrow \infty$, $g=1,2$. Moreover, we assume $0 < p < 1$ and that there exists no $x$ such that $\mathbb{P}(X_{11}=x)=1$ or $\mathbb{P}(X_{21}=x)=1$, i.e., excluding the degenerate cases of completely separated samples and one-point distributions.

To obtain asymptotic standard normal test statistics, it remains to define suitable estimators of $\mathbb{V}(\widehat{p})$, which is the purpose of the next subsection.

\subsection{Variance estimators}
\label{s:wt}
Under the assumption of equal distributions, i.e., $F=F_{1}=F_{2}$, the variance estimand $\mathbb{V}(\widehat{p})$ takes the following form,
\begin{equation*}
\sigma_{WMW}^{2}=\frac{\sigma_{R}^{2}}{Nn_{1}n_{2}},
\text{ with }
\sigma^{2}_{R}=N\{(N-2)\int F^{2}dF-\frac{N-3}{4} \}-\frac{N}{4}\int (F^{+}-F^{-}) dF.
\end{equation*}
If $F=F_{1}=F_{2}$ holds, a consistent and unbiased estimator of the variance $\sigma_{WMW}^{2}$ is given by
\begin{equation}\label{vareWMW}
\widehat{\sigma}^{2}_{WMW}=\frac{\widehat{\sigma}_{R}^{2}}{Nn_{1}n_{2}},
\text{ with }
\widehat{\sigma}^{2}_{R}=
\frac{N^{3}}{N-1}( \int \widehat{F}^{2}d\widehat{F} - \nicefrac{1}{4} )
=
\frac{1}{N}
\sum_{g=1}^{2} \sum_{i=1}^{n_{g}} ( R_{gi} - \frac{N+1}{2} )^{2},
\end{equation}
resulting in $T_{WMW}= (\widehat{p}-\nicefrac{1}{2})/\widehat{\sigma}_{WMW} \xrightarrow[]{\, \mathcal{D} \,} \mathcal{N}(0,1)$, which is nothing but the Wilcoxon-Mann-Whitney test allowing for ties \citep{np}. In the context of this manuscript, we feel it more tenable to regard the Wilcoxon-Mann-Whitney test as a way of testing the null hypothesis formulated in terms of $p$, i.e., $H_{0}:p=\nicefrac{1}{2}$, whereas $F_{1}=F_{2}$ amounts to an additional assumption on the model under the null.

As for arbitrary distributions $F_{1}$ and $F_{2}$, \cite{bamber} as well as \citep{brunnerpreprint} provide a formula of the variance estimand $\mathbb{V}(\widehat{p})$, which reads in our notation as
\begin{equation*}
\sigma^{2}_{N} =
\frac{\tau_{0}+(n_{2}-1)\tau_{1}+(n_{1}-1)\tau_{2}-(n_{1}+n_{2}-1)p^{2}}{n_{1}n_{2}},
\end{equation*}
where 
$\tau_{0} = p - \nicefrac{1}{4}\cdot\int (F_{1}^{+}-F_{1}^{-})dF_{2}$, $\tau_{1} = \int (1-F_{2})^{2}dF_{1}$, and $\tau_{2} = \int F_{1}^{2}dF_{2}$. 

\cite{bamber} as well as \cite{brunnerpreprint} propose an unbiased variance estimator of $\sigma^{2}_{N}$ as well, namely,
\begin{equation}\label{vareunb}
\widehat{\sigma}_{N}^{2}
=
\frac{
n_{2}\widehat{\tau}_{1}
+
n_{1}\widehat{\tau}_{2}
-\widehat{\tau}_{0}
-
(n_{1}+n_{2}-1)\widehat{p}^{2}
}
{(n_{1}-1)(n_{2}-1)},
\end{equation}
with
$\widehat{\tau}_{0} = \widehat{p} - \nicefrac{1}{4}\cdot\int (\widehat{F}_{1}^{+}-\widehat{F}_{1}^{-})d\widehat{F}_{2}$, $\widehat{\tau}_{1} = \int (1-\widehat{F}_{2})^{2}d\widehat{F}_{1}$, and $\widehat{\tau}_{2} = \int \widehat{F}_{1}^{2}d\widehat{F}_{2}$. 
For a computationally more efficient expression of $\widehat{\sigma}_{N}^{2}$ in terms of ranks, see \cite{brunnerpreprint}.

\cite{bm} derived an estimator of $\sigma_{N}^{2}$ similar in structure to the variance estimator of the two-sample $t$-test under heteroskedasticity, i.e.,
\begin{equation}\label{varebm}
\widehat{\sigma}_{BM}^{2}
= \frac{\widehat{\sigma}^{2}_{1}}{n_{1}}
+
\frac{\widehat{\sigma}^{2}_{2}}{n_{2}}
,
\text{ where }
\widehat{\sigma}^{2}_{1}
=
\frac{n_{1}}{n_{1}-1}(\widehat{\tau}_{1} - \widehat{p}^{2})
\text{ and }
\widehat{\sigma}^{2}_{2}
=
\frac{n_{2}}{n_{2}-1}(\widehat{\tau}_{2} - \widehat{p}^{2}),
\end{equation}
with $\widehat{\tau}_{1}$ and $\widehat{\tau}_{2}$ as given in \eqref{vareunb}. For a computationally more efficient rank representation of $\widehat{\sigma}_{1}^{2}$ and $\widehat{\sigma}_{2}^{2}$, see \cite{bm}. Note that the estimator $\widehat{\sigma}_{BM}^{2}$ is identical to the one given in \cite{delong}.

\cite{pm} propose yet another estimator for $\sigma_{N}^{2}$, which they refer to as exact, i.e.,
\begin{equation}\label{varepm}
\widehat{\sigma}_{PM}^{2}
=
\frac{\widehat{p}(1-\widehat{p})+(n_{2}-1)\widehat{\sigma}_{1}^{2}
+ (n_{1}-1)\widehat{\sigma}_{2}^{2}
}{n_{1}n_{2}},
\end{equation}
with $\widehat{\sigma}_{1}^{2}$ and $\widehat{\sigma}_{2}^{2}$ as just defined in \eqref{varebm}.

In the Introduction, we have vaguely hinted at the consistency of the variance estimators. More precisely, the dominating terms $\widehat{\tau}_{1}$, $\widehat{\tau}_{2}$, $\widehat{p}^2$ are consistent for $\tau_{1}$, $\tau_{2}$, $p^2$ and they converge, as weighted with the sample sizes in $\widehat{\sigma}^{2}_{N}$ \eqref{vareunb}, $\widehat{\sigma}^{2}_{BM}$ \eqref{varebm}, $\widehat{\sigma}^{2}_{PM}$ \eqref{varepm}, to zero in probability with the same speed.

\subsection{Shirahata's formulas for continuous distributions}
\label{s:df}
Assuming that ties cannot occur almost surely, \cite{shirahata} discusses the following four estimators, i.e., an unbiased one, a bootstrap estimator, an estimator by \cite{fligner}, and a jackknife estimator, which in our notation read as
\begin{align*}
\widehat{\sigma}_{U}^{2}
&=
\frac{n_{2}\int (1-\widehat{F}_{2}^{-})^{2}d\widehat{F}_{1}+n_{1}\int (\widehat{F}_{1}^{+})^{2}d\widehat{F}_{2} - \int \widehat{F}_{1}^{+}d\widehat{F}_{2}- (n_{1}+n_{2}-1)(\int \widehat{F}_{1}^{+}d\widehat{F}_{2})^{2}}{(n_{1}-1)(n_{2}-1)},
\\
\widehat{\sigma}_{B}^{2}
&=
\frac{(n_{2}-1)\int (1-\widehat{F}_{2}^{-})^{2}d\widehat{F}_{1}+(n_{1}-1)\int (\widehat{F}_{1}^{+})^{2}d\widehat{F}_{2}+\int \widehat{F}_{1}^{+}d\widehat{F}_{2}-(n_{1}+n_{2}-1)(\int \widehat{F}_{1}^{+}d\widehat{F}_{2})^{2}}{n_{1}n_{2}},
\\
\widehat{\sigma}_{FP}^{2}
&=
\frac{\int (1-\widehat{F}_{2}^{-})^{2}d\widehat{F}_{1}}{n_{1}}+\frac{\int (\widehat{F}_{1}^{+})^{2}d\widehat{F}_{2}}{n_{2}} -
\frac{\int \widehat{F}_{1}^{+}d\widehat{F}_{2}+(n_{1}+n_{2}+1)(\int \widehat{F}_{1}^{+}d\widehat{F}_{2})^{2}}{n_{1}n_{2}},
\\
\widehat{\sigma}_{J}^{2}
&=
\frac{\int (1-\widehat{F}_{2}^{-})^{2}d\widehat{F}_{1}}{n_{1}-1}
+
\frac{\int (\widehat{F}_{1}^{+})^{2}d\widehat{F}_{2}}{n_{2}-1}
-
\frac{ (n_{1}+n_{2}-2)(\int \widehat{F}_{1}^{+}d\widehat{F}_{2})^{2}}{(n_{1}-1)(n_{2}-1)}.
\end{align*}
In case of continuous distributions, however, it follows
$\int \widehat{F}_{1}^{+}d\widehat{F}_{2} = 
\int \widehat{F}_{1}d\widehat{F}_{2} = \widehat{p}=\widehat{\tau}_{0}$,
$\int (1-\widehat{F}_{2}^{-})^{2}d\widehat{F}_{1}=\int (1-\widehat{F}_{2})^{2}d\widehat{F}_{1}=\widehat{\tau}_{1}$,
$\int (\widehat{F}_{1}^{+})^{2}d\widehat{F}_{2}=\int \widehat{F}_{1}^{2}d\widehat{F}_{2}=\widehat{\tau}_{2}$,
so that then we can express the variance estimators as follows
\begin{align*}
\widehat{\sigma}_{U}^{2}
&=
\frac{n_{2}\widehat{\tau}_{1}+n_{1}\widehat{\tau}_{2} - \widehat{\tau}_{0}- (n_{1}+n_{2}-1)\widehat{p}^{2}}{(n_{1}-1)(n_{2}-1)}
=\widehat{\sigma}^{2}_{N},
\\
\widehat{\sigma}_{B}^{2}
&=
\frac{(n_{2}-1)\widehat{\tau}_{1}+(n_{1}-1)\widehat{\tau}_{2} + \widehat{\tau}_{0}- (n_{1}+n_{2}-1)\widehat{p}^{2}}{n_{1}n_{2}},
\\
\widehat{\sigma}_{FP}^{2}
&=
\frac{\widehat{\tau}_{1}}{n_{1}}
+
\frac{\widehat{\tau}_{2}}{n_{2}}
- 
\frac{\widehat{\tau}_{0}+ (n_{1}+n_{2}+1)\widehat{p}^{2}}{n_{1}n_{2}},
\\
\widehat{\sigma}_{J}^{2}
&=
\frac{\widehat{\tau}_{1}}{n_{1}-1}
+
\frac{\widehat{\tau}_{2}}{n_{2}-1}
-
\frac{ (n_{1}+n_{2}-2)\widehat{p}^{2}}{(n_{1}-1)(n_{2}-1)}=\widehat{\sigma}_{BM}^{2}.
\end{align*}
\subsection{Degrees of freedom}
\label{s:df2}
Analogous to the parametric Behrens-Fisher problem, \cite{bm} propose a Satterthwaite-Smith-Welch $t$-approximation \citep{satter,smith,welch} for small samples with degrees of freedom estimated by
\begin{equation}\label{eq:df}
df=\frac
{
\{
\widehat{\sigma}_{1}^{2}/{n_{1}}
+
\widehat{\sigma}_{2}^{2}/{n_{2}}
\}^{2}
}
{
\widehat{\sigma}_{1}^{4}/\{n_{1}^{2}(n_{1}-1)\}
+
\widehat{\sigma}_{2}^{4}/ \{n_{2}^{2}(n_{2}-1)\}
},
\end{equation}
where $\widehat{\sigma}_{1}^{2}$ and $\widehat{\sigma}_{2}^{2}$ are defined as in \eqref{varebm}. In addition, we will consider the degrees of freedom
\begin{align}
df_{1} &= \frac{
\{
\widehat{\sigma}_{1}^{2}/({n_{1}-1})
+
\widehat{\sigma}_{2}^{2}/({n_{2}-1})
\}^{2}
}
{
\widehat{\sigma}_{1}^{4}/\{(n_{1}-1)^{2}(n_{1}-2)\}
+
\widehat{\sigma}_{2}^{4}/\{(n_{2}-1)^{2}(n_{2}-2)\}
}, \label{eq:df1}
\\
df_{2} &= \frac{
\{
\widehat{\sigma}_{1}^{2}/({n_{1}-2})
+
\widehat{\sigma}_{2}^{2}/({n_{2}-2})
\}^{2}
}
{
\widehat{\sigma}_{1}^{4}/\{(n_{1}-2)^{2}(n_{1}-3)\}
+
\widehat{\sigma}_{2}^{4}/\{(n_{2}-2)^{2}(n_{2}-3)\}
},\label{eq:df2}
\\
df_{3} &= \frac{
2
}
{
1/(n_{1}-1)
+
1/(n_{2}-1)}, \label{eq:df3}
\\
df_{4} &= \frac{\widehat{\sigma}_{N}^{4}}
{ \widehat{\sigma}_{1\mid N}^{4}/(n_{1}-1) + \widehat{\sigma}_{2\mid N}^{4}/(n_{2}-1)},\label{eq:df4}
\end{align}
with $\widehat{\sigma}_{N}^{2}$ as in \eqref{vareunb} and 
$\widehat{\sigma}_{1\mid N}^{2}=\frac{n_{2}\widehat{\tau}_{1}-\nicefrac{1}{2}\cdot\widehat{\tau}_{0}-(n_{2}-\nicefrac{1}{2})\widehat{p}^{2}}{(n_{1}-1)(n_{2}-1)}$, 
$\widehat{\sigma}_{2\mid N}^{2}=\frac{n_{1}\widehat{\tau}_{2}-\nicefrac{1}{2}\cdot\widehat{\tau}_{0}-(n_{1}-\nicefrac{1}{2})\widehat{p}^{2}}{(n_{1}-1)(n_{2}-1)}$.

The intuition behind using \eqref{eq:df1} and \eqref{eq:df2} in small samples is similar to \eqref{eq:df}, we merely assume that there were one (or two) fewer observations in each of the two groups, with $\widehat{\sigma}_{1}^{2}$ and $\widehat{\sigma}_{2}^{2}$ remaining unchanged as in \eqref{eq:df}. On the other hand, formulas \eqref{eq:df3} and \eqref{eq:df4} are 
loosely based on a Box-type \citep{box,np} approximation as regards the unbiased variance estimator.

Another way to address the liberal behaviour of the tests is to employ a variance stabilising transformation, such as the \emph{logit} function, or a permutation approach as described in Section \ref{s:tests} \citep{np,asen}.
\subsection{Test statistics}
\label{s:tests}
Collecting the test statistics with regard to the null hypothesis $H_{0}:p=\nicefrac{1}{2}$ allowing for ties, we have
\begin{align}
T_{WMW} &= (\widehat{p}-\nicefrac{1}{2})/\widehat{\sigma}_{WMW}, \label{eq:testwmw} \\
T_{N} &= (\widehat{p}-\nicefrac{1}{2})/\widehat{\sigma}_{N}, \label{eq:testunb} \\
T_{BM} &= (\widehat{p}-\nicefrac{1}{2})/\widehat{\sigma}_{BM}, \label{eq:testbm} \\
T_{PM} &= (\widehat{p}-\nicefrac{1}{2})/\widehat{\sigma}_{PM}. \label{eq:testpm} 
\end{align}
For the Wilcoxon-Mann-Whitney test \eqref{eq:testwmw} we use the standard normal distribution to compute $p$-values, for the other tests, \eqref{eq:testunb} to \eqref{eq:testpm}, a central $t$-distribution with degrees of freedom given as in \eqref{eq:df} to \eqref{eq:df4}. As already alluded to, we will additionally consider the following test statistics based on a \emph{logit} transformation using the delta method, i.e.,
\begin{align}
T_{N}^{logit}  &= \widehat{p}(1-\widehat{p})\cdot\ln\{ \widehat{p}/ (1-\widehat{p} ) \}/\widehat{\sigma}_{N}, \label{eq:testunblogit} \\
T_{BM}^{logit} &= \widehat{p}(1-\widehat{p})\cdot\ln\{ \widehat{p}/ (1-\widehat{p} ) \}/\widehat{\sigma}_{BM}, \label{eq:testbmlogit} \\
T_{PM}^{logit} &= \widehat{p}(1-\widehat{p})\cdot\ln\{ \widehat{p}/ (1-\widehat{p} ) \}/\widehat{\sigma}_{PM}. \label{eq:testpmlogit} 
\end{align}
As with the Wilcoxon-Mann-Whitney test, we employ the standard normal distribution to obtain $p$-values for \eqref{eq:testunblogit} to \eqref{eq:testpmlogit}.

Moreover, we will make use of the studentised permutation approach suggested by \cite{asen}. To this end, we randomly allocate $n_{1}$ out of the entire $N=n_{1}+n_{2}$ observations from the whole sample as originating from the first distribution $F_{1}$, with the remaining $n_{2}$ responses regarded as having been drawn from $F_{2}$. Repeating this procedure, say $n_{perm}=10\,000$ times, and computing one of the test statistics as in \eqref{eq:testunb} to \eqref{eq:testpmlogit} each time, we obtain a permutation distribution on which to base rejection of the null hypothesis.
More formally, we relabel the entire data $(X_{11},\dots,X_{1n_{1}},X_{21},\dots,X_{2n_{2}})\eqqcolon (X_{1},\dots,X_{N})$ and define a random variable $\pi$ uniformly distributed on $S_{N}$ which is the set of all permutations of $1,\dots,N$. For a particular data set at hand, we then use the permuted pooled sample $(X_{\pi(1)},\dots,X_{\pi(N)})$ -- with the first $n_{1}$ and last $n_{2}$ components considered as belonging to samples 1 and 2 respectively -- to compute \eqref{eq:testunb} to \eqref{eq:testpmlogit}, yielding the permuted versions
\begin{equation}\label{eq:permtests}
\widetilde{T}_{N}, \, \, \widetilde{T}_{BM}, \, \, \widetilde{T}_{PM}, \, \, \widetilde{T}_{N}^{logit}, \, \, \widetilde{T}_{BM}^{logit}, \, \, \widetilde{T}_{PM}^{logit}.
\end{equation}
With $T_{N}$ denoting the test statistic as in \eqref{eq:testunb} based on the original data and $\widetilde{T}_{N}^{1}, \dots, \widetilde{T}_{N}^{n_{perm}}$ the corresponding test statistics of the $n_{perm}$ random permutations, we calculate the two-sided $p$-value as follows,
\begin{equation*}
2\min(p_{1},p_{2}), \, \text{ where } \, p_{1}=\frac{1}{n_{perm}}\sum_{k=1}^{n_{perm}}\mathbf{1}( \widetilde{T}_{N}^{k} \leq T_{N} )
\, \text{ and } \, p_{2}=\frac{1}{n_{perm}}\sum_{\ell=1}^{n_{perm}}\mathbf{1}( \widetilde{T}_{N}^{\ell} \geq T_{N} ).
\end{equation*}
As for the other test statistics, the permutation based $p$-values are computed in a completely analogous manner.

\section{Simulations} %
\label{s:sim}
As the methods treated in Section \ref{s:model} are of asymptotic nature, we explore their applicability for finite sample sizes in a range of scenarios. In that regard, we consider the null hypothesis $H_{0}:p=\nicefrac{1}{2}$ against $H_{1}:p \neq \nicefrac{1}{2}$ at a two-sided nominal significance level of $\alpha=0.05$. We first present simulation results for the asymptotic tests as defined in \eqref{eq:testwmw} to \eqref{eq:testpmlogit}. As far as the test statistics \eqref{eq:testunb} to \eqref{eq:testpm} are concerned, we only report rejection rates for degrees of freedom $df_{2}$ \eqref{eq:df2} in the main manuscript as they outperformed the other versions. Simulations of permutation tests as in \eqref{eq:permtests} being computationally much more expensive, we restrict our focus to some select scenarios as outlined in Section \ref{sec:simperm}.

In extreme cases as alluded to earlier, some variance estimates might actually turn out to be zero or negative as would be the case in two completely separated samples. Since this happened very rarely and has virtually no bearing on the results, we relegate the discussion of exception handling to the appendix.

\subsection{Asymptotic tests}\label{sec:simasymp}
\begin{figure}
\centerline{\includegraphics[scale=0.54]{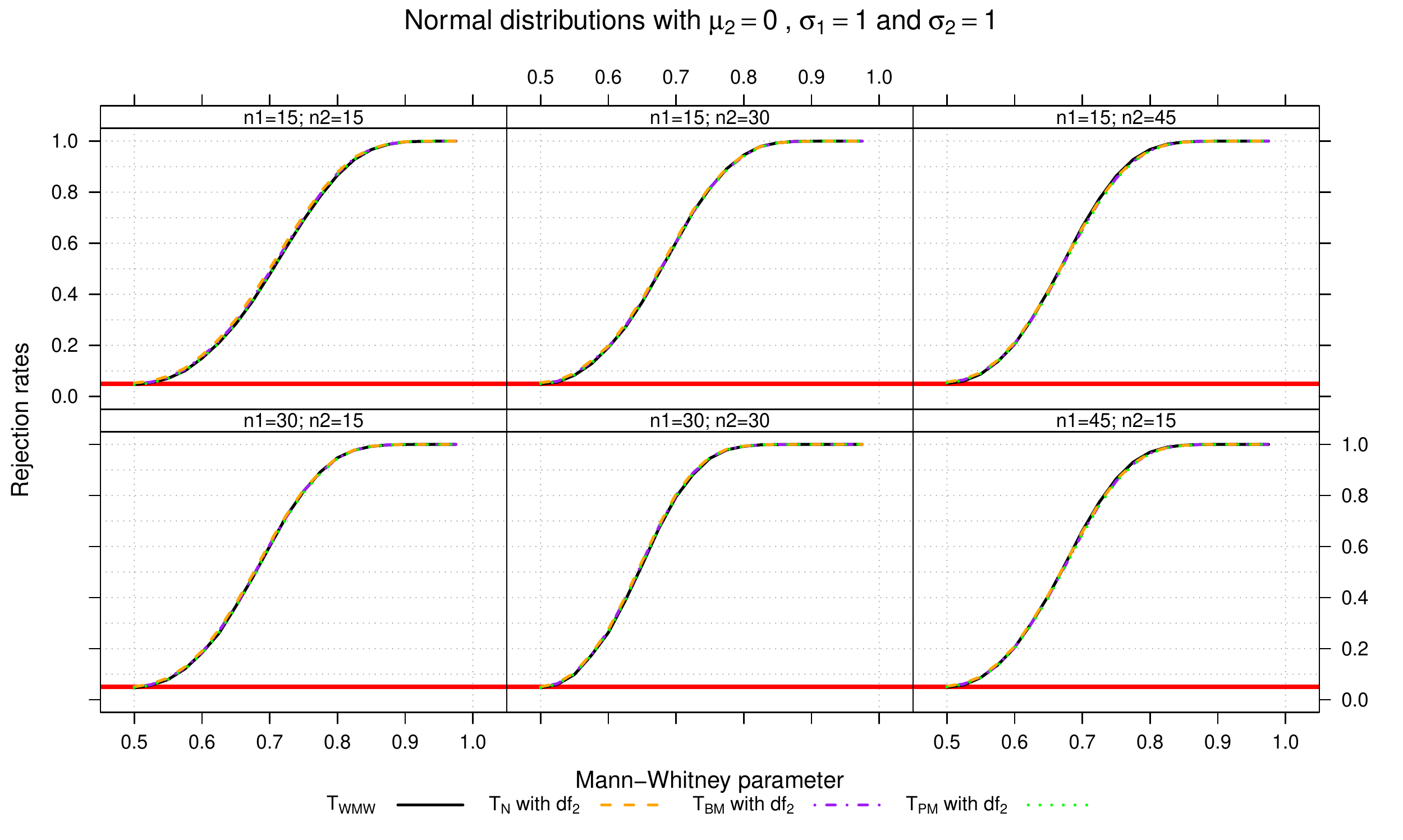}}
\vspace{12pt}
\centerline{\includegraphics[scale=0.54]{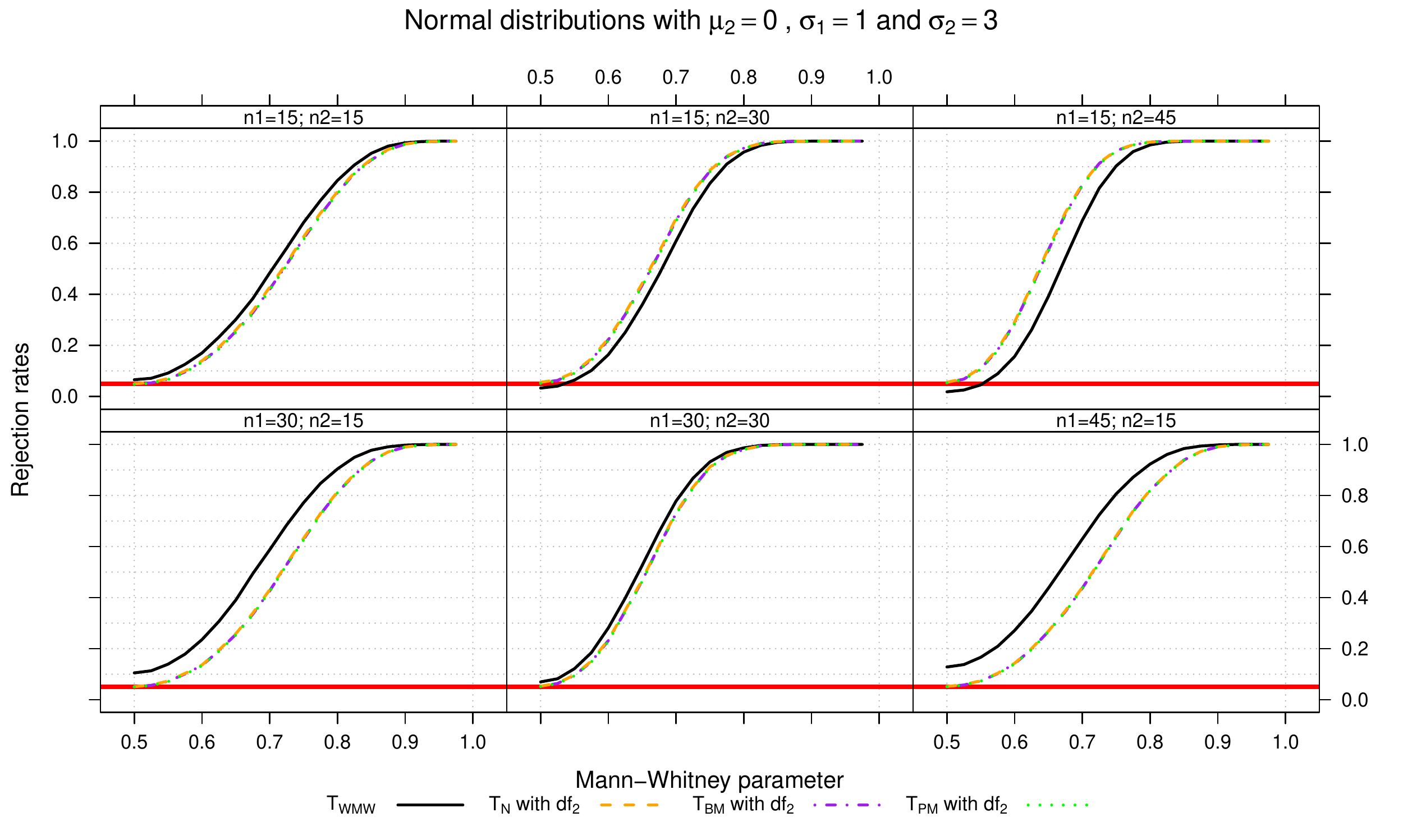}}
\caption{Power graphs for normal distributions based on 10\,000 simulation runs}{}
\label{f:normal1}
\end{figure}

First we generate data from normal distributions, namely $X_{gi} \overset{iid}{\sim} \mathcal{N}(\mu_{g},\sigma_{g}^{2})$, $g=1,2$, $i=1,\dots,n_{g}$.
To gauge the type I error rate of the different tests, we set $\mu_{1}=\mu_{2}=0$ and perform 100\,000 simulation runs for each scenario, giving rise to a Monte Carlo error of about $0.0006$ 
based on a 95\%-precision interval for a nominal significance level of $\alpha=0.05$. The results depicted in Table \ref{t:normal1} indicate that Perme and Manveski's test $T_{PM}$ \eqref{eq:testpm} with $df_{2}$ degrees of freedom \eqref{eq:df2} best maintains the nominal significance level, especially for $\min(n_{1},n_{2})\geq 15$, although the difference from $T_{N}$ \eqref{eq:testunb} as well as $T_{BM}$ \eqref{eq:testbm} is not particularly pronounced. However, in the heteroskedastic settings, the Wilcoxon-Mann-Whitney test $T_{WMW}$ \eqref{eq:testwmw} is generally either far too liberal or far too conservative depending on sample size allocation. More precisely, if more patients are allocated to the arm producing less dispersed outcomes, then $T_{WMW}$ \eqref{eq:testwmw} becomes too liberal, and too conservative otherwise.
While \emph{logit} based tests, \eqref{eq:testunblogit} to \eqref{eq:testpmlogit}, virtually never exceed the nominal significance level, they exhibit a somewhat conservative tendency in many cases. In that regard, we only present power graphs for the tests $T_{WMW}$ \eqref{eq:testwmw} and $T_{N}$ \eqref{eq:testunb}, $T_{BM}$ \eqref{eq:testbm}, $T_{PM}$ \eqref{eq:testpm} with $df_{2}$ degrees of freedom \eqref{eq:df2} as set forth in Figure \ref{f:normal1}. Unlike before, the power graphs are based on only 10\,000 simulation runs per scenario.

\begin{table}
\centering
\caption{Type I error rates for normal distributions $F_{1}=\mathcal{N}(0,\sigma_{1}^{2})$ and $F_{2}=\mathcal{N}(0,\sigma_{2}^{2})$ based on 100\,000 replications at a two-sided nominal significance level of $\alpha=0.05$ as regards the test statistics $T_{WMW}$ \eqref{eq:testwmw}; $T_{N}$ \eqref{eq:testunb}, $T_{BM}$ \eqref{eq:testbm}, $T_{PM}$ \eqref{eq:testpm} with degrees of freedom $df_{2}$ \eqref{eq:df2}; $T_{N}^{Logit}$ \eqref{eq:testunblogit}, $T_{BM}^{Logit}$ \eqref{eq:testbmlogit}, $T_{PM}^{Logit}$ \eqref{eq:testpmlogit}}
\label{t:normal1}
\begin{tabular}{rrrrccccccc}
  \hline
  $n_{1}$ & $n_{2}$ & $\sigma_{1}$ & $\sigma_{2}$ & $T_{WMW} $ & $T_{N}$ & $T_{BM}$ & $T_{PM}$ & $T_{N}^{Logit}$ & $T_{BM}^{Logit}$ & $T_{PM}^{Logit}$ \\ 
  \hline
   7 &  7 & 1 & 1 & 0.05318 & 0.05527 & 0.04796 & 0.04304 & 0.02886 & 0.02318 & 0.01860 \\ 
  10 &  7 & 1 & 1 & 0.04348 & 0.05428 & 0.05003 & 0.04725 & 0.03545 & 0.02981 & 0.02441 \\ 
   7 & 10 & 1 & 1 & 0.04290 & 0.05399 & 0.04975 & 0.04708 & 0.03460 & 0.02899 & 0.02386 \\ 
  10 & 10 & 1 & 1 & 0.05320 & 0.05696 & 0.05225 & 0.04856 & 0.03583 & 0.02993 & 0.02710 \\ 
  15 & 15 & 1 & 1 & 0.05072 & 0.05651 & 0.05290 & 0.05012 & 0.04067 & 0.03691 & 0.03435 \\ 
  30 & 15 & 1 & 1 & 0.04906 & 0.05417 & 0.05183 & 0.05001 & 0.04498 & 0.04234 & 0.04050 \\ 
  15 & 30 & 1 & 1 & 0.04911 & 0.05431 & 0.05207 & 0.05004 & 0.04504 & 0.04240 & 0.04044 \\ 
  30 & 30 & 1 & 1 & 0.04950 & 0.05306 & 0.05138 & 0.04978 & 0.04510 & 0.04316 & 0.04163 \\ 
  15 & 45 & 1 & 1 & 0.04891 & 0.05340 & 0.05167 & 0.05040 & 0.04841 & 0.04624 & 0.04461 \\ 
  15 & 60 & 1 & 1 & 0.04889 & 0.05192 & 0.05044 & 0.04945 & 0.04957 & 0.04784 & 0.04641 \\ 
  15 & 75 & 1 & 1 & 0.04959 & 0.05292 & 0.05144 & 0.05046 & 0.05219 & 0.05059 & 0.04944 \\ 
  45 & 15 & 1 & 1 & 0.04945 & 0.05329 & 0.05150 & 0.05002 & 0.04835 & 0.04644 & 0.04489 \\ 
  60 & 15 & 1 & 1 & 0.04930 & 0.05262 & 0.05120 & 0.04995 & 0.05003 & 0.04844 & 0.04697 \\ 
  75 & 15 & 1 & 1 & 0.04918 & 0.05164 & 0.05055 & 0.04955 & 0.05100 & 0.04978 & 0.04859 \\ \hline
   7 &  7 & 1 & 3 & 0.07223 & 0.04572 & 0.04206 & 0.03917 & 0.02785 & 0.02442 & 0.02135 \\ 
  10 &  7 & 1 & 3 & 0.08066 & 0.04509 & 0.04320 & 0.04175 & 0.03273 & 0.02984 & 0.02716 \\ 
   7 & 10 & 1 & 3 & 0.04122 & 0.05091 & 0.04734 & 0.04540 & 0.03058 & 0.02694 & 0.02365 \\ 
  10 & 10 & 1 & 3 & 0.07141 & 0.05172 & 0.04893 & 0.04705 & 0.03310 & 0.02961 & 0.02835 \\ 
  15 & 15 & 1 & 3 & 0.06833 & 0.05235 & 0.05036 & 0.04926 & 0.03822 & 0.03577 & 0.03434 \\ 
  30 & 15 & 1 & 3 & 0.10568 & 0.05111 & 0.04995 & 0.04919 & 0.04008 & 0.03887 & 0.03810 \\ 
  15 & 30 & 1 & 3 & 0.03163 & 0.05299 & 0.05111 & 0.04973 & 0.04355 & 0.04140 & 0.04013 \\ 
  30 & 30 & 1 & 3 & 0.07001 & 0.05308 & 0.05218 & 0.05124 & 0.04572 & 0.04468 & 0.04372 \\ 
  15 & 45 & 1 & 3 & 0.01618 & 0.05185 & 0.04994 & 0.04859 & 0.04490 & 0.04314 & 0.04196 \\ 
  15 & 60 & 1 & 3 & 0.00965 & 0.05130 & 0.04978 & 0.04853 & 0.04642 & 0.04473 & 0.04332 \\ 
  15 & 75 & 1 & 3 & 0.00616 & 0.05263 & 0.05114 & 0.04978 & 0.04871 & 0.04720 & 0.04593 \\ 
  45 & 15 & 1 & 3 & 0.12749 & 0.05153 & 0.05080 & 0.05041 & 0.04236 & 0.04145 & 0.04075 \\ 
  60 & 15 & 1 & 3 & 0.14104 & 0.05116 & 0.05044 & 0.05000 & 0.04224 & 0.04145 & 0.04093 \\ 
  75 & 15 & 1 & 3 & 0.14588 & 0.05068 & 0.05009 & 0.04973 & 0.04233 & 0.04171 & 0.04134 \\ \hline
   7 &  7 & 1 & 5 & 0.08821 & 0.03694 & 0.03496 & 0.03319 & 0.02367 & 0.02168 & 0.02000 \\ 
  10 &  7 & 1 & 5 & 0.09850 & 0.03648 & 0.03563 & 0.03504 & 0.02702 & 0.02500 & 0.02380 \\ 
   7 & 10 & 1 & 5 & 0.04932 & 0.04573 & 0.04335 & 0.04229 & 0.02883 & 0.02666 & 0.02399 \\ 
  10 & 10 & 1 & 5 & 0.08485 & 0.04618 & 0.04453 & 0.04376 & 0.03061 & 0.02890 & 0.02805 \\ 
  15 & 15 & 1 & 5 & 0.08132 & 0.05108 & 0.04987 & 0.04928 & 0.03690 & 0.03543 & 0.03458 \\ 
  30 & 15 & 1 & 5 & 0.12597 & 0.05022 & 0.04948 & 0.04907 & 0.03803 & 0.03722 & 0.03669 \\ 
  15 & 30 & 1 & 5 & 0.03419 & 0.05207 & 0.05079 & 0.04985 & 0.04309 & 0.04149 & 0.04059 \\ 
  30 & 30 & 1 & 5 & 0.08136 & 0.05257 & 0.05186 & 0.05128 & 0.04465 & 0.04387 & 0.04345 \\ 
  15 & 45 & 1 & 5 & 0.01548 & 0.05039 & 0.04890 & 0.04800 & 0.04333 & 0.04222 & 0.04151 \\ 
  15 & 60 & 1 & 5 & 0.00770 & 0.05111 & 0.04984 & 0.04898 & 0.04580 & 0.04473 & 0.04347 \\ 
  15 & 75 & 1 & 5 & 0.00431 & 0.05116 & 0.05003 & 0.04899 & 0.04731 & 0.04634 & 0.04530 \\ 
  45 & 15 & 1 & 5 & 0.15064 & 0.05092 & 0.05044 & 0.05015 & 0.03894 & 0.03833 & 0.03791 \\ 
  60 & 15 & 1 & 5 & 0.16777 & 0.05023 & 0.04993 & 0.04979 & 0.03828 & 0.03769 & 0.03731 \\ 
  75 & 15 & 1 & 5 & 0.17407 & 0.05004 & 0.04986 & 0.04961 & 0.03914 & 0.03882 & 0.03852 \\ 
   \hline
\end{tabular}
\end{table}

\begin{figure}
\centerline{\includegraphics[scale=0.65]{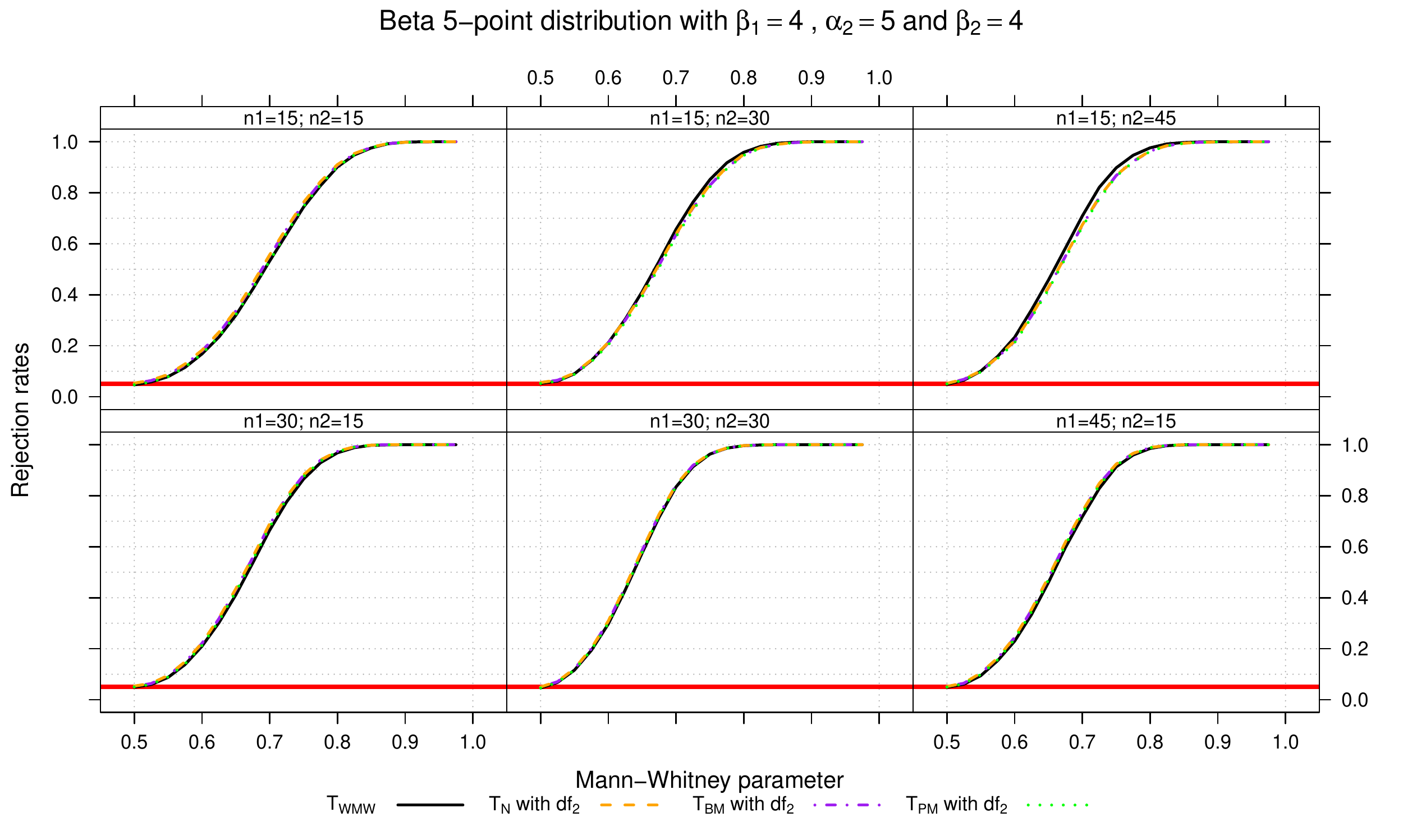}}
\vspace{12pt}
\centerline{\includegraphics[scale=0.65]{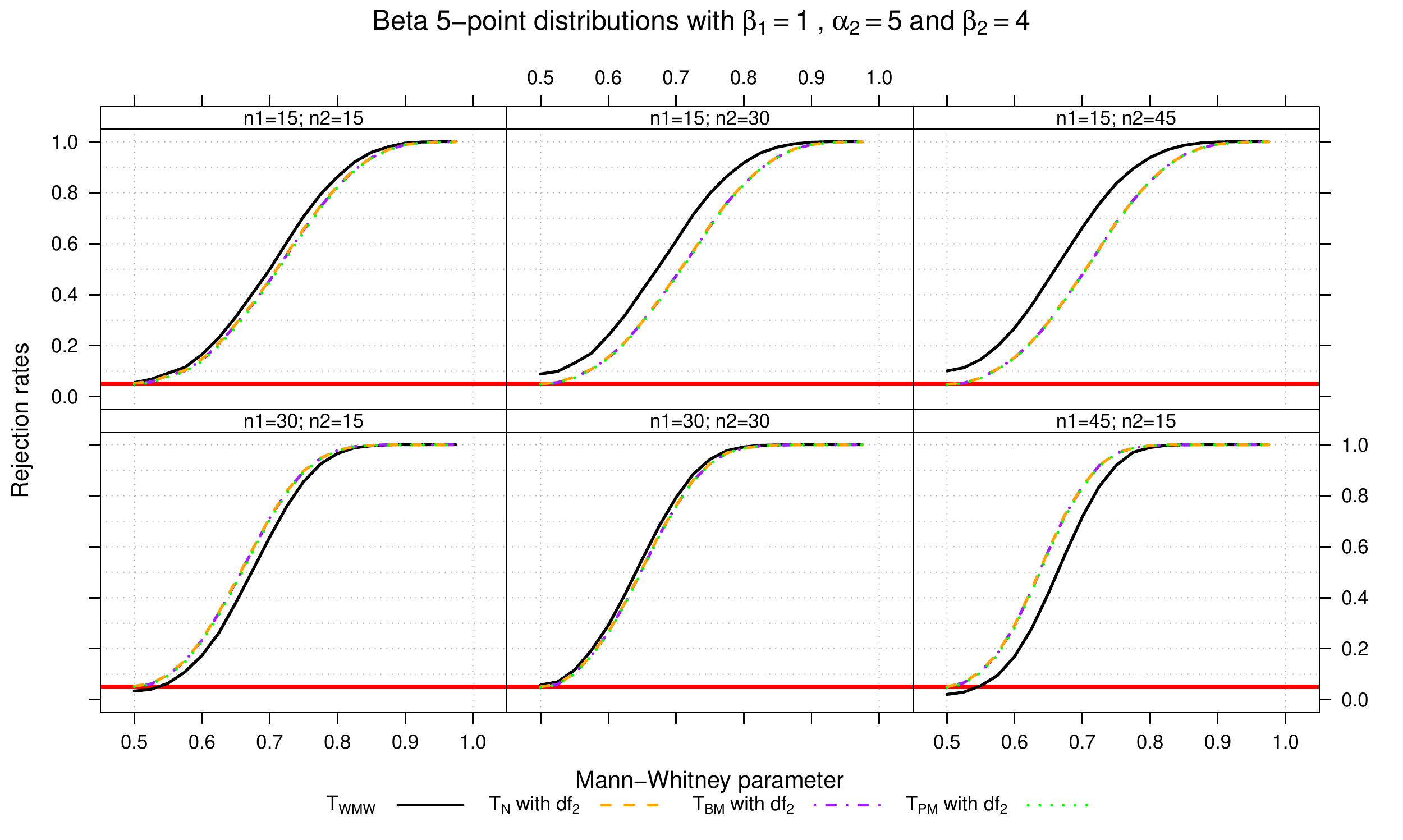}}
\caption{Power graphs for Beta 5-point distributions based on 10\,000 simulation runs}{}
\label{f:ordinal1}
\end{figure}

Now we choose an ordinal 5-point-distribution with categories $\mathcal{C}_{1} < \mathcal{C}_{2} < \mathcal{C}_{3} < \mathcal{C}_{4} < \mathcal{C}_{5}$. As in \cite{tm}, the probabilities of each category occurring are derived through a latent Beta distribution, i.e., we consider Beta random variables $Y_{gi} \overset{iid}{\sim} \mathcal{B}(\alpha_{g},\beta_{g})$, $g=1,2$, $i=1,\dots,n_{g}$, with shape parameters $\alpha_{g},\beta_{g}>0$, such that the expectation and variance of $Y_{gi}$ are given by
\begin{align*}
\mathbb{E}(Y_{gi}) &= \frac{\alpha_{g}}{\alpha_{g}+\beta_{g}},
& 
\mathbb{V}(Y_{gi}) &= \frac{\alpha_{g}\beta_{g}}{ (\alpha_{g}+\beta_{g})^{2} (\alpha_{g} + \beta_{g} +1)}.
\end{align*}
Then we discretise $Y_{gi}$ to the random variable $X_{gi}$, $g=1,2$, $i=1,\dots,n_{g}$, as follows
\begin{align*}
X_{gi}&= \mathcal{C}_{k} \text{ if } Y_{gi} \in \left[0.2(k-1),0.2k\right[ \text{ for } k=1,\dots,5.
\end{align*}
Consequently, the probability mass function of $X_{gi}$ is nothing but
\begin{align*}
\mathbb{P}(X_{gi}=\mathcal{C}_{k})&= \mathbb{P}(0.2(k-1) \leq Y_{gi} < 0.2k) \text{ for } k=1,\dots,5.
\end{align*}
Analogous to the normal setting, we consider a homo- and a heteroskedastic scenario as outlined in Table \ref{t:ordinal1}. As before, $T_{PM}$ \eqref{eq:testpm} with $df_{2}$ degrees of freedom \eqref{eq:df2} best controls the nominal type I error rate. Moreover, the power graphs in Figure \ref{f:ordinal1} based on 10\,000 simulation runs show a pattern similar to the normal scenarios as well.
\begin{table}
\centering
\caption{Type I error rates for the 5-point distributions with latent $F_{1}=\mathcal{B}(\alpha_{1},\beta_{1})$ and $F_{2}=\mathcal{B}(5,4)$ based on 100\,000 replications at a two-sided nominal significance level of $\alpha=0.05$ as regards the test statistics $T_{WMW}$ \eqref{eq:testwmw}; $T_{N}$ \eqref{eq:testunb}, $T_{BM}$ \eqref{eq:testbm}, $T_{PM}$ \eqref{eq:testpm} with degrees of freedom $df_{2}$ \eqref{eq:df2}; $T_{N}^{Logit}$ \eqref{eq:testunblogit}, $T_{BM}^{Logit}$ \eqref{eq:testbmlogit}, $T_{PM}^{Logit}$ \eqref{eq:testpmlogit}}
\label{t:ordinal1}
\begin{tabular}{rrccccccccc}
  \hline
  $n_{1}$ & $n_{2}$ & $\alpha_{1}$ & $\beta_{1}$ & $T_{WMW} $ & $T_{N}$ & $T_{BM}$ & $T_{PM}$ & $T_{N}^{Logit}$ & $T_{BM}^{Logit}$ & $T_{PM}^{Logit}$ \\ 
  \hline
   7 &  7 & 5      & 4 & 0.04611 & 0.05628 & 0.05045 & 0.04129 & 0.03990 & 0.03889 & 0.02723 \\ 
  10 &  7 & 5      & 4 & 0.04761 & 0.05391 & 0.05298 & 0.04431 & 0.04309 & 0.04004 & 0.03127 \\ 
   7 & 10 & 5      & 4 & 0.04769 & 0.05457 & 0.05350 & 0.04426 & 0.04320 & 0.04016 & 0.03184 \\ 
  10 & 10 & 5      & 4 & 0.04832 & 0.05450 & 0.05316 & 0.04798 & 0.04072 & 0.03944 & 0.03263 \\ 
  15 & 15 & 5      & 4 & 0.04875 & 0.05440 & 0.05315 & 0.04910 & 0.04330 & 0.04208 & 0.03741 \\ 
  30 & 15 & 5      & 4 & 0.04814 & 0.05212 & 0.05140 & 0.04825 & 0.04591 & 0.04513 & 0.04194 \\ 
  15 & 30 & 5      & 4 & 0.04902 & 0.05391 & 0.05315 & 0.04990 & 0.04743 & 0.04668 & 0.04344 \\ 
  30 & 30 & 5      & 4 & 0.04857 & 0.05175 & 0.05115 & 0.04875 & 0.04577 & 0.04521 & 0.04266 \\ 
  15 & 45 & 5      & 4 & 0.04923 & 0.05258 & 0.05201 & 0.04961 & 0.04996 & 0.04937 & 0.04675 \\ 
  15 & 60 & 5      & 4 & 0.05026 & 0.05292 & 0.05248 & 0.05041 & 0.05288 & 0.05231 & 0.05004 \\ 
  15 & 75 & 5      & 4 & 0.04959 & 0.05263 & 0.05228 & 0.05055 & 0.05426 & 0.05396 & 0.05188 \\ 
  45 & 15 & 5      & 4 & 0.04898 & 0.05186 & 0.05144 & 0.04909 & 0.04942 & 0.04892 & 0.04610 \\ 
  60 & 15 & 5      & 4 & 0.04868 & 0.05194 & 0.05142 & 0.04940 & 0.05191 & 0.05136 & 0.04899 \\ 
  75 & 15 & 5      & 4 & 0.04856 & 0.05090 & 0.05054 & 0.04906 & 0.05263 & 0.05217 & 0.05033 \\ \hline
   7 &  7 & 1.2071 & 1 & 0.05763 & 0.05126 & 0.04755 & 0.04387 & 0.03552 & 0.03366 & 0.02719 \\ 
  10 &  7 & 1.2071 & 1 & 0.04264 & 0.05281 & 0.05086 & 0.04506 & 0.03508 & 0.03298 & 0.02710 \\ 
   7 & 10 & 1.2071 & 1 & 0.07446 & 0.05213 & 0.05046 & 0.04612 & 0.04219 & 0.03910 & 0.03381 \\ 
  10 & 10 & 1.2071 & 1 & 0.05798 & 0.05432 & 0.05293 & 0.04872 & 0.03863 & 0.03638 & 0.03171 \\ 
  15 & 15 & 1.2071 & 1 & 0.05579 & 0.05146 & 0.05031 & 0.04752 & 0.04035 & 0.03912 & 0.03606 \\ 
  30 & 15 & 1.2071 & 1 & 0.03218 & 0.05345 & 0.05251 & 0.05029 & 0.04528 & 0.04408 & 0.04154 \\ 
  15 & 30 & 1.2071 & 1 & 0.08897 & 0.05188 & 0.05131 & 0.04977 & 0.04529 & 0.04446 & 0.04248 \\ 
  30 & 30 & 1.2071 & 1 & 0.05827 & 0.05120 & 0.05065 & 0.04917 & 0.04508 & 0.04436 & 0.04264 \\ 
  15 & 45 & 1.2071 & 1 & 0.10304 & 0.05027 & 0.04976 & 0.04876 & 0.04582 & 0.04528 & 0.04399 \\ 
  15 & 60 & 1.2071 & 1 & 0.11529 & 0.05152 & 0.05116 & 0.05029 & 0.04857 & 0.04813 & 0.04702 \\ 
  15 & 75 & 1.2071 & 1 & 0.12079 & 0.05096 & 0.05064 & 0.05008 & 0.04851 & 0.04815 & 0.04739 \\ 
  45 & 15 & 1.2071 & 1 & 0.02028 & 0.05296 & 0.05196 & 0.04966 & 0.04718 & 0.04624 & 0.04375 \\ 
  60 & 15 & 1.2071 & 1 & 0.01493 & 0.05145 & 0.05075 & 0.04874 & 0.04806 & 0.04725 & 0.04507 \\ 
  75 & 15 & 1.2071 & 1 & 0.01190 & 0.05180 & 0.05111 & 0.04917 & 0.05009 & 0.04916 & 0.04721 \\ 
   \hline
\end{tabular}
\end{table}

\subsection{Permutation tests}\label{sec:simperm}
As for the type I error rate of the permutation tests \eqref{eq:permtests} based on the approach proposed by \cite{asen}, we examine some of the scenarios as set out in Tables \ref{t:normal1} and \ref{t:ordinal1} using $n_{perm}=10\,000$ random permutations per simulation run. Bearing in mind that 10\,000 simulation runs for each setting give rise to a Monte Carlo error of $0.002$ for a two-sided nominal significance level of $\alpha=0.05$, it still seems fair to us to observe in light of the results depicted in Tables \ref{t:perm1} and \ref{t:perm2} that Perme and Manevski's original $T_{PM}$ \eqref{eq:testpm} with degrees of freedom $df_{2}$ \eqref{eq:df2} better maintains the nominal significance level on the whole.

More results as regards similar settings as in \cite{asen}, i.e., exponential and binomial distributions, as well as some power scenarios for normal and 5-point-distributions are provided in the appendix.

 


\begin{table}
\centering
\caption{Type I error rates for normal distributions $F_{1}=\mathcal{N}(0,\sigma_{1}^{2})$ and $F_{2}=\mathcal{N}(0,\sigma_{2}^{2})$ at a two-sided nominal significance level of $\alpha=0.05$ for the studentised permutation tests as given in \eqref{eq:permtests} based on 10\,000 random permutations for each of the 10\,000 replications}
\label{t:perm1}
\begin{tabular}{rrcccccccc}
  \hline
  $n_{1}$ & $n_{2}$ & $\sigma_{1}$ & $\sigma_{1}$ & $\widetilde{T}_{N}$ & $\widetilde{T}_{BM}$ & $\widetilde{T}_{PM}$ & $\widetilde{T}_{N}^{Logit}$ & $\widetilde{T}_{BM}^{Logit}$ & $\widetilde{T}_{PM}^{Logit}$ \\ 
  \hline
   7 &  7 & 1 & 1 & 0.0486 & 0.0492 & 0.0492 & 0.0492 & 0.0482 & 0.0485 \\ 
   7 & 10 & 1 & 1 & 0.0473 & 0.0484 & 0.0486 & 0.0471 & 0.0485 & 0.0484 \\ 
  10 &  7 & 1 & 1 & 0.0489 & 0.0501 & 0.0502 & 0.0501 & 0.0509 & 0.0498 \\ 
  10 & 10 & 1 & 1 & 0.0505 & 0.0507 & 0.0507 & 0.0510 & 0.0505 & 0.0504 \\ 
  15 & 15 & 1 & 1 & 0.0507 & 0.0506 & 0.0507 & 0.0500 & 0.0503 & 0.0505 \\ 
  15 & 30 & 1 & 1 & 0.0525 & 0.0525 & 0.0523 & 0.0526 & 0.0526 & 0.0526 \\ 
  30 & 15 & 1 & 1 & 0.0494 & 0.0494 & 0.0495 & 0.0496 & 0.0496 & 0.0494 \\ 
  30 & 30 & 1 & 1 & 0.0526 & 0.0525 & 0.0525 & 0.0523 & 0.0523 & 0.0525 \\ 
  15 & 45 & 1 & 1 & 0.0529 & 0.0528 & 0.0529 & 0.0520 & 0.0525 & 0.0525 \\ 
  45 & 15 & 1 & 1 & 0.0509 & 0.0510 & 0.0507 & 0.0508 & 0.0510 & 0.0512 \\ \hline
   7 &  7 & 1 & 3 & 0.0477 & 0.0546 & 0.0553 & 0.0381 & 0.0388 & 0.0396 \\ 
   7 & 10 & 1 & 3 & 0.0421 & 0.0439 & 0.0466 & 0.0415 & 0.0435 & 0.0445 \\ 
  10 &  7 & 1 & 3 & 0.0602 & 0.0638 & 0.0651 & 0.0410 & 0.0414 & 0.0418 \\ 
  10 & 10 & 1 & 3 & 0.0563 & 0.0579 & 0.0603 & 0.0494 & 0.0514 & 0.0533 \\ 
  15 & 15 & 1 & 3 & 0.0520 & 0.0529 & 0.0538 & 0.0468 & 0.0485 & 0.0500 \\ 
  15 & 30 & 1 & 3 & 0.0436 & 0.0445 & 0.0448 & 0.0474 & 0.0480 & 0.0485 \\ 
  30 & 15 & 1 & 3 & 0.0548 & 0.0561 & 0.0567 & 0.0436 & 0.0453 & 0.0469 \\ 
  30 & 30 & 1 & 3 & 0.0521 & 0.0531 & 0.0535 & 0.0499 & 0.0512 & 0.0516 \\ 
  15 & 45 & 1 & 3 & 0.0430 & 0.0431 & 0.0432 & 0.0484 & 0.0487 & 0.0490 \\ 
  45 & 15 & 1 & 3 & 0.0542 & 0.0550 & 0.0558 & 0.0403 & 0.0412 & 0.0420 \\
   \hline
\end{tabular}
\end{table}

\begin{table}
\centering
\caption{Type I error rates for the 5-point distributions with latent $F_{1}=\mathcal{B}(\alpha_{1},\beta_{1})$ and $F_{2}=\mathcal{B}(5,4)$ at a two-sided nominal significance level of $\alpha=0.05$ for the studentised permutation tests as given in \eqref{eq:permtests} based on 10\,000 random permutations for each of the 10\,000 replications}
\label{t:perm2}
\begin{tabular}{rrcccccccc}
  \hline
  $n_{1}$ & $n_{2}$ & $\alpha_{1}$ & $\beta_{1}$ & $\widetilde{T}_{N}$ & $\widetilde{T}_{BM}$ & $\widetilde{T}_{PM}$ & $\widetilde{T}_{N}^{Logit}$ & $\widetilde{T}_{BM}^{Logit}$ & $\widetilde{T}_{PM}^{Logit}$ \\ 
  \hline
   7 &  7 & 5      & 4 & 0.0237 & 0.0235 & 0.0235 & 0.0225 & 0.0223 & 0.0224 \\ 
   7 & 10 & 5      & 4 & 0.0306 & 0.0307 & 0.0309 & 0.0316 & 0.0309 & 0.0313 \\ 
  10 &  7 & 5      & 4 & 0.0275 & 0.0276 & 0.0279 & 0.0271 & 0.0270 & 0.0273 \\ 
  10 & 10 & 5      & 4 & 0.0342 & 0.0337 & 0.0338 & 0.0336 & 0.0338 & 0.0341 \\ 
  15 & 15 & 5      & 4 & 0.0421 & 0.0421 & 0.0420 & 0.0418 & 0.0419 & 0.0419 \\ 
  15 & 30 & 5      & 4 & 0.0464 & 0.0464 & 0.0467 & 0.0469 & 0.0468 & 0.0463 \\ 
  30 & 15 & 5      & 4 & 0.0407 & 0.0407 & 0.0406 & 0.0412 & 0.0412 & 0.0410 \\ 
  30 & 30 & 5      & 4 & 0.0490 & 0.0489 & 0.0489 & 0.0487 & 0.0488 & 0.0490 \\ 
  15 & 45 & 5      & 4 & 0.0481 & 0.0479 & 0.0478 & 0.0483 & 0.0483 & 0.0483 \\ 
  45 & 15 & 5      & 4 & 0.0464 & 0.0464 & 0.0465 & 0.0466 & 0.0464 & 0.0462 \\ \hline
   7 &  7 & 1.2071 & 1 & 0.0365 & 0.0366 & 0.0385 & 0.0319 & 0.0321 & 0.0329 \\ 
   7 & 10 & 1.2071 & 1 & 0.0474 & 0.0475 & 0.0483 & 0.0413 & 0.0414 & 0.0422 \\ 
  10 &  7 & 1.2071 & 1 & 0.0388 & 0.0401 & 0.0405 & 0.0391 & 0.0398 & 0.0421 \\ 
  10 & 10 & 1.2071 & 1 & 0.0486 & 0.0487 & 0.0492 & 0.0450 & 0.0450 & 0.0468 \\ 
  15 & 15 & 1.2071 & 1 & 0.0531 & 0.0535 & 0.0543 & 0.0502 & 0.0507 & 0.0521 \\ 
  15 & 30 & 1.2071 & 1 & 0.0575 & 0.0576 & 0.0586 & 0.0522 & 0.0525 & 0.0541 \\ 
  30 & 15 & 1.2071 & 1 & 0.0501 & 0.0501 & 0.0504 & 0.0524 & 0.0528 & 0.0528 \\ 
  30 & 30 & 1.2071 & 1 & 0.0566 & 0.0570 & 0.0571 & 0.0554 & 0.0559 & 0.0562 \\ 
  15 & 45 & 1.2071 & 1 & 0.0534 & 0.0536 & 0.0544 & 0.0460 & 0.0462 & 0.0468 \\ 
  45 & 15 & 1.2071 & 1 & 0.0440 & 0.0442 & 0.0440 & 0.0475 & 0.0477 & 0.0476 \\ 
   \hline
\end{tabular}
\end{table}
\newpage
\section{Discussion} %
\label{s:discuss}
In this manuscript, we reviewed different variance estimators for the Mann-Whitney parameter and, more generally, different ways of how to approximate its sampling distribution in small samples. To stick to the unbiased variance estimator appears to be somewhat ill-advised. Indeed, in almost all scenarios Perme and Manevski's $T_{PM}$ \eqref{eq:testpm} with degrees of freedom $df_{2}$ \eqref{eq:df2} seems preferable in terms of controlling the type I error rate. Of course, Perme and Manevski's variance estimator is not unbiased and the particular choice of degrees of freedom lack a sound theoretical justification, even if they are consistent.

In heteroskedastic settings, the Wilcoxon-Mann-Whitney test $T_{WMW}$ \eqref{eq:testwmw} performs poorly, particularly in case of unequal sample sizes, a pattern which also very slightly emerges when using the permutation approach \eqref{eq:permtests} by \cite{asen}.

As far as group sequential models for the Mann-Whitney parameter are concerned, it would be interesting to examine whether the test statistics, in particular $T_{PM}$ \eqref{eq:testpm} with $df_{2}$ \eqref{eq:df2}, would equally well maintain one-sided nominal significance levels close to zero and up to 0.025. With that caveat in mind, we would further like to point out that $T_{PM}$ \eqref{eq:testpm} with $df_{2}$ \eqref{eq:df2} works very well for sample sizes $\min(n_{1},n_{2})\geq 15$ for a range of different distributions and tends to be somewhat conservative in smaller samples.

\newpage
\bibliographystyle{abbrvnat}
\bibliography{references}  






\newpage
\begin{center}
{\LARGE\bf Appendix}
\end{center}

This appendix consists of two main parts. First, we discuss the different variance estimators, translate them into our notation and briefly outline our approach to the Box-type degrees of freedom as regards the unbiased variance estimator. Second, we focus on the simulations, dealing with exception handling and providing more detailed results and results from settings not considered in the main manuscript.

\section*{Part I -- Variance estimators and degrees of freedom}
First we extend our notation of nonparametric theory, then we discuss the variance estimand and its different estimators. Lastly, we briefly explain how we arrived at the Box-type formulas of the degrees of freedom.

\subsection*{General notation}

Let $X$ denote a random variable representing ordered categorical or real data, defined on some probability space $(\Omega, \mathcal{A}, \mathbb{P})$ . Then we define for each possible value $x$ the following versions of distribution functions
\begin{align*}
F^{+}(x) &= \mathbb{P}(X\leq x), \\
F^{-}(x) &= \mathbb{P}(X< x), \\
F(x) &= \mathbb{P}(X < x) + \nicefrac{1}{2}\cdot\mathbb{P}(X=x) 
.
\end{align*}
Now suppose we have a sample of observations $X_{1},\dots,X_{n}\overset{iid}{\sim} F(x)$. The empirical distribution functions then take the form
\begin{align*}
\widehat{F}^{+}(x) &= \frac{1}{n}\sum_{i=1}^{n}c^{+}(x,X_{i}), \quad  c^{+}(x,X_{i})= 
\begin{cases}
 0 & \text{ if } x < X_{i} \\
 1 & \text{ if } x \geq X_{i}
\end{cases}
,
\\
\widehat{F}^{-}(x) &= \frac{1}{n}\sum_{i=1}^{n}c^{-}(x,X_{i}), \quad  c^{-}(x,X_{i})= 
\begin{cases}
 0 & \text{ if } x  \leq X_{i} \\
 1 & \text{ if } x > X_{i}
\end{cases}
,
\\
\widehat{F}(x) &= \frac{1}{n}\sum_{i=1}^{n} c(x,X_{i}), \quad c(x,X_{i})= 
\begin{cases}
 0 & \text{ if }  x  < X_{i} \\
\nicefrac{1}{2} & \text{ if } x =X_{i} \\
 1 & \text{ if } x > X_{i}
\end{cases}
.
\end{align*}
Now we define the nonparametric relative effect and give some of its properties, which we will use in the derivations later on.

Let $X_{ij}\sim F_{i}(x)$, $i=1,2$, $j=1,\dots,n_{i}$ be independent random variables. Then the nonparametric relative effect is given by
\begin{align*}
p = \mathbb{P}(X_{1j}<X_{2j'}) + \nicefrac{1}{2}\cdot\mathbb{P}(X_{1j}=X_{2j'}) = \int F_{1}dF_{2}
\end{align*}
for all $j=1,\dots,n_{1}$ and $j'=1,\dots,n_{2}$. In particular, if $F=F_{1}=F_{2}$, then $\int FdF=\nicefrac{1}{2}$.

Let $X_{ij}\sim F_{i}(x)$, $i=1,\dots,d$, $j=1,\dots,n_{i}$ be independent real-valued random variables. Then for all $i,i'=1,\dots,d$ as well as $j=1,\dots,n_{i}$ and $j'=1,\dots,n_{i'}$ it holds
\begin{align*}
\mathbb{E}(\widehat{F}_{i}(x))=\mathbb{E}( c(x,X_{ij})) &= F_{i}(x), \\
\mathbb{E}(\widehat{F}_{i}(X_{i'j'}))=\mathbb{E}( c(X_{i'j'},X_{ij})) &= \int F_{i}dF_{i'}.
\end{align*}

We will also be using the survival functions
\begin{align*}
S^{+}(x) &= \mathbb{P}(X\geq x), \\
S^{-}(x) &= \mathbb{P}(X> x), \\
S(x) &= \mathbb{P}(X > x) + \nicefrac{1}{2}\cdot\mathbb{P}(X=x),
\end{align*}
with their empirical counterparts $\widehat{S}^{+}(x)$, $\widehat{S}^{-}(x)$, and $\widehat{S}(x)$ defined accordingly. Note that
\begin{align*}
\mathbb{E}(\widehat{S}_{i}(x))=\mathbb{E}( c(X_{ij},x)) &= S_{i}(x), \\
\mathbb{E}(\widehat{S}_{i}(X_{i'j'}))=\mathbb{E}( c(X_{ij},X_{i'j'})) &= \int S_{i}dF_{i'}.
\end{align*}

\subsection*{Variance estimand}
In our notation, we have for the general case of arbitrary $F_{1}$ and $F_{2}$,
\begin{align*}
\sigma^{2}_{N}&=\mathbb{V}(\widehat{p}) =
\frac{\tau_{0}+(n_{2}-1)\tau_{1}+(n_{1}-1)\tau_{2}-(n_{1}+n_{2}-1)p^{2}}{n_{1}n_{2}},
\end{align*}
where 
\begin{align*}
\tau_{0} &= \int F_{1}dF_{2} - \nicefrac{1}{4}\cdot \underbrace{\int (F_{1}^{+}-F_{1}^{-})dF_{2}}_{=\mathbb{P}(X_{1}=X_{2})\eqqcolon\beta}=p - \nicefrac{1}{4}\cdot \beta,
\\
\tau_{1} &= \int S_{2}^{2}dF_{1} = \int (1-F_{2})^{2}dF_{1} ,\\
\tau_{2} &= \int F_{1}^{2}dF_{2}. 
\end{align*}
If both distributions coincide, i.e., $F_{1}=F_{2}$, and are continuous, then it holds $\tau_{0}=\nicefrac{1}{2}$, $\tau_{1}=\tau_{2}=\nicefrac{1}{3}$, yielding $\sigma^{2}_{N}=\frac{n_{1}+n_{2}+1}{12n_{1}n_{2}}$.

In the unpublished preprint of \cite{brunnerpreprint}, we find formula (1.9), i.e.,
\begin{align*}
\sigma^{2}_{N}&=\frac{(n_{2}-1)\sigma_{1}^{2} + (n_{1}-1) \sigma_{2}^{2} + p(1-p) - \nicefrac{1}{4}\cdot\beta }{n_{1}n_{2}},
\end{align*}
where 
\begin{align*}
\sigma_{1}^{2}&= \mathbb{V}(F_{2}(X_{11}))= \int \{F_{2}-(1-p)\}^{2}dF_{1} = \int (S_{2}-p)^{2}dF_{1} = \int S_{2}^{2}dF_{1} - 2p\int S_{2}dF_{1} + p^{2}
= \tau_{1} - p^{2},\\
\sigma_{2}^{2}&=\mathbb{V}(F_{1}(X_{21}))=
\int (F_{1}-p)^{2}dF_{2} = \int F_{1}^{2}dF_{2} - 2p\int F_{1}dF_{2} + p^{2}
= \tau_{2} - p^{2}.
\end{align*}
So it should be evident that both definitions of $\sigma_{N}^{2}$ are equivalent.

Bamber's [\citeyear{bamber}] definition of $\sigma_{N}^{2}$, which he calls $\sigma_{a}^{2}$, is equivalent as well. Assuming that $X$ refers to sample 1 and $Y$ to sample $2$, it holds $B_{YYX} = 4 \tau_{1} - 4p + 1$ as well as $B_{XXY}=4 \tau_{2} - 4p + 1$. With $F_{X}=F_{1}$ and $F_{Y}=F_{2}$ and taking $\mathbb{P}(Y_{1},Y_{2}<X)$ to mean $\mathbb{P}(Y_{1} < X, Y_{2} < X)$, we can deduce
\begin{align*}
B_{YYX}
&=
\mathbb{P}(Y_{1} < X, Y_{2} < X) + \mathbb{P}(X < Y_{1}, X < Y_{2}) - 2 \mathbb{P}( Y_{1} < X < Y_{2})
\\
&=
\int \mathbb{P}(Y_{1} < x,Y_{2} < x)dF_{X}(x) 
+
\int \mathbb{P}(Y_{1} > x,Y_{2} > x)dF_{X}(x) 
-
2
\int \mathbb{P}(Y_{1} < x,Y_{2} > x)dF_{X}(x) 
\\
&=
\int \mathbb{P}(Y_{1} < x)\mathbb{P}(Y_{2} < x)dF_{X}(x) 
+
\int \mathbb{P}(Y_{1} > x)\mathbb{P}(Y_{2} > x)dF_{X}(x) 
-
2
\int \mathbb{P}(Y_{1} < x)\mathbb{P}(Y_{2} > x)dF_{X}(x) 
\\
&=
\int \{F_{Y}^{-}(x)\}^{2}dF_{X}(x)
+
\int \{S_{Y}^{-}(x)\}^{2}dF_{X}(x)
-
2\int \{F_{Y}^{-}(x)S_{Y}^{-}(x)\}dF_{X}(x)
\\
&=
\int \{S_{Y}^{-}(x)-F_{Y}^{-}(x)\}^{2}dF_{X}(x)
\\
&=
\int \{S_{Y}^{-}(x)+\nicefrac{1}{2}\cdot\mathbb{P}(Y=x)-F_{Y}^{-}(x)-\nicefrac{1}{2}\cdot\mathbb{P}(Y=x)\}^{2}dF_{X}(x)
\\
&=
\int \{S_{Y}(x)-F_{Y}(x)\}^{2}dF_{X}(x)
\\
&=
\int \{2S_{Y}(x)-1\}^{2}dF_{X}(x)
\\
&
=
4\int S_{Y}^{2}dF_{X} - 4 \int S_{Y}dF_{X} +1
\\
&=
4\int S_{2}^{2}dF_{1} - 4 \int S_{2}dF_{1} +1
\\
&=
4\tau_{1}-4p +1.
\end{align*}
By the same token, it holds 
\begin{align*}
B_{XXY}
&=
\mathbb{P}(X_{1} < Y, X_{2} < Y) + \mathbb{P}(Y < X_{1}, Y < X_{2}) - 2 \mathbb{P}( X_{1} < Y < X_{2})
\\
&=
\int \mathbb{P}(X_{1} < y,X_{2} < y)dF_{Y}(y) 
+
\int \mathbb{P}(X_{1} > y,X_{2} > y)dF_{Y}(y) 
-
2
\int \mathbb{P}(X_{1} < y,X_{2} > y)dF_{Y}(y) 
\\
&=
\int \mathbb{P}(X_{1} < y)\mathbb{P}(X_{2} < y)dF_{Y}(y) 
+
\int \mathbb{P}(X_{1} > y)\mathbb{P}(X_{2} > y)dF_{Y}(y) 
-
2
\int \mathbb{P}(X_{1} < y)\mathbb{P}(X_{2} > y)dF_{Y}(y) 
\\
&=
\int \{F_{X}^{-}(y)\}^{2}dF_{Y}(y)
+
\int \{S_{X}^{-}(y)\}^{2}dF_{Y}(y)
-
2\int \{F_{X}^{-}(y)S_{X}^{-}(y)\}dF_{Y}(y)
\\
&=
\int \{F_{X}^{-}(y)-S_{X}^{-}(y)\}^{2}dF_{Y}(y)
\\
&=
\int \{S_{X}^{-}(y)+\nicefrac{1}{2}\cdot\mathbb{P}(X=y)-F_{X}^{-}(y)-\nicefrac{1}{2}\cdot\mathbb{P}(X=y)\}^{2}dF_{Y}(y)
\\
&=
\int \{F_{X}(y)-S_{X}(y)\}^{2}dF_{Y}(y)
\\
&=
\int \{2F_{X}(y)-1\}^{2}dF_{Y}(y)
\\
&
=
4\int F_{X}^{2}dF_{Y} - 4 \int F_{X}dF_{Y} +1
\\
&=
4\int F_{1}^{2}dF_{2} - 4 \int F_{1}dF_{2} +1
\\
&=
4\tau_{2}-4p +1.
\end{align*}
Note that in our notation $N_{X}=n_{1}$ and $N_{Y}=n_{2}$ as well as $A=p$, so that
\begin{align*}
\sigma_{a}^{2}
&=
\frac{1}{4N_{X}N_{Y}}
\left\{
\mathbb{P}(X\neq Y)
+
(N_{X}-1)B_{XXY}
+
(N_{Y}-1)B_{YYX}
-4(N_{X}+N_{Y}-1)(A-\nicefrac{1}{2})^{2}\right\}
\\
&=
\frac{1}{4n_{1}n_{2}}
\left\{
1-\beta
+
(n_{1}-1)(4\tau_{2}-4p +1)
+
(n_{2}-1)(4\tau_{1}-4p +1)
-4(n_{1}+n_{2}-1)(p-\nicefrac{1}{2})^{2}\right\}
\\
&=
\frac{(n_{1}-1)\tau_{2}+(n_{2}-1)\tau_{1}-(n_{1}+n_{2}-1)p^{2}}{n_{1}n_{2}}
\\
&\phantom{xxxxxxxxxxxxxxx}
+
\frac{1-\beta - (n_{1}+n_{2}-2)(4p-1)+4(n_{1}+n_{2}-1)(p-\nicefrac{1}{4})}{4n_{1}n_{2}}
\\
&=
\frac{(n_{1}-1)\tau_{2}+(n_{2}-1)\tau_{1}-(n_{1}+n_{2}-1)p^{2}}{n_{1}n_{2}}
+
\frac{4p-\beta}{4n_{1}n_{2}}
\\
&=
\frac{\tau_{0}+(n_{1}-1)\tau_{2}+(n_{2}-1)\tau_{1}-(n_{1}+n_{2}-1)p^{2}}{n_{1}n_{2}}
=
\sigma^{2}_{N}.
\end{align*}

As for Perme and Manevski [\citeyear{pm}], they define
\begin{align*}
\mathbb{V}(\widehat{\theta})
&=
\frac{\theta(1-\theta)}{mn}+\frac{n-1}{nm}\mathbb{V}(S_{Y}(X)) + \frac{m-1}{mn}\mathbb{V}(S_{X}(Y)),
\end{align*}
which in our notation should read as
\begin{align*}
\mathbb{V}(\widehat{p})
&=
\frac{p(1-p)}{n_{1}n_{2}}+\frac{n_{2}-1}{n_{1}n_{2}}\mathbb{V}(S_{2}^{-}(X_{11})) + \frac{n_{2}-1}{n_{1}n_{2}}\mathbb{V}(S_{1}^{-}(X_{21})),
\end{align*}
However, this formula assumes that $F_{X}$ and $F_{Y}$ are both continuous. Perme and Manevski say as much in the supplementary material, i.e., ``For better clarity of all the derivations, we shall assume that both $F_{X}$ and $F_{Y}$ are continuous (the extension of formulae to the case of ties is then straightforward)''. In the main paper they state ``This work focuses on continuous random variables $X$ and $Y$. In practice, the data may often be documented on a discrete scale and thus ties can occur. Therefore, we shall always extend the definition to the case of ties.'' Nonetheless, Perme and Manevski seemingly do not explicitly set out what they consider to be the variance estimand in case of ties.
%
In any event, in case of continuity we have
\begin{align*}
\mathbb{V}(S_{2}^{-}(X_{11}))
&=
\mathbb{V}(S_{2}(X_{11}))
=
\mathbb{V}(1-F_{2}(X_{11}))
=
\mathbb{V}(F_{2}(X_{11}))
=
\sigma_{1}^{2},
\\
\mathbb{V}(S_{1}^{-}(X_{21}))
&=
\mathbb{V}(S_{1}(X_{21}))
=
\mathbb{V}(1-F_{1}(X_{21}))
=
\mathbb{V}(F_{1}(X_{21}))
=
\sigma_{2}^{2},
\\
\beta &= 0.
\end{align*}
Thus Perme and Manevski's formula is equal to our definition of $\sigma^{2}_{N}$ so long as both distributions $F_{1}$ and $F_{2}$ are continuous.

\section*{Variance estimation}

Our plug-in estimators for $\tau_{0}$, $\tau_{1}$, $\tau_{2}$, $p^{2}$ are given by
\begin{align*}
\widehat{\tau}_{0}
&= \widehat{p} -\nicefrac{1}{4}\cdot\widehat{\beta}, \text{ with } \, \widehat{\beta}=
\frac{1}{n_{1}}\frac{1}{n_{2}}\sum_{j=1}^{n_{2}}\sum_{i=1}^{n_{1}} \mathbb{I}(X_{2j}=X_{1i}),
\quad 
 \mathbb{I}(X_{2j}=X_{1i}) =
\begin{cases}
1 & \text{ if } X_{2j}=X_{1i} \\
0 & \text{ if } X_{2j}\neq X_{1i} 
\end{cases},
\\
\widehat{\tau}_{1} &=
\int\widehat{S}_{2}^{2}d\widehat{F}_{1}=
\frac{1}{n_{1}}\sum_{i=1}^{n_{1}} \{\widehat{S}_{2}(X_{1i})\}^{2}
=
\frac{1}{n_{1}}\sum_{i=1}^{n_{1}} \{ \frac{1}{n_{2}}\sum_{j=1}^{n_{2}} c(X_{2j},X_{1i}) \}^{2},
\\
\widehat{\tau}_{2} &=
\int\widehat{F}_{1}^{2}d\widehat{F}_{2}=
\frac{1}{n_{2}}\sum_{j=1}^{n_{2}} \{\widehat{F}_{1}(X_{2j})\}^{2}
=
\frac{1}{n_{2}}\sum_{j=1}^{n_{2}} \{ \frac{1}{n_{1}}\sum_{i=1}^{n_{1}} c(X_{2j},X_{1i}) \}^{2},
\\
\widehat{p}^{2}&=(\int\widehat{F}_{1}d\widehat{F}_{2})^{2}=\left(\frac{1}{n_{1}}\frac{1}{n_{2}}\sum_{j=1}^{n_{2}}\sum_{i=1}^{n_{1}}c(X_{2j},X_{1i})\right)^{2}.
\end{align*}
It can readily be seen that $\mathbb{E}(\widehat{\tau}_{0})=\tau_{0}$ since $\widehat{p}$ and $\widehat{\beta}$ are unbiased.
As regards $\widehat{\tau}_{1}$ and $\widehat{\tau}_{2}$, we find
\begin{align*}
\mathbb{E}( \widehat{\tau}_{1})
&=
\frac{1}{n_{1}}\sum_{i=1}^{n_{1}} \{ \frac{1}{n_{2}}\frac{1}{n_{2}}\sum_{j=1}^{n_{2}}  \sum_{j'=1}^{n_{2}} \mathbb{E} ( c(X_{2j},X_{1i}) c(X_{2j'},X_{1i}) ) \}
\\
&=
\frac{1}{n_{1}}\frac{1}{n_{2}}\frac{1}{n_{2}}\sum_{i=1}^{n_{1}}  \sum_{j=1}^{n_{2}}  \sum_{\substack{j'=1\\j'\neq j}}^{n_{2}} \mathbb{E} ( c(X_{2j},X_{1i}) c(X_{2j'},X_{1i}) ) 
+ \frac{1}{n_{1}}\frac{1}{n_{2}}\frac{1}{n_{2}}\sum_{i=1}^{n_{1}}  \sum_{j=1}^{n_{2}} \mathbb{E} ( c(X_{2j},X_{1i}) c(X_{2j},X_{1i}) ) 
\\
&= \frac{n_{2}-1}{n_{2}}\int S_{2}^{2}dF_{1} + \frac{1}{n_{2}}\mathbb{E}(c(X_{21},X_{11})^{2})
= \frac{n_{2}-1}{n_{2}}\tau_{1} + \frac{1}{n_{2}}\tau_{0}
.
\end{align*}
In a similar vein, it follows
\begin{align*}
\mathbb{E}( \widehat{\tau}_{2} )
&=
\frac{1}{n_{2}}\sum_{j=1}^{n_{2}} \{ \frac{1}{n_{1}}\frac{1}{n_{1}}\sum_{i=1}^{n_{1}} \sum_{i'=1}^{n_{1}} \mathbb{E} ( c(X_{2j},X_{1i}) c(X_{2j},X_{1i'}) ) \}
\\
&=
\frac{1}{n_{2}}\frac{1}{n_{1}}\frac{1}{n_{1}}\sum_{j=1}^{n_{2}}  \sum_{i=1}^{n_{1}} \sum_{\substack{i'=1\\i' \neq i}}^{n_{1}} \mathbb{E} ( c(X_{2j},X_{1i}) c(X_{2j},X_{1i'}) ) 
+ \frac{1}{n_{2}}\frac{1}{n_{1}}\frac{1}{n_{1}}\sum_{j=1}^{n_{2}}  \sum_{i=1}^{n_{1}} \mathbb{E} ( c(X_{2j},X_{1i}) c(X_{2j},X_{1i}) ) 
\\
&= \frac{n_{1}-1}{n_{1}}\int F_{1}^{2}dF_{2} + \frac{1}{n_{1}}\mathbb{E}(c(X_{21},X_{11})^{2})
= \frac{n_{1}-1}{n_{1}}\tau_{2} + \frac{1}{n_{1}}\tau_{0}
.
\end{align*}
As for $\widehat{p}^{2}$, we now look to
\begin{align*}
\mathbb{E}( \widehat{p}^{2} )
&=
\mathbb{E}( \frac{1}{n_{1}}\frac{1}{n_{2}}
\sum_{j=1}^{n_{2}}\sum_{i=1}^{n_{1}}c(X_{2j},X_{1i})\frac{1}{n_{1}}\frac{1}{n_{2}}
\sum_{j'=1}^{n_{2}}\sum_{i'=1}^{n_{1}} c(X_{2j'},X_{1i'})  )
\\
&=
\frac{1}{n_{1}}\frac{1}{n_{2}}
\frac{1}{n_{1}}\frac{1}{n_{2}}
\sum_{j=1}^{n_{2}}\sum_{i=1}^{n_{1}}
\sum_{j'=1}^{n_{2}}\sum_{i'=1}^{n_{1}} \mathbb{E}(\underbrace{c(X_{2j},X_{1i})}_{\eqqcolon \zeta_{ij}} \underbrace{c(X_{2j'},X_{1i'})}_{\eqqcolon \zeta_{i'j'}} )
\end{align*}
Looking at the quadruple sum produces
\begin{align*}
&\sum_{j'=1}^{n_{2}}\sum_{i'=1}^{n_{1}}
\sum_{j=1}^{n_{2}}\sum_{i=1}^{n_{1}}
\mathbb{E}(\zeta_{ij}\zeta_{i'j'})
\\
&\hspace{10pt}=
\sum_{j'=1}^{n_{2}}\sum_{i'=1}^{n_{1}}
\sum_{\substack{j=1\\j\neq j'}}^{n_{2}}\sum_{\substack{i=1\\i\neq i'}}^{n_{1}}
\mathbb{E}(\zeta_{ij}\zeta_{i'j'})
+
\sum_{j'=1}^{n_{2}}\sum_{i'=1}^{n_{1}}
\sum_{\substack{j=1\\j\neq j'}}^{n_{2}}
\mathbb{E}(\zeta_{i'j}\zeta_{i'j'})
+
\sum_{j'=1}^{n_{2}}\sum_{i'=1}^{n_{1}}
\sum_{\substack{i=1\\i\neq i'}}^{n_{1}}
\mathbb{E}(\zeta_{ij'}\zeta_{i'j'})
+
\sum_{j'=1}^{n_{2}}\sum_{i'=1}^{n_{1}}
\mathbb{E}(\zeta_{i'j'}\zeta_{i'j'}),
\\
&\hspace{10pt}=
n_{1}n_{2}(n_{2}-1)(n_{1}-1) \{\int F_{1}dF_{2} \}^{2} 
+
n_{1}n_{2}(n_{2}-1) \int S_{2}^{2}dF_{1} 
+
n_{1}n_{2}(n_{1}-1) \int F_{1}^{2}dF_{2} 
+
n_{1}n_{2}\mathbb{E}( \zeta_{11}^{2} ).
\end{align*}
Therefore, we have
\begin{align*}
\mathbb{E}( \widehat{p}^{2} )
=
\frac{(n_{2}-1)(n_{1}-1)}{n_{1}n_{2}} p^{2}
+
\frac{n_{2}-1}{n_{1}n_{2}}\tau_{1}
+
\frac{n_{1}-1}{n_{1}n_{2}}\tau_{2}
+
\frac{1}{n_{1}n_{2}}\tau_{0}.
\end{align*}
An unbiased estimator of $\sigma_{N}^{2}$ should then take the form
\begin{align*}
\widehat{\sigma}^{2}_{N}
=
\frac{
n_{2}\widehat{\tau}_{1}
+
n_{1}\widehat{\tau}_{2}
-\widehat{\tau}_{0}
-
(n_{1}+n_{2}-1)\widehat{p}^{2}
}
{(n_{1}-1)(n_{2}-1)}.
\end{align*}
To check the unbiasedness of $\sigma_{N}^{2}$, consider
\begin{align*}
(n_{1}&-1)(n_{2}-1)\mathbb{E}(\widehat{\sigma}_{N}^{2})
\\
&=
n_{2}\mathbb{E}(\widehat{\tau}_{1})
+
n_{1}\mathbb{E}(\widehat{\tau}_{2})
-\mathbb{E}(\widehat{\tau}_{0})
-
(n_{1}+n_{2}-1)\mathbb{E}(\widehat{p}^{2})
\\
&=
(n_{2}-1)\tau_{1}+\tau_{0}
+
(n_{1}-1)\tau_{2}+\tau_{0}
-
\tau_{0}
\\
&\hspace{90pt}
-
\frac{n_{1}+n_{2}-1}{n_{1}n_{2}}
\{(n_{1}-1)(n_{2}-1)p^{2}+(n_{2}-1)\tau_{1}+(n_{1}-1)\tau_{2}+\tau_{0}\}
\\
&=
\left(1 - \frac{n_{1}+n_{2}-1}{n_{1}n_{2}} \right)\left\{
\tau_{0}+
(n_{2}-1)\tau_{1}
+
(n_{1}-1)\tau_{2}\right\}
-
\frac{(n_{1}-1)(n_{2}-1)}{n_{1}n_{2}} (n_{1}+n_{2}-1)p^{2}.
\end{align*}
Now since $\left(1 - \frac{n_{1}+n_{2}-1}{n_{1}n_{2}} \right)=\frac{(n_{1}-1)(n_{2}-1)}{n_{1}n_{2}}$, it follows that
$\mathbb{E}(\widehat{\sigma}_{N}^{2})=\sigma^{2}_{N}$.

\subsubsection*{Brunner form of the unbiased variance estimator}
Now we want to have a closer look at the estimator in (2.39) derived as in the unpublished preprint of \cite{brunnerpreprint},
\begin{align*}
\widehat{\sigma}_{N}^{2}
&=
\frac{1}{n_{1}(n_{1}-1)n_{2}(n_{2}-1)}
\left(
\sum_{i=1}^{2}
\sum_{k=1}^{n_{i}}
\left( R_{ik}-R_{ik}^{(i)} - \left[\bar{R}_{i\bullet}-\frac{n_{i}+1}{2}\right] \right)^{2}
- n_{1}n_{2}\left[\widehat{\theta}(1-\widehat{\theta})-\frac{1}{4}\widehat{\beta} \right] \right),
\end{align*}
where $\widehat{\theta}=\int\widehat{F}_{2}d\widehat{F}_{1}=1-\widehat{p}$ and $\widehat{\beta}=\frac{1}{n_{1}n_{2}}\sum_{j=2}^{n_{2}}\sum_{i=1}^{n_{1}}\mathbb{I}(X_{2j}=X_{1i})$.
First recall the following identities as regards the rank representations,
\begin{align*}
n_{2}\widehat{F}_{2}(X_{1i}) &= R_{1i} - R_{1i}^{(1)}, 
& 1-\widehat{p}=\int\widehat{F}_{2}d\widehat{F}_{1} 
&= \frac{1}{n_{1}}\sum_{i=1}^{n_{1}}\widehat{F}_{2}(X_{1i}) = \frac{1}{n_{2}}( \bar{R}_{1\bullet}-\frac{n_{1}+1}{2}),\\
n_{1}\widehat{F}_{1}(X_{2j}) &= R_{2j} - R_{2j}^{(2)}, 
&
\widehat{p}=\int\widehat{F}_{1}d\widehat{F}_{2} 
&= \frac{1}{n_{2}}\sum_{j=1}^{n_{2}}\widehat{F}_{1}(X_{2j}) = \frac{1}{n_{1}}( \bar{R}_{2\bullet}-\frac{n_{2}+1}{2}),
\end{align*}
where $R_{1i}^{(1)}$ and $R_{2j}^{(2)}$ denote the so-called internal ranks with respect to the first and second sample. That is to say $R_{1i}^{(1)}$ is the mid-rank of $X_{1i}$ among $X_{11},\dots,X_{1n_{1}}$ whereas $R_{2j}^{(2)}$ is the mid-rank of $X_{2j}$ among $X_{21},\dots,X_{2n_{2}}$.

Together with the equality 
\begin{align*}
\int \widehat{F}_{2}^{2}d\widehat{F}_{1}- (\int \widehat{F}_{2}d\widehat{F}_{1})^{2}
&=
\int (1-\widehat{S}_{2})^{2}d\widehat{F}_{1}- (1-\int \widehat{S}_{2}d\widehat{F}_{1})^{2}
\\
&=
1+\int\widehat{S}_{2}^{2}d\widehat{F}_{1} - 2\int\widehat{S}_{2}d\widehat{F}_{1} - \{1+(\int \widehat{S}_{2}d\widehat{F}_{1})^{2}- 2\int\widehat{S}_{2}d\widehat{F}_{1} \}
\\
&=
\int\widehat{S}_{2}^{2}d\widehat{F}_{1}-(\int \widehat{S}_{2}d\widehat{F}_{1})^{2} \\&= 
\int\widehat{S}_{2}^{2}d\widehat{F}_{1} - \widehat{p}^{2},
\end{align*} 
this produces
\begin{align*}
\sum_{i=1}^{n_{1}}
&\left( R_{1i}-R_{1i}^{(1)} - \left[\bar{R}_{1\bullet}-\frac{n_{1}+1}{2}\right] \right)^{2}
\\
&=
\sum_{i=1}^{n_{1}}\left(
n_{2}\widehat{F}_{2}(X_{1i}) - n_{2}\int\widehat{F}_{2}d\widehat{F}_{1}\right)^{2}
= n_{2}^{2}\sum_{i=1}^{n_{1}}\left(\widehat{F}_{2}(X_{1i}) -\int\widehat{F}_{2}d\widehat{F}_{1}\right)^{2}
\\
&= 
n_{2}^{2}n_{1}\frac{1}{n_{1}}\sum_{i=1}^{n_{1}}\left(\widehat{F}_{2}^{2}(X_{1i}) -2\widehat{F}_{2}(X_{1i}) \int\widehat{F}_{2}d\widehat{F}_{1}+(\int\widehat{F}_{2}d\widehat{F}_{1})^{2}\right)
\\
&=
n_{2}^{2}n_{1} \{\int \widehat{F}_{2}^{2}d\widehat{F}_{1}- (\int \widehat{F}_{2}d\widehat{F}_{1})^{2}  \}
=
n_{2}^{2}n_{1} \{\int \widehat{S}_{2}^{2}d\widehat{F}_{1}- \widehat{p}^{2}  \}
= 
n_{2}^{2}n_{1} (\widehat{\tau}_{1}-\widehat{p}^{2}),
\end{align*}
\begin{align*}
\sum_{j=1}^{n_{2}}
&\left( R_{2j}-R_{2j}^{(2)} - \left[\bar{R}_{2\bullet}-\frac{n_{2}+1}{2}\right] \right)^{2}
\\
&=
\sum_{j=1}^{n_{2}}\left(
n_{1}\widehat{F}_{1}(X_{2j}) - n_{1}\int\widehat{F}_{1}d\widehat{F}_{2}\right)^{2}
= n_{1}^{2}\sum_{j=1}^{n_{2}}\left(\widehat{F}_{1}(X_{2j}) -\int\widehat{F}_{1}d\widehat{F}_{2}\right)^{2}
\\
&= 
n_{1}^{2}n_{2}\frac{1}{n_{2}}\sum_{j=1}^{n_{2}}\left(\widehat{F}_{1}^{2}(X_{2j}) -2\widehat{F}_{1}(X_{2j}) \int\widehat{F}_{1}d\widehat{F}_{2}+(\int\widehat{F}_{1}d\widehat{F}_{2})^{2}\right)
\\
&=
n_{1}^{2}n_{2} \{\int \widehat{F}_{1}^{2}d\widehat{F}_{2}- (\int \widehat{F}_{1}d\widehat{F}_{2})^{2}  \}
=
n_{1}^{2}n_{2} \{\int \widehat{F}_{1}^{2}d\widehat{F}_{2}- \widehat{p}^{2}  \}
=
n_{1}^{2}n_{2} ( \widehat{\tau}_{2} - \widehat{p}^{2} ),
\\
\widehat{\theta}(1&-\widehat{\theta})-\frac{1}{4}\widehat{\beta} 
\\
&= \widehat{p}(1-\widehat{p})-\frac{1}{4}\widehat{\beta}
=
\widehat{p}-\frac{1}{4}\widehat{\beta}-\widehat{p}^{2}
=
\widehat{\tau}_{0}-\widehat{p}^{2}.
\end{align*}
Now putting everything together we find
\begin{align*}
\widehat{\sigma}^{2}_{N}
&=
\frac{
n_{2}^{2}n_{1} (\widehat{\tau}_{1}-\widehat{p}^{2})
+
n_{1}^{2}n_{2} ( \widehat{\tau}_{2} - \widehat{p}^{2})
-
n_{1}n_{2}( \widehat{\tau}_{0}-\widehat{p}^{2})
}{n_{1}(n_{1}-1)n_{2}(n_{2}-1)}
\\
&=
\frac{
n_{2}(\widehat{\tau}_{1}-\widehat{p}^{2})
+
n_{1}( \widehat{\tau}_{2} - \widehat{p}^{2})
-
\widehat{\tau}_{0}+\widehat{p}^{2}
}
{(n_{1}-1)(n_{2}-1)}
\\
&=
\frac{
n_{2}\widehat{\tau}_{1}
+
n_{1}\widehat{\tau}_{2}
-\widehat{\tau}_{0}
-
(n_{1}+n_{2}-1)\widehat{p}^{2}
}
{(n_{1}-1)(n_{2}-1)}.
\end{align*}
\subsubsection*{Bamber form of the unbiased variance estimator}
As to Bamber's [\citeyear{bamber}] notation, he labels observations from sample 1 as $X_{1},\dots,X_{N_{X}}$ and from sample 2 as $Y_{1},\dots,Y_{N_{Y}}$. He then goes on to define an estimator for $B_{YYX}$, that is
\begin{align*}
b_{YYX}
&=
p(Y_{1},Y_{2}<X) + p(X<Y_{1},Y_{2}) - 2p(Y_{1}<X<Y_{2}),
\end{align*}
where
\begin{align*}
p(Y_{1},Y_{2}<X)
&=
\frac{1}{N_{X}N_{Y}(N_{Y}-1)}
\sum_{i=1}^{N_{X}}\sum_{j=1}^{N_{Y}}\sum_{\substack{j'=1\\j'\neq j}}^{N_{Y}}
\mathbb{I}(Y_{j}< X_{i})
\mathbb{I}(Y_{j'}< X_{i})
\\
&=
\frac{1}{N_{X}N_{Y}(N_{Y}-1)}
\left\{
\sum_{i=1}^{N_{X}}\left(\sum_{j=1}^{N_{Y}}
\mathbb{I}(Y_{j}< X_{i})\right)^{2}
-
\sum_{i=1}^{N_{X}}\sum_{j=1}^{N_{Y}}
\mathbb{I}(Y_{j}< X_{i})\right\}
\\
&=
\frac{N_{Y}}{N_{X}(N_{Y}-1)}
\sum_{i=1}^{N_{X}}\{\widehat{F}_{Y}^{-}(X_{i})\}^{2}
-
\frac{\sum_{i=1}^{N_{X}}\sum_{j=1}^{N_{Y}}
\mathbb{I}(Y_{j}< X_{i})}{N_{X}N_{Y}(N_{Y}-1)},
\\
p(X<Y_{1},Y_{2})
&=
\frac{1}{N_{X}N_{Y}(N_{Y}-1)}
\sum_{i=1}^{N_{X}}\sum_{j=1}^{N_{Y}}\sum_{\substack{j'=1\\j'\neq j}}^{N_{Y}}
\mathbb{I}(Y_{j}> X_{i})
\mathbb{I}(Y_{j'}> X_{i})
\\
&=
\frac{1}{N_{X}N_{Y}(N_{Y}-1)}
\left\{
\sum_{i=1}^{N_{X}}\left(\sum_{j=1}^{N_{Y}}
\mathbb{I}(Y_{j}> X_{i})\right)^{2}
-
\sum_{i=1}^{N_{X}}\sum_{j=1}^{N_{Y}}
\mathbb{I}(Y_{j}> X_{i})\right\}
\\
&=
\frac{N_{Y}}{N_{X}(N_{Y}-1)}
\sum_{i=1}^{N_{X}}\{\widehat{S}_{Y}^{-}(X_{i})\}^{2}
-
\frac{\sum_{i=1}^{N_{X}}\sum_{j=1}^{N_{Y}}
\mathbb{I}(Y_{j}> X_{i})}{N_{X}N_{Y}(N_{Y}-1)},
\end{align*}
\begin{align*}
p(Y_{1}<X<Y_{2})
&=
\frac{1}{N_{X}N_{Y}(N_{Y}-1)}
\sum_{i=1}^{N_{X}}\sum_{j=1}^{N_{Y}}\sum_{\substack{j'=1\\j'\neq j}}^{N_{Y}}
\mathbb{I}(Y_{j}< X_{i})
\mathbb{I}(Y_{j'}> X_{i})
\\
&=
\frac{1}{N_{X}N_{Y}(N_{Y}-1)}
\left\{
\sum_{i=1}^{N_{X}}\sum_{j=1}^{N_{Y}}\sum_{j'=1}^{N_{Y}}
\mathbb{I}(Y_{j}< X_{i})
\mathbb{I}(Y_{j'}> X_{i})
-
\sum_{i=1}^{N_{X}}\sum_{j=1}^{N_{Y}}
\mathbb{I}(Y_{j}< X_{i})\mathbb{I}(Y_{j}> X_{i})\right\}
\\
&=
\frac{N_{Y}}{N_{X}(N_{Y}-1)}
\sum_{i=1}^{N_{X}}\{\widehat{F}_{Y}^{-}(X_{i})\widehat{S}_{Y}^{-}(X_{i})\}.
\end{align*}
Further note that he calls $u_{X}=\sum_{j=1}^{N_{Y}}
\mathbb{I}(Y_{j}< X)$ and $v_{X}=\sum_{j=1}^{N_{Y}}
\mathbb{I}(Y_{j}> X)$.

Similar to before, it holds for each $i\in\{1,\dots,N_{X}\}$ that
\begin{align*}
&\{\widehat{S}_{Y}^{-}(X_{i})\}^{2}+\{\widehat{F}_{Y}^{-}(X_{i})\}^{2}-2\widehat{F}_{Y}^{-}(X_{i})\widehat{S}_{Y}^{-}(X_{i})
\\
&\phantom{xxx}
=
\{\widehat{S}_{Y}^{-}(X_{i})-\widehat{F}_{Y}^{-}(X_{i})\}^{2}
=
\{\widehat{S}_{Y}^{-}(X_{i})+\frac{1}{2}\sum_{j=1}^{N_{X}}\mathbb{I}(Y_{j}=X_{i})-\widehat{F}_{Y}^{-}-\frac{1}{2}\sum_{j=1}^{N_{X}}\mathbb{I}(Y_{j}=X_{i})\}^{2}
\\
&\phantom{xxx}
=
\{\widehat{S}_{Y}(X_{i})-\widehat{F}_{Y}(X_{i})\}^{2}
=
\{2\widehat{S}_{Y}(X_{i})-1\}^{2}
\\
&\phantom{xxx}
=
4\widehat{S}_{Y}^{2}(X_{i})-4\widehat{S}_{Y}(X_{i})+1.
\end{align*}
Therefore it follows,
\begin{align*}
b_{YYX} &= \frac{N_{Y}}{N_{Y}-1}\{4\int \widehat{S}_{Y}^{2}d\widehat{F}_{X} - 4\int \widehat{S}_{Y}d\widehat{F}_{X} + 1\} - \frac{1-\widehat{\beta}}{N_{Y}-1}
\\
&=
\frac{n_{2}}{n_{2}-1}\{4\int \widehat{S}_{2}^{2}d\widehat{F}_{1} - 4\int \widehat{S}_{2}d\widehat{F}_{1} + 1\} - \frac{1-\widehat{\beta}}{n_{2}-1}
\\
&=
\frac{n_{2}}{n_{2}-1}\{4\widehat{\tau}_{1} - 4\widehat{p} + 1\} - \frac{1-\widehat{\beta}}{n_{2}-1}.
\end{align*}
By the same arguments, it should then hold
\begin{align*}
b_{XXY} &= 
\frac{n_{1}}{n_{1}-1}\{4\widehat{\tau}_{2} - 4\widehat{p} + 1\} - \frac{1-\widehat{\beta}}{n_{1}-1}.
\end{align*}
Now putting everything together we again find
\begin{align*}
s_{a}^{2}
&=
\frac
{p(X\neq Y)
+
(N_{X}-1)b_{XXY}
+
(N_{Y}-1)b_{YYX}
-4(N_{X}+N_{Y}-1)(a-\nicefrac{1}{2})^{2}}
{4(N_{X}-1)(N_{Y}-1)}
\\
&=
\frac
{
1-\widehat{\beta}
+
n_{1}(4\widehat{\tau}_{2}-4\widehat{p} +1) - 1+\widehat{\beta}
+
n_{2}(4\widehat{\tau}_{1}-4\widehat{p} +1) - 1+\widehat{\beta}
-4(n_{1}+n_{2}-1)(\widehat{p}-\nicefrac{1}{2})^{2}
}
{4(n_{1}-1)(n_{2}-1)}
\\
&=
\frac{n_{1}\widehat{\tau}_{2}+n_{2}\widehat{\tau}_{1}-(n_{1}+n_{2}-1)\widehat{p}^{2}}{(n_{1}-1)(n_{2}-1)}
\\
&\phantom{xxxxxxxxxxxxxxx}
+
\frac{\widehat{\beta} - 4\widehat{p}(n_{1}+n_{2})+4\widehat{p}(n_{1}+n_{2}-1)+n_{1}+n_{2}-1 - (n_{1}+n_{2}-1)}{4(n_{1}-1)(n_{2}-1)}
\\
&=
\frac{n_{1}\widehat{\tau}_{2}+n_{2}\widehat{\tau}_{1}-(n_{1}+n_{2}-1)\widehat{p}^{2}}{(n_{1}-1)(n_{2}-1)}+
\frac{\widehat{\beta}-4\widehat{p}}{4(n_{1}-1)(n_{2}-1)}
\\
&=
\frac{n_{1}\widehat{\tau}_{2}+n_{2}\widehat{\tau}_{1}-(n_{1}+n_{2}-1)\widehat{p}^{2}-\widehat{\tau}_{0}}{(n_{1}-1)(n_{2}-1)}
=
\widehat{\sigma}^{2}_{N}.
\end{align*}
So we see that Bamber's definition of the estimator $\widehat{\sigma}^{2}_{N}$ is equivalent as well.

\subsubsection*{Perme and Manevski's estimators}
As to Perme and Manevski's [\citeyear{pm}] notation, they have $X_{1},\dots,X_{m}$ for sample one and $Y_{1},\dots,Y_{n}$ for sample 2. They give two estimators
\begin{align*}
\widehat{\mathbb{V}}_{DL}(\widehat{\theta})
&=
\frac{1}{m}
\widehat{\mathbb{V}}(S_{Y}(X)) 
+
\frac{1}{n}
\widehat{\mathbb{V}}(S_{X}(Y)),
\\
\widehat{\mathbb{V}}_{DLe}(\widehat{\theta})
&=
\frac{\widehat{\theta}(1-\widehat{\theta})}{mn}
+
\frac{n-1}{mn}
\widehat{\mathbb{V}}(S_{Y}(X)) 
+
\frac{m-1}{mn}
\widehat{\mathbb{V}}(S_{X}(Y)),
\end{align*}
where 
\begin{align*}
\widehat{\mathbb{V}}(S_{Y}(X)) &=
\frac{1}{m-1}
\sum_{i=1}^{m}(V_{i\bullet}-\widehat{\theta})^{2},
\\
\widehat{\mathbb{V}}(S_{X}(Y)) &=
\frac{1}{n-1}\sum_{j=1}^{n}(V_{\bullet j}-\widehat{\theta})^{2},
\end{align*}
with $V_{i\bullet}=\frac{1}{n}\sum_{j=1}^{n}V_{ij}$, $V_{\bullet j}=\frac{1}{m}\sum_{i=1}^{m}V_{ij}$ where $V_{ij}=c(Y_{j},X_{i})$ according to our notation using the normalised version of the count function.

So these quantities should equal
\begin{align*}
V_{i\bullet} &= \frac{1}{n}\sum_{j=1}^{n}V_{ij}
= \frac{1}{n}\sum_{j=1}^{n}c(Y_{j},X_{i}) = \widehat{S}_{Y}(X_{i})=\widehat{S}_{2}(X_{1i}), \\
V_{\bullet j} &= \frac{1}{m}\sum_{i=1}^{m}V_{ij}
= \frac{1}{m}\sum_{i=1}^{m}c(Y_{j},X_{i}) = \widehat{F}_{X}(Y_{j})=\widehat{F}_{1}(X_{2j}).
\end{align*}
Therefore, it follows
\begin{align*}
\widehat{\mathbb{V}}(S_{Y}(X)) &=
\frac{1}{n_{1}-1} \sum_{i=1}^{n_{1}}(\widehat{S}_{2}(X_{1i}) - \widehat{p})^{2}
=
\frac{1}{n_{1}-1} \sum_{i=1}^{n_{1}}(\widehat{S}_{2}^{2}(X_{1i}) - 2\widehat{p}\widehat{S}_{2}(X_{1i}) +\widehat{p}^{2})
\\
&=
\frac{n_{1}}{n_{1}-1}
( \int \widehat{S}^{2}_{2}d\widehat{F}_{1}
- 2\widehat{p} \int \widehat{S}_{2}d\widehat{F}_{1}
+\widehat{p}^{2})
\\
&=
\frac{n_{1}}{n_{1}-1}(\widehat{\tau}_{1}-\widehat{p}^{2})=\widehat{\sigma}^{2}_{1},
\\
\widehat{\mathbb{V}}(S_{X}(Y)) &=
\frac{1}{n_{2}-1} \sum_{j=1}^{n_{2}}(\widehat{F}_{1}(X_{2j}) - \widehat{p})^{2}
=
\frac{1}{n_{2}-1} \sum_{j=1}^{n_{2}}(\widehat{F}_{1}^{2}(X_{2j}) - 2\widehat{p}\widehat{F}_{1}(X_{2j}) +\widehat{p}^{2})
\\
&=
\frac{n_{2}}{n_{2}-1}
( \int \widehat{F}^{2}_{1}d\widehat{F}_{2}
- 2\widehat{p} \int \widehat{F}_{1}d\widehat{F}_{2}
+\widehat{p}^{2})
\\
&=
\frac{n_{2}}{n_{2}-1}(\widehat{\tau}_{2}-\widehat{p}^{2})=\widehat{\sigma}^{2}_{2}.
\end{align*}
Equipped with these results it follows that
\begin{align*}
\widehat{\mathbb{V}}_{DL}(\widehat{\theta})
&=
\widehat{\mathbb{V}}_{DL}(\widehat{p})
= \frac{\widehat{\sigma}_{1}^{2}}{n_{1}}
+
\frac{\widehat{\sigma}_{2}^{2}}{n_{2}}
=
\frac{\widehat{\tau}_{1}-\widehat{p}^{2}}{n_{1}-1}
+
\frac{\widehat{\tau}_{2}-\widehat{p}^{2}}{n_{2}-1}= \widehat{\sigma}_{BM}^{2},
\end{align*}
in other words, the variance estimator proposed by DeLong et al. (\citeyear{delong}) is identical to Brunner and Munzel's (\citeyear{bm}).
As for the estimator Perme and Manevski call \emph{exact}, we have
\begin{align*}
\widehat{\mathbb{V}}_{DLe}(\widehat{\theta})
=
\widehat{\mathbb{V}}_{DLe}(\widehat{p})
&=
\frac{\widehat{p}(1-\widehat{p})+(n_{2}-1)\widehat{\sigma}_{1}^{2}
+ (n_{1}-1)\widehat{\sigma}_{2}^{2}
}{n_{1}n_{2}}
\\
&=
\frac{\widehat{p}(1-\widehat{p})}{n_{1}n_{2}}
+
\frac{(n_{2}-1)(\widehat{\tau}_{1}-\widehat{p}^{2})}{n_{2}(n_{1}-1)}
+
\frac{(n_{1}-1)(\widehat{\tau}_{2}-\widehat{p}^{2})}{n_{1}(n_{2}-1)}=\widehat{\sigma}^{2}_{PM}.
\end{align*}


\subsubsection*{Shirahata's formulas}
Shirahata [\citeyear{shirahata}] only considers continuous distributions, assuming two independent samples $X_{1},\dots,X_{m} \sim F(x)$ and $Y_{1},\dots,Y_{n}\sim G(x)$. To ease the translation of his formulas into our notation, we exchange the samples, i.e, we set $(X_{1},\dots,X_{m}) = (X_{21},\dots,X_{2n_{2}})$ and $(Y_{1},\dots,Y_{n}) = (X_{11},\dots,X_{1n_{1}})$. Hence $F(x)=F_{2}(x)$ and $G(x)=F_{1}(x)$.

Moreover Shirahata uses the count function $u(x)=1$ or $0$ according as $x\geq 0$ or $x<0$. Hence we have the following equivalence
\begin{align*}
u(X_{i}-Y_{j})
=
\left.
\begin{cases}
1 &\text{ if } X_{i}-Y_{i}\geq 0 \\
0 &\text{ if } X_{i}-Y_{i}< 0 
\end{cases}
\right\}
=
\left.
\begin{cases}
1 &\text{ if } X_{i}\geq Y_{i} \\
0 &\text{ if } X_{i} < Y_{i} 
\end{cases}
\right\}
=
c^{+}(X_{i},Y_{j})
=
c^{+}(X_{2i},X_{1j}).
\end{align*}
Hence the quantities in Section 2 in Shirahata should read in our notation as
\begin{align*}
\zeta_{11}&=\theta= 
\mathbb{E}(u(X_{1}-Y_{1}))
=
\mathbb{E}(c^{+}(X_{21},X_{11}))
=
\int F_{1}^{+}dF_{2},
\\
\zeta_{21}&=
\mathbb{E}(u(X_{1}-Y_{1})u(X_{2}-Y_{1}))
=
\mathbb{E}(c^{+}(X_{21},X_{11})c^{+}(X_{22},X_{11}) )
\\
&=
\int \mathbb{E}(c^{+}(X_{21},x)c^{+}(X_{22},x) )d F_{1}(x)
\\
&=
\int (S_{2}^{+})^{2}d F_{1},
\\
\zeta_{12}&=
\mathbb{E}(u(X_{1}-Y_{1})u(X_{1}-Y_{2}))
=
\mathbb{E}(c^{+}(X_{21},X_{11})c^{+}(X_{21},X_{12}) )
\\
&=
\int \mathbb{E}(c^{+}(x,X_{11})c^{+}(x,X_{12}) )d F_{2}(x)
\\
&=
\int (F_{1}^{+})^{2}d F_{2},
\\
\zeta_{22}&=\theta^{2}= (\int F_{1}^{+}dF_{2})^{2}.
\end{align*}
So in case of continuous distributions we have $\zeta_{11}=p=\tau_{0}$, $\zeta_{21}=\int S_{2}^{2}d F_{1}=\tau_{1}$, $\zeta_{22}=\int F_{1}^{2}d F_{2}=\tau_{2}$, and $\zeta_{22}=p^{2}$. Hence in case of no ties, it follows that their formula of the theoretical variance coincides with ours,
\begin{align*}
\sigma^{2}
&=
\frac{1}{mn}\{\zeta_{11}+(m-1)\zeta_{21}+(n-1)\zeta_{12}- (m+n-1)\zeta_{22}\}
\\
&=
\frac{1}{n_{1}n_{2}}\{\tau_{0}+(n_{2}-1)\tau_{1}+(n_{1}-1)\tau_{2}- (n_{1}+n_{2}-1)p^{2}\} = \sigma_{N}^{2}.
\end{align*}

As for the estimator, Shirahata considers the quantities $B$, $C^{2}$ and $D^{2}$, which we will now translate into our notation, i.e.,
\begin{align*}
B&=\sum_{i=1}^{m}\sum_{j=1}^{n}u(X_{i}-Y_{j})
=\sum_{i=1}^{n_{2}}\sum_{j=1}^{n_{1}}c^{+}(X_{2i},X_{1j})
=
\sum_{i=1}^{n_{2}} n_{1}\widehat{F}_{1}^{+}(X_{2i})
= n_{1}n_{2}\int \widehat{F}_{1}^{+}d\widehat{F}_{2},
\\
C^{2}
&=
\sum_{j=1}^{n}\left(\sum_{i=1}^{m}u(X_{i}-Y_{j})\right)^{2}
=
\sum_{j=1}^{n_{1}}\left(\sum_{i=1}^{n_{2}}c^{+}(X_{2i},X_{1j})\right)^{2}
=
\sum_{j=1}^{n_{1}}n_{2}^{2}\{\widehat{S}_{2}^{+}(X_{1j})\}^{2}
\\
&=
n_{1}n_{2}^{2}\int (\widehat{S}_{2}^{+})^{2}d\widehat{F}_{1},
\\
D^{2}
&=
\sum_{i=1}^{m}\left(\sum_{j=1}^{n}u(X_{i}-Y_{j})\right)^{2}
=
\sum_{i=1}^{n_{2}}\left(\sum_{j=1}^{n_{1}}c^{+}(X_{2i},X_{1j})\right)^{2}
=
\sum_{i=1}^{n_{2}}n_{1}^{2}\{\widehat{F}_{1}^{+}(X_{2i})\}^{2}
\\
&=
n_{1}^{2}n_{2}\int (\widehat{F}_{1}^{+})^{2}d\widehat{F}_{2}.
\end{align*}
Shirahata considers a range of variance estimators, the first one being the unbiased one,
\begin{align*}
\widehat{\sigma}_{U}^{2}
&=
\frac{1}{m(m-1)n(n-1)}\left(-\frac{m+n-1}{mn}B^{2} - B + C^{2}+D^{2} \right)
\\
&=
\frac{1}{n_{1}(n_{1}-1)n_{2}(n_{2}-1)}\left(-\frac{n_{1}+n_{2}-1}{n_{1}n_{2}}n_{1}^{2}n_{2}^{2}(\int \widehat{F}_{1}^{+}d\widehat{F}_{2})^{2} - n_{1}n_{2}\int \widehat{F}_{1}^{+}d\widehat{F}_{2}\right. \\
&\left.\phantom{xxxxxxxxxxxxxxxx\frac{n_{1}+n_{2}}{n_{1}n_{2}}} + n_{1}n_{2}^{2}\int (\widehat{S}_{2}^{+})^{2}d\widehat{F}_{1}+n_{1}^{2}n_{2}\int (\widehat{F}_{1}^{+})^{2}d\widehat{F}_{2} \right)
\\
&=
\frac{n_{2}\int (\widehat{S}_{2}^{+})^{2}d\widehat{F}_{1}+n_{1}\int (\widehat{F}_{1}^{+})^{2}d\widehat{F}_{2} - \int \widehat{F}_{1}^{+}d\widehat{F}_{2}- (n_{1}+n_{2}-1)(\int \widehat{F}_{1}^{+}d\widehat{F}_{2})^{2}}{(n_{1}-1)(n_{2}-1)}.
\end{align*}
Their bootstrap estimator is given by
\begin{align*}
\widehat{\sigma}_{B}^{2}
&=
\frac{1}{m^{2}n^{2}}\left(-\frac{m+n-1}{mn}B^{2} + B + \frac{m-1}{m}C^{2}+\frac{n-1}{n}D^{2} \right)
\\
&=
\frac{1}{n_{1}^{2}n_{2}^{2}}\left(-\frac{n_{1}+n_{2}-1}{n_{1}n_{2}}n_{1}^{2}n_{2}^{2}(\int \widehat{F}_{1}^{+}d\widehat{F}_{2})^{2} + n_{1}n_{2}\int \widehat{F}_{1}^{+}d\widehat{F}_{2}\right. \\
&\left.\phantom{xxxxxxxxx} + \frac{n_{2}-1}{n_{2}}n_{1}n_{2}^{2}\int (\widehat{S}_{2}^{+})^{2}d\widehat{F}_{1}+\frac{n_{1}-1}{n_{1}}n_{1}^{2}n_{2}\int (\widehat{F}_{1}^{+})^{2}d\widehat{F}_{2} \right)
\\
&=
\frac{(n_{2}-1)\int (\widehat{S}_{2}^{+})^{2}d\widehat{F}_{1}+(n_{1}-1)\int (\widehat{F}_{1}^{+})^{2}d\widehat{F}_{2}+\int \widehat{F}_{1}^{+}d\widehat{F}_{2}-(n_{1}+n_{2}-1)(\int \widehat{F}_{1}^{+}d\widehat{F}_{2})^{2}}{n_{1}n_{2}}.
\end{align*}
The simple version of the Fligner and Policello [\citeyear{fligner}] estimator then takes the form
\begin{align*}
\widehat{\sigma}_{FP}^{2}
&=
\frac{1}{m^{2}n^{2}}\left(-\frac{m+n+1}{mn}B^{2} - B + C^{2}+D^{2} \right)
\\
&=
\frac{n_{2}\int (\widehat{S}_{2}^{+})^{2}d\widehat{F}_{1}+n_{1}\int (\widehat{F}_{1}^{+})^{2}d\widehat{F}_{2} - \int \widehat{F}_{1}^{+}d\widehat{F}_{2}- (n_{1}+n_{2}+1)(\int \widehat{F}_{1}^{+}d\widehat{F}_{2})^{2}}{n_{1}n_{2}}
\\
&=
\frac{\int (\widehat{S}_{2}^{+})^{2}d\widehat{F}_{1}}{n_{1}}+\frac{\int (\widehat{F}_{1}^{+})^{2}d\widehat{F}_{2}}{n_{2}} -
\frac{\int \widehat{F}_{1}^{+}d\widehat{F}_{2}+(n_{1}+n_{2}+1)(\int \widehat{F}_{1}^{+}d\widehat{F}_{2})^{2}}{n_{1}n_{2}}
.
\end{align*}
Lastly they consider the jackknife estimator
\begin{align*}
\widehat{\sigma}^{2}_{J}
&=
\frac{1}{m(m-1)n(n-1)}\left(-\frac{m+n-2}{mn}B^{2} - \frac{m-1}{m}C^{2}+\frac{n-1}{n}D^{2} \right)
\\
&=
\frac{1}{n_{1}(n_{1}-1)n_{2}(n_{2}-1)}\left(-\frac{n_{1}+n_{2}-1}{n_{1}n_{2}}n_{1}^{2}n_{2}^{2}(\int \widehat{F}_{1}^{+}d\widehat{F}_{2})^{2} \right. \\
&\left.\phantom{xxxxxxxxxxxxxxxxxxxx} + \frac{n_{2}-1}{n_{2}}n_{1}n_{2}^{2}\int (\widehat{S}_{2}^{+})^{2}d\widehat{F}_{1}+\frac{n_{1}-1}{n_{1}}n_{1}^{2}n_{2}\int (\widehat{F}_{1}^{+})^{2}d\widehat{F}_{2} \right)
\\
&=
\frac{\int (\widehat{S}_{2}^{+})^{2}d\widehat{F}_{1}}{n_{1}-1}
+
\frac{\int (\widehat{F}_{1}^{+})^{2}d\widehat{F}_{2}}{n_{2}-1}
-
\frac{ (n_{1}+n_{2}-2)(\int \widehat{F}_{1}^{+}d\widehat{F}_{2})^{2}}{(n_{1}-1)(n_{2}-1)}.
\end{align*}

\subsection*{Box-type approximation of degrees of freedom}
To begin with, we summarise the pertinent results of Box-type [\citeyear{box}] degrees of freedom as developed in Chapter 7.5.1.2 of Brunner et al. [\citeyear{np}].

First, we consider independent normal random variables $X_{11},\dots,X_{1n_{1}} \sim \mathcal{N}(\mu_{1},\sigma_{1}^{2})$ as well as
$X_{21},\dots,X_{2n_{2}} \sim \mathcal{N}(\mu_{2},\sigma_{2}^{2})$, $N=n_{1}+n_{2}$, $\bar{\mathbf{X}}=(\sum_{i=1}^{n_{1}}X_{1i}/n_{1} \; \; \;\sum_{j=1}^{n_{2}}X_{2j}/n_{2})^{\top}$. Then we have
\begin{align*}
\mathbf{S}_{N} &= \mathbb{C}\mathrm{ov}(\sqrt{N} \bar{\mathbf{X}}) = \bigoplus_{i=1}^{2} N\sigma_{i}^{2}/n_{i}
= N
\begin{pmatrix}
\sigma_{1}^{2}/n_{1} & 0 \\
0 & \sigma_{2}^{2}/n_{2}  \\
\end{pmatrix}, \\
\widehat{\mathbf{S}}_{N} &=  N
\begin{pmatrix}
\widehat{\sigma}_{1}^{2}/n_{1} & 0 \\
0 & \widehat{\sigma}_{2}^{2}/n_{2}  \\
\end{pmatrix},
\end{align*}
where $\widehat{\sigma}_{g}^{2}=\frac{1}{n_{g}-1}\sum_{i=1}^{n_{g}}(X_{gi}-\bar{X}_{g\bullet})^{2}$, $g=1,2$.
Furthermore we will need 
\begin{align*}
\mathbf{\Lambda} = 
\left(
\begin{pmatrix}
n_{1} & 0 \\
0 & n_{2}  \\
\end{pmatrix}
-
\begin{pmatrix}
1 & 0 \\
0 & 1  \\
\end{pmatrix}
\right)^{-1}
=
\begin{pmatrix}
1/(n_{1}-1) & 0 \\
0 & 1/(n_{2}-1)  \\
\end{pmatrix}.
\end{align*}
As for the contrast matrix (centering matrix), we have
\begin{align*}
\mathbf{C} = 
\begin{pmatrix}
1 & 0 \\
0 & 1 \\
\end{pmatrix}
- \frac{1}{2}
\begin{pmatrix}
1 & 1 \\
1 & 1  \\
\end{pmatrix}
=
\frac{1}{2}
\begin{pmatrix}
1 & -1 \\
-1 & 1  \\
\end{pmatrix},
\end{align*}
resulting in 
\begin{align*}
\mathbf{T}&
=
\mathbf{C}^{\top}(\mathbf{C}\mathbf{C}^{\top})^{-}\mathbf{C}
=
\mathbf{C},
&
\mathbf{D}_T &= \frac{1}{2} \begin{pmatrix}
1 & 0 \\
0 & 1 \\
\end{pmatrix},
\end{align*}
since $\mathbf{C}$ is positive semi-definite, symmetric and idempotent.

Then the approximate degrees of freedom for the two sample $t$ test are given by
\begin{align*}
f_{0}&= \left[{\rm tr}\left(\mathbf{D}_{T}\mathbf{S}_{N}\right) \right]^{2} / {\rm tr}\left(\mathbf{D}_{T}^{2}\mathbf{S}_{N}^{2}\mathbf{\Lambda}\right),
&
\widehat{f}_{0}&=
\left[{\rm tr}\left(\mathbf{D}_{T}\widehat{\mathbf{S}}_{N}\right) \right]^{2} / {\rm tr}\left(\mathbf{D}_{T}^{2}\widehat{\mathbf{S}}_{N}^{2}\mathbf{\Lambda}\right),
\end{align*}
which then simplifies to
\begin{align*}
\mathbf{D}_{T}\widehat{\mathbf{S}}_{N}
&=\frac{N}{2} 
\begin{pmatrix}
1 & 0 \\
0 & 1 
\end{pmatrix}
\begin{pmatrix}
\widehat{\sigma}_{1}^{2}/n_{1} & 0 \\
0 & \widehat{\sigma}_{2}^{2}/n_{2}  
\end{pmatrix}
=\frac{N}{2} 
\begin{pmatrix}
\widehat{\sigma}_{1}^{2}/n_{1} & 0 \\
0 & \widehat{\sigma}_{2}^{2}/n_{2}  
\end{pmatrix},
\\
\left[{\rm tr}\left(\mathbf{D}_{T}\widehat{\mathbf{S}}_{N}\right) \right]^{2}
&=\frac{N^{2}}{4} \left(\widehat{\sigma}_{1}^{2}/n_{1} + \widehat{\sigma}_{2}^{2}/n_{2} \right)^{2},
\\
\mathbf{D}_{T}^{2}\widehat{\mathbf{S}}_{N}^{2}\mathbf{\Lambda}
&=
\frac{N^{2}}{4} 
\begin{pmatrix}
1 & 0 \\
0 & 1 
\end{pmatrix}
\begin{pmatrix}
\widehat{\sigma}_{1}^{4}/n_{1}^{2} & 0 \\
0 & \widehat{\sigma}_{2}^{4}/n_{2}^{2}  
\end{pmatrix}
\begin{pmatrix}
1/(n_{1}-1) & 0 \\
0 & 1/(n_{2}-1)  \\
\end{pmatrix}
\\
{\rm tr}\left(\mathbf{D}_{T}^{2}\widehat{\mathbf{S}}_{N}^{2}\mathbf{\Lambda}\right)
&=\frac{N^{2}}{4}\left( \frac{\widehat{\sigma}_{1}^{4}}{n_{1}^{2}(n_{1}-1)} + \frac{\widehat{\sigma}_{2}^{4}}{n_{2}^{2}(n_{2}-1)} \right)
,
\end{align*}
yielding
$\widehat{f}_{0}
=
\left(\widehat{\sigma}_{1}^{2}/n_{1} + \widehat{\sigma}_{2}^{2}/n_{2} \right)^{2}
/
\left( \frac{\widehat{\sigma}_{1}^{4}}{n_{1}^{2}(n_{1}-1)} + \frac{\widehat{\sigma}_{2}^{4}}{n_{2}^{2}(n_{2}-1)} \right)$.

For the degrees of freedom proposed by Brunner and Munzel, the derivation is completely analogous. To this end, consider
$X_{11},\dots,X_{1n_{1}} \sim F_{1}$ as well as
$X_{21},\dots,X_{2n_{2}} \sim F_{2}$, $N=n_{1}+n_{2}$, $\bar{\mathbf{X}}=(\sum_{i=1}^{n_{1}}F_{2}(X_{1i})/n_{1} \; \; \;\sum_{j=1}^{n_{2}}F_{1}(X_{2j})/n_{2})^{\top}$. Then we have
\begin{align*}
\mathbf{S}_{N} &= \mathbb{C}\mathrm{ov}(\sqrt{N} \bar{\mathbf{X}}) = \bigoplus_{i=1}^{2} N\sigma_{i}^{2}/n_{i}
= N
\begin{pmatrix}
\sigma_{1}^{2}/n_{1} & 0 \\
0 & \sigma_{2}^{2}/n_{2}  \\
\end{pmatrix}, \\
\widehat{\mathbf{S}}_{N} &=  N
\begin{pmatrix}
\widehat{\sigma}_{1}^{2}/n_{1} & 0 \\
0 & \widehat{\sigma}_{2}^{2}/n_{2}  \\
\end{pmatrix},
\end{align*}
with variances $\sigma_{1}^{2}=\mathbb{V}(F_{2}(X_{11}))=\tau_{1}-p^{2}$, $\sigma_{2}^{2}=\mathbb{V}(F_{1}(X_{21}))=\tau_{2}-p^{2}$, and estimators
$\widehat{\sigma}_{1}^{2}=\frac{n_{1}}{n_{1}-1}(\widehat{\tau}_{1}-\widehat{p}^{2})$ and $\widehat{\sigma}_{2}^{2}=\frac{n_{2}}{n_{2}-1}(\widehat{\tau}_{2}-\widehat{p}^{2})$, yielding $\widehat{f}_{0}
=
\left(\widehat{\sigma}_{1}^{2}/n_{1} + \widehat{\sigma}_{2}^{2}/n_{2} \right)^{2}
/
\left( \frac{\widehat{\sigma}_{1}^{4}}{n_{1}^{2}(n_{1}-1)} + \frac{\widehat{\sigma}_{2}^{4}}{n_{2}^{2}(n_{2}-1)} \right)$.

The idea behind the newly adjusted degrees of freedom is to derive them similarly as for the Brunner-Munzel [\citeyear{bm}] test, but using the empirical distribution functions to compute the mean vector instead of the theoretical ones, i.e., $\bar{\mathbf{X}}=(\sum_{i=1}^{n_{1}}\widehat{F}_{2}(X_{1i})/n_{1} \; \; \;\sum_{j=1}^{n_{2}}\widehat{F}_{1}(X_{2j})/n_{2})^{\top}$.

To obtain the entries of $\mathbf{S}_{N} = \mathbb{C}\mathrm{ov}(\sqrt{N} \bar{\mathbf{X}})$, we will first consider the following quantities
\begin{align*}
\psi_{1}^{2}
&=\mathbb{V}(\widehat{F}_{2}(X_{11}))
=\mathbb{V}(1-\widehat{S}_{2}(X_{11}))
=\mathbb{V}(\widehat{S}_{2}(X_{11}))
=
\mathbb{E}((\widehat{S}_{2}(X_{11}))^{2})
-
(\mathbb{E}(\widehat{S}_{2}(X_{11})))^{2},
\\
\psi_{1\mid1}
&=\mathbb{C}\mathrm{ov}(\widehat{F}_{2}(X_{11}),\widehat{F}_{2}(X_{12}))
=\mathbb{C}\mathrm{ov}(\widehat{S}_{2}(X_{11}),\widehat{S}_{2}(X_{12})),
\\
\psi_{2}^{2}
&=\mathbb{V}(\widehat{F}_{1}(X_{21}))
=
\mathbb{E}((\widehat{F}_{1}(X_{21}))^{2})
-
(\mathbb{E}(\widehat{F}_{1}(X_{21})))^{2},
\\
\psi_{2\mid2}
&=\mathbb{C}\mathrm{ov}(\widehat{F}_{1}(X_{21}),\widehat{F}_{1}(X_{21})),
\\
\psi_{12} &=
\mathbb{C}\mathrm{ov}(\widehat{F}_{2}(X_{11}),\widehat{F}_{1}(X_{21}))
=
-
\mathbb{C}\mathrm{ov}(\widehat{S}_{2}(X_{11}),\widehat{F}_{1}(X_{21}))
\\
&=
-\mathbb{E}(\widehat{S}_{2}(X_{11}),\widehat{F}_{1}(X_{21}))+
\mathbb{E}(\widehat{S}_{2}(X_{11}))\mathbb{E}(\widehat{F}_{1}(X_{21})).
\end{align*}
So we now consider
\begin{align*}
\mathbb{E}(\widehat{S}_{2}(X_{11}))
&=
\frac{1}{n_{2}}\sum_{\ell=1}^{n_{2}}\mathbb{E}(c(X_{2\ell},X_{11}))=\int F_{1}dF_{2}=p,
\\
\mathbb{E}(\widehat{F}_{1}(X_{21}))
&=
\frac{1}{n_{1}}\sum_{k=1}^{n_{1}}\mathbb{E}(c(X_{21},X_{1k}))=\int F_{1}dF_{2}=p.
\end{align*}
Moreover,
\begin{align*}
\mathbb{E}((\widehat{S}_{2}(X_{11}))^{2})
&=
\frac{1}{n_{2}^{2}}\sum_{\ell=1}^{n_{2}}\sum_{\ell'=1}^{n_{2}}\mathbb{E}(c(X_{2\ell},X_{11})c(X_{2\ell'},X_{11}))
\\
&=
\frac{1}{n_{2}^{2}}\sum_{\ell=1}^{n_{2}}\sum_{\substack{\ell'=1\\\ell'\neq\ell}}^{n_{2}}\mathbb{E}(c(X_{2\ell},X_{11})c(X_{2\ell'},X_{11}))
+
\frac{1}{n_{2}^{2}}\sum_{\ell=1}^{n_{2}}\mathbb{E}(c(X_{2\ell},X_{11})c(X_{2\ell},X_{11}))
\\
&=
\frac{n_{2}-1}{n_{2}}\int S_{2}^{2}dF_{1}+\frac{1}{n_{2}}\mathbb{E}(c(X_{21},X_{11})^{2})=
\frac{n_{2}-1}{n_{2}}\tau_{1}+\frac{1}{n_{2}}\tau_{0},
\\
\mathbb{E}(\widehat{S}_{2}(X_{11})\widehat{S}_{2}(X_{12}))
&=
\frac{1}{n_{2}^{2}}\sum_{\ell=1}^{n_{2}}\sum_{\ell'=1}^{n_{2}}\mathbb{E}(c(X_{2\ell},X_{11})c(X_{2\ell'},X_{12}))
\\
&=
\frac{1}{n_{2}^{2}}\sum_{\ell=1}^{n_{2}}\sum_{\substack{\ell'=1\\\ell'\neq\ell}}^{n_{2}}\mathbb{E}(c(X_{2\ell},X_{11})c(X_{2\ell'},X_{12}))
+
\frac{1}{n_{2}^{2}}\sum_{\ell=1}^{n_{2}}\mathbb{E}(c(X_{2\ell},X_{11})c(X_{2\ell},X_{12}))
\\
&=
\frac{n_{2}-1}{n_{2}}p^{2} - \frac{1}{n_{2}}\int F_{1}^{2}dF_{2}=
\frac{n_{2}-1}{n_{2}}p^{2}+\frac{1}{n_{2}}\tau_{2},
\end{align*}
\begin{align*}
\mathbb{E}((\widehat{F}_{1}(X_{21}))^{2})
&=
\frac{1}{n_{1}^{2}}\sum_{k=1}^{n_{1}}\sum_{k'=1}^{n_{1}}\mathbb{E}(c(X_{21},X_{1k})c(X_{21},X_{1k'}))
\\
&=
\frac{1}{n_{1}^{2}}\sum_{k=1}^{n_{1}}\sum_{\substack{k'=1\\k'\neq k}}^{n_{1}}\mathbb{E}(c(X_{21},X_{1k})c(X_{21},X_{1k'}))
+
\frac{1}{n_{1}^{2}}\sum_{k=1}^{n_{1}}\mathbb{E}(c(X_{21},X_{1k})c(X_{21},X_{1k}))
\\
&=
\frac{n_{1}-1}{n_{1}}\int F_{1}^{2}dF_{2}+\frac{1}{n_{2}}\mathbb{E}(c(X_{21},X_{11})^{2})=
\frac{n_{1}-1}{n_{1}}\tau_{2}+\frac{1}{n_{1}}\tau_{0},
\\
\mathbb{E}(\widehat{F}_{1}(X_{21})\widehat{F}_{1}(X_{22}))
&=
\frac{1}{n_{1}^{2}}\sum_{k=1}^{n_{1}}\sum_{k'=1}^{n_{1}}\mathbb{E}(c(X_{21},X_{1k})c(X_{22},X_{1k'}))
\\
&=
\frac{1}{n_{1}^{2}}\sum_{k=1}^{n_{1}}\sum_{\substack{k'=1\\k'\neq k}}^{n_{1}}\mathbb{E}(c(X_{21},X_{1k})c(X_{22},X_{1k'}))
+
\frac{1}{n_{1}^{2}}\sum_{k=1}^{n_{1}}\mathbb{E}(c(X_{21},X_{1k})c(X_{22},X_{1k}))
\\
&=
\frac{n_{1}-1}{n_{1}}p^{2}dF_{2}+\frac{1}{n_{2}}\int S_{2}^{2}dF_{1}=
\frac{n_{1}-1}{n_{1}}p_{2}+\frac{1}{n_{1}}\tau_{1},
\end{align*}
as well as
\begin{align*}
\mathbb{E}(\widehat{S}_{2}(X_{11}),\widehat{F}_{1}(X_{21}))
&=
\frac{1}{n_{1}n_{2}}\sum_{k=1}^{n_{1}}\sum_{\ell=1}^{n_{2}}\mathbb{E}(c(X_{2\ell},X_{11})c(X_{21},X_{1k}))
\\
&=
\frac{1}{n_{1}n_{2}}\sum_{k=2}^{n_{1}}\sum_{\ell=2}^{n_{2}}\mathbb{E}(c(X_{2\ell},X_{11})c(X_{21},X_{1k}))
+
\frac{1}{n_{1}n_{2}}\mathbb{E}(c(X_{21},X_{11})c(X_{21},X_{11}))
\\
&\phantom{xx} +
\frac{1}{n_{1}n_{2}}\sum_{k=2}^{n_{1}}\mathbb{E}(c(X_{21},X_{11})c(X_{21},X_{1k}))
+
\frac{1}{n_{1}n_{2}}\sum_{\ell=2}^{n_{2}}\mathbb{E}(c(X_{2\ell},X_{11})c(X_{21},X_{11}))
\\
&=
\frac{(n_{1}-1)(n_{2}-1)p^{2}+\tau_{0}+(n_{1}-1)\tau_{2}+(n_{2}-1)\tau_{1}}{n_{1}n_{2}}.
\end{align*}
Collecting terms, we have
\begin{align*}
\psi_{1}^{2}
&=\frac{n_{2}-1}{n_{2}}\tau_{1}+\frac{1}{n_{2}}p - \frac{1}{4n_{2}}\beta - p^{2}
=\frac{1}{n_{2}}\left[(n_{2}-1)(\tau_{1}-p^{2})+p-p^{2}-\frac{1}{4}\beta\right]
\\
&=
\frac{1}{n_{2}}\left[(n_{2}-1)\sigma_{1}^{2}+p(1-p)-\frac{1}{4}\beta\right],
\\
\psi_{2}^{2}
&=\frac{n_{1}-1}{n_{1}}\tau_{2}+\frac{1}{n_{1}}p - \frac{1}{4n_{1}}\beta - p^{2}
=\frac{1}{n_{1}}\left[(n_{1}-1)(\tau_{2}-p^{2})+p-p^{2}-\frac{1}{4}\beta\right]
\\
&=
\frac{1}{n_{1}}\left[(n_{1}-1)\sigma_{2}^{2}+p(1-p)-\frac{1}{4}\beta\right],
\\
\psi_{1\mid1}
&=
\frac{n_{2}-1}{n_{2}}p^{2}+\frac{1}{n_{2}}\tau_{2}-p^{2}= \frac{1}{n_{2}}(\tau_{2}-p^{2})=\frac{1}{n_{2}}\sigma_{2}^{2},
\\
\psi_{2\mid2}
&=
\frac{n_{1}-1}{n_{1}}p^{2}+\frac{1}{n_{1}}\tau_{1}-p^{2}= \frac{1}{n_{1}}(\tau_{1}-p^{2})=\frac{1}{n_{1}}\sigma_{1}^{2},
\end{align*}
which are the same terms as in equations (2.32), (2.33), (2.34), and (2.35) of Brunner et al. [\citeyear{brunnerpreprint}] multiplied by either $n_{2}^{2}$ or $n_{1}^{2}$. Now we turn to
\begin{align*}
\psi_{12}
&=p^{2} - \frac{(n_{1}-1)(n_{2}-1)p^{2}+\tau_{0}+(n_{1}-1)\tau_{2}+(n_{2}-1)\tau_{1}}{n_{1}n_{2}}
\\
&=
-\frac{(n_{1}-1)\tau_{2}+(n_{2}-1)\tau_{1}+\tau_{0}-(n_{1}+n_{2}-1)p^{2}}{n_{1}n_{2}}=-\sigma_{N}^{2}=-\mathbb{V}(\widehat{p}),
\end{align*}
Now, since
\begin{align*}
\mathbb{V}( \frac{1}{n_{1}}\sum_{i=1}^{n_{1}} \widehat{F}_{2}(X_{1i}))
&=
\frac{1}{n_{1}^{2}} \sum_{i=1}^{n_{1}} \mathbb{V}(\widehat{F}_{2}(X_{1i}))
+
\frac{1}{n_{1}^{2}} \sum_{k=1}^{n_{1}} \sum_{\substack{k'=1\\k'\neq k}}^{n_{1}}
\mathbb{C}\mathrm{ov}(\widehat{F}_{2}(X_{1k}),\widehat{F}_{2}(X_{1k'}))
\\
&=
\frac{1}{n_{1}^{2}} n_{1} \psi_{1}^{2}
+
\frac{1}{n_{1}^{2}} n_{1}(n_{1}-1)\psi_{1\mid1}
\\
&=
\frac{1}{n_{1}n_{2}}\left[(n_{2}-1)\sigma_{1}^{2}+p(1-p)-\frac{1}{4}\beta\right]+\frac{n_{1}-1}{n_{1}n_{2}}\sigma_{2}^{2}=\sigma_{N}^{2},
\end{align*}
\begin{align*}
\mathbb{V}( \frac{1}{n_{2}}\sum_{j=1}^{n_{2}} \widehat{F}_{1}(X_{2j}))
&=
\frac{1}{n_{2}^{2}} \sum_{j=1}^{n_{2}} \mathbb{V}(\widehat{F}_{1}(X_{2j}))
+
\frac{1}{n_{2}^{2}} \sum_{\ell=1}^{n_{2}} \sum_{\substack{\ell'=1\\\ell'\neq \ell}}^{n_{2}}
\mathbb{C}\mathrm{ov}(\widehat{F}_{1}(X_{2\ell}),\widehat{F}_{1}(X_{2\ell'}))
\\
&=
\frac{1}{n_{2}^{2}} n_{2} \psi_{2}^{2}
+
\frac{1}{n_{2}^{2}} n_{2}(n_{2}-1)\psi_{2\mid2}
\\
&=
\frac{1}{n_{1}n_{2}}\left[(n_{1}-1)\sigma_{2}^{2}+p(1-p)-\frac{1}{4}\beta\right]+\frac{n_{2}-1}{n_{1}n_{2}}\sigma_{1}^{2}=\sigma_{N}^{2},
\end{align*}
and finally
\begin{align*}
\mathbb{C}\mathrm{ov}(\frac{1}{n_{1}}\sum_{i=1}^{n_{1}} \widehat{F}_{2}(X_{1i}),\frac{1}{n_{2}}\sum_{j=1}^{n_{2}} \widehat{F}_{1}(X_{2j}))
=
\frac{1}{n_{1}n_{2}}\sum_{i=1}^{n_{1}}\sum_{j=1}^{n_{2}}
\mathbb{C}\mathrm{ov}(\widehat{F}_{2}(X_{1i}),\widehat{F}_{1}(X_{2j}))=\frac{n_{1}n_{2}}{n_{1}n_{2}}\psi_{12}=-\sigma_{N}^{2}.
\end{align*}
Therefore, we have
\begin{align*}
\mathbf{S}_{N} &= \mathbb{C}\mathrm{ov}(\sqrt{N} \bar{\mathbf{X}}) = N\sigma_{N}^{2}
\begin{pmatrix}
1 & -1 \\
-1 & 1  \\
\end{pmatrix}, &
\widehat{\mathbf{S}}_{N} &=  N\widehat{\sigma}_{N}^{2}
\begin{pmatrix}
1 & -1 \\
-1 & 1  \\
\end{pmatrix}.
\end{align*}
If we are to derive the degrees of freedom in an analogous manner as before, we should get
\begin{align*}
\mathbf{D}_{T}\widehat{\mathbf{S}}_{N}
&=\frac{N\widehat{\sigma}_{N}^{2}}{2} 
\begin{pmatrix}
1 & 0 \\
0 & 1 
\end{pmatrix}
\begin{pmatrix}
1 & -1\\
-1& 1
\end{pmatrix}
=\frac{N\widehat{\sigma}_{N}^{2}}{2} 
\begin{pmatrix}
1 & -1\\
-1& 1
\end{pmatrix}
\\
\left[{\rm tr}\left(\mathbf{D}_{T}\widehat{\mathbf{S}}_{N}\right) \right]^{2}
&=N^{2}\widehat{\sigma}_{N}^{4},
\\
\mathbf{D}_{T}^{2}\widehat{\mathbf{S}}_{N}^{2}\mathbf{\Lambda}
&=
\frac{N^{2}\widehat{\sigma}_{N}^{4}}{4} 
\begin{pmatrix}
1 & 0 \\
0 & 1 
\end{pmatrix}
\begin{pmatrix}
2 & -2\\
-2& 2
\end{pmatrix}
\begin{pmatrix}
1/(n_{1}-1) & 0 \\
0 & 1/(n_{2}-1)  \\
\end{pmatrix}
\\
&=
\frac{N^{2}\widehat{\sigma}_{N}^{4}}{2} 
\begin{pmatrix}
1/(n_{1}-1) & -1/(n_{2}-1) \\
-1/(n_{1}-1) & 1/(n_{2}-1)  \\
\end{pmatrix}
\\
{\rm tr}\left(\mathbf{D}_{T}^{2}\widehat{\mathbf{S}}_{N}^{2}\mathbf{\Lambda}\right)
&=\frac{N^{2}\widehat{\sigma}_{N}^{4}}{2}\left( \frac{1}{n_{1}-1} + \frac{1}{n_{2}-1} \right)
,
\end{align*}
yielding
$\widehat{f}_{0} 
=
2
/
\left( \frac{1}{n_{1}-1} + \frac{1}{n_{2}-1} \right)$.

Furthermore, we now consider a simple heuristic alternative which may be viewed as a middle ground of the Brunner-Munzel approach and the approach just developed. To this end, recall the Brunner form of the unbiased variance estimator, i.e.,
\begin{align*}
\widehat{\sigma}_{N}^{2}
&=
\frac{1}{n_{1}(n_{1}-1)n_{2}(n_{2}-1)}
\left(
\sum_{i=1}^{2}
\sum_{k=1}^{n_{i}}
\left( R_{ik}-R_{ik}^{(i)} - \left[\bar{R}_{i\bullet}-\frac{n_{i}+1}{2}\right] \right)^{2}
- n_{1}n_{2}\left[\widehat{\theta}(1-\widehat{\theta})-\frac{1}{4}\widehat{\beta} \right] \right).
\end{align*}
We now split this unbiased estimator into two ``symmetric'' parts,
\begin{align*}
\widehat{\sigma}_{1\mid N}^{2}
&=
\frac{1}{n_{1}(n_{1}-1)n_{2}(n_{2}-1)}
\left(
\sum_{k=1}^{n_{1}}
\left( R_{1k}-R_{1k}^{(1)} - \left[\bar{R}_{1\bullet}-\frac{n_{1}+1}{2}\right] \right)^{2}
- \frac{1}{2}n_{1}n_{2}\left[\widehat{\theta}(1-\widehat{\theta})-\frac{1}{4}\widehat{\beta} \right] \right),\\
\widehat{\sigma}_{2\mid N}^{2}
&=
\frac{1}{n_{1}(n_{1}-1)n_{2}(n_{2}-1)}
\left(
\sum_{\ell=1}^{n_{2}}
\left( R_{2\ell}-R_{2\ell}^{(2)} - \left[\bar{R}_{2\bullet}-\frac{n_{2}+1}{2}\right] \right)^{2}
- \frac{1}{2}n_{1}n_{2}\left[\widehat{\theta}(1-\widehat{\theta})-\frac{1}{4}\widehat{\beta} \right] \right),
\end{align*}
such that $\widehat{\sigma}_{1\mid N}^{2}+\widehat{\sigma}_{2\mid N}^{2}=\widehat{\sigma}_{N}^{2}$. Moreover, we artificially set the covariances zero so that now we have
\begin{align*}
\widehat{\mathbf{S}}_{N} &=  N
\begin{pmatrix}
\widehat{\sigma}_{1\mid N}^{2} & 0 \\
0 & \widehat{\sigma}_{2\mid N}^{2}   \\
\end{pmatrix}.
\end{align*}
With this artificial covariance matrix, we now compute the degrees of freedom similar to before,
\begin{align*}
\mathbf{D}_{T}\widehat{\mathbf{S}}_{N}
&=\frac{N}{2} 
\begin{pmatrix}
1 & 0 \\
0 & 1 
\end{pmatrix}
\begin{pmatrix}
\widehat{\sigma}_{1\mid N}^{2} & 0 \\
0 & \widehat{\sigma}_{2\mid N}^{2}   \\
\end{pmatrix}
=\frac{N}{2} 
\begin{pmatrix}
\widehat{\sigma}_{1\mid N}^{2} & 0 \\
0 & \widehat{\sigma}_{2\mid N}^{2}   \\
\end{pmatrix},
\\
\left[{\rm tr}\left(\mathbf{D}_{T}\widehat{\mathbf{S}}_{N}\right) \right]^{2}
&=\frac{N^{2}}{4} \left(\widehat{\sigma}_{1\mid N}^{2} + \widehat{\sigma}_{2\mid N}^{2}\right)^{2},
\\
\mathbf{D}_{T}^{2}\widehat{\mathbf{S}}_{N}^{2}\mathbf{\Lambda}
&=
\frac{N^{2}}{4} 
\begin{pmatrix}
1 & 0 \\
0 & 1 
\end{pmatrix}
\begin{pmatrix}
\widehat{\sigma}_{1\mid N}^{4} & 0 \\
0 & \widehat{\sigma}_{2\mid N}^{4}   \\
\end{pmatrix}
\begin{pmatrix}
1/(n_{1}-1) & 0 \\
0 & 1/(n_{2}-1)  \\
\end{pmatrix}
\\
{\rm tr}\left(\mathbf{D}_{T}^{2}\widehat{\mathbf{S}}_{N}^{2}\mathbf{\Lambda}\right)
&=\frac{N^{2}}{4}\left( \widehat{\sigma}_{1\mid N}^{4}/(n_{1}-1) + \widehat{\sigma}_{2\mid N}^{4}/(n_{2}-1) \right)
,
\end{align*}
yielding
$\widehat{f}_{0}
=
\left(\widehat{\sigma}_{1\mid N}^{2} + \widehat{\sigma}_{2\mid N}^{2}\right)^{2}
/
\left( \widehat{\sigma}_{1\mid N}^{4}/(n_{1}-1) + \widehat{\sigma}_{2\mid N}^{4}/(n_{2}-1) \right) = \widehat{\sigma}_{N}^{4}/
\left( \widehat{\sigma}_{1\mid N}^{4}/(n_{1}-1) + \widehat{\sigma}_{2\mid N}^{4}/(n_{2}-1) \right)$.

\section*{Part II -- Simulations}
Now, we will briefly discuss how we dealt with cases where the variance estimates turned out to be zero or negative.
Thereafter, we will present simulation results in more detail and for more settings than in the main manuscript.

\subsection*{Exception handling}

The fact that $\widehat{\sigma}^{2}_{WMW}\leq 0$ can only occur when all outcomes for patients on both treatment arms coincide, that is to say,
$$x_{11} = \dots = x_{1n_{1}} = x_{21} = \dots = x_{2n_{2}},$$
yielding $\int \widehat{F}^{2}d\widehat{F}=\nicefrac{1}{4}$ and consequently $\widehat{\sigma}_{WMW}^{2}=0$.
With the Mann-Whitney parameter $p$ remaining unchanged, we then pretended that the last observation was different,
$$x_{11} = \dots = x_{1n_{1k}} = x_{21} = \dots \neq x_{2n_{2k}},$$
yielding $\widehat{\sigma}^{2}_{WMW}=\nicefrac{1}{4n_{1}n_{2}}$ and thus $T_{WMW}=0$ leading to nonrejection of the null hypothesis. We likewise set $T_{N}$ and $T_{BM}$ to zero as well in such cases, although we replaced the variances in a different manner, i.e., we always used $\max(\nicefrac{1}{n_{1}^{2}n_{2}^{2}} \, , \, \widehat{\sigma}^{2}_{N})$ instead of $\widehat{\sigma}^{2}_{N}$ and $\max(\nicefrac{1}{n_{1}^{2}n_{2}^{2}} \, , \, \widehat{\sigma}^{2}_{BM})$ instead of $\widehat{\sigma}^{2}_{BM}$. Note that if all values in both samples coincide, we would have $\widehat{p}^{2}=\widehat{\tau}_{0}=\widehat{\tau}_{1}=\widehat{\tau}_{2}=\nicefrac{1}{4}$, yielding $\widehat{\sigma}_{N}^{2}=\widehat{\sigma}_{BM}^{2}=0$. Moreover, $\widehat{\sigma}_{PM}^{2}=\nicefrac{1}{4n_{1}n_{2}}$.

These lower bounds for the unbiased and Brunner-Munzel variance estimates are motivated by the opposite degenerate case, i.e., completely separated samples without ties such as
$$x_{11} < \dots < x_{1n_{1k}} < x_{21} < \dots < x_{2n_{2k}} \, \, \, \text{ or } \, \, \, x_{11} > \dots > x_{1n_{1k}} > x_{21} > \dots > x_{2n_{2k}},$$
such that either $\widehat{p}^{2}=\widehat{\tau}_{0}=\widehat{\tau}_{1}=\widehat{\tau}_{2}=1$ or $\widehat{p}^{2}=\widehat{\tau}_{0}=\widehat{\tau}_{1}=\widehat{\tau}_{2}=0$, producing $\widehat{\sigma}_{N}^{2}=\widehat{\sigma}_{BM}^{2}=\sigma_{PM}^{2}=0$. Taking a similar approach as before (see also Brunner et al. \citeyear{np} and \citeyear{brunnerpreprint}), we then pretended the sample was slightly different, i.e.,
$$x_{11} < \dots < x_{1(n_{1}-1)} < x_{21} < x_{1n_{1}} < x_{22} < \dots < x_{2n_{2k}},$$
or 
$$x_{11} > \dots > x_{1(n_{1}-1)} > x_{21} > x_{1n_{1}} > x_{22} > \dots > x_{2n_{2k}},$$
yielding a slight change in the effect estimate $\widehat{p}=1-\nicefrac{1}{n_{1}n_{2}}$ or $\widehat{p}=\nicefrac{1}{n_{1}n_{2}}$ respectively. In this changed settings, we would have $\widehat{\sigma}_{N}^{2}=\widehat{\sigma}_{BM}^{2}=\nicefrac{1}{n_{1}^{2}n_{2}^{2}}$. As regards the \emph{logit} transformation, we employed the changed Mann-Whitney effect estimates as well so that the resulting test statistics would not be undefined. As for Perme and Manveski's [\citeyear{pm}] approach in completely separated samples, we used their new ``shift method'' to construct confidence intervals for $p$ and rejected the null hypothesis if and only if the number $\nicefrac{1}{2}$ was not an element of this confidence interval.

As far as the degrees of freedom in separated samples are concerned, we assumed $\widehat{\sigma}_{1}^{2}=\widehat{\sigma}_{2}^{2}>0$, giving rise to $df=\{ N^{2} (n_{1}-1) (n_{2}-1) \} / \{ n_{1}^{2}(n_{1}-1) + n_{2}^{2}(n_{2}-1)\}$. The other degrees of freedom were chosen accordingly.

However, since we only considered scenarios with $\min(n_{1},n_{2})\geq 7$, one might just as well have opted to always reject the null hypothesis in case of completely separated samples and never to reject it when all values coincide -- no matter which test statistic is at issue. It would have virtually never resulted in a different decision.

\subsection*{More simulation results -- Tables and Graphs}

\begin{figure}[ht!]
\centerline{\includegraphics[scale=0.65]{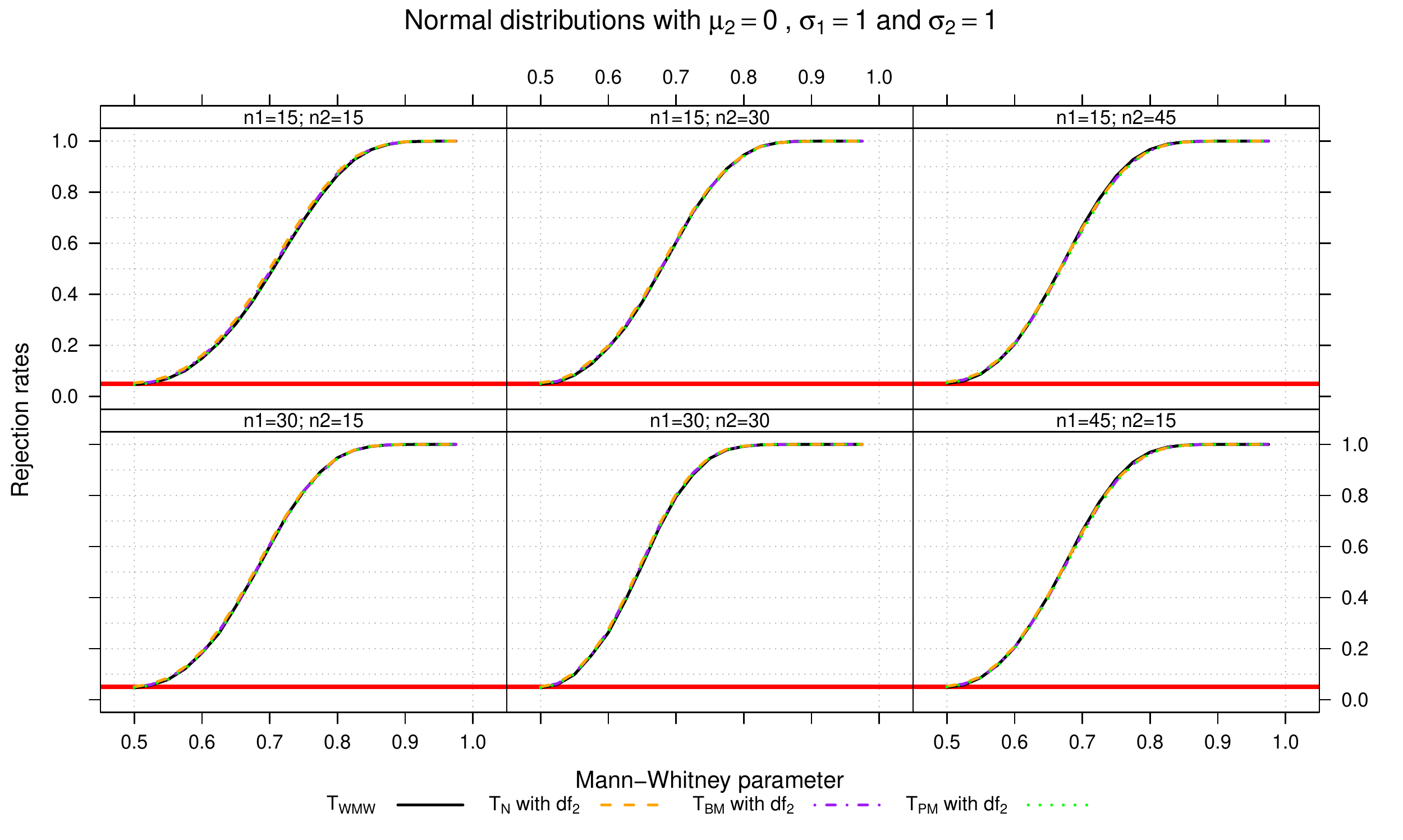}}
\caption{Power graphs for normal distributions based on 10\,000 simulation runs}{}
\label{fig:app1}
\end{figure}

\begin{figure}[ht!]
\centerline{\includegraphics[scale=0.65]{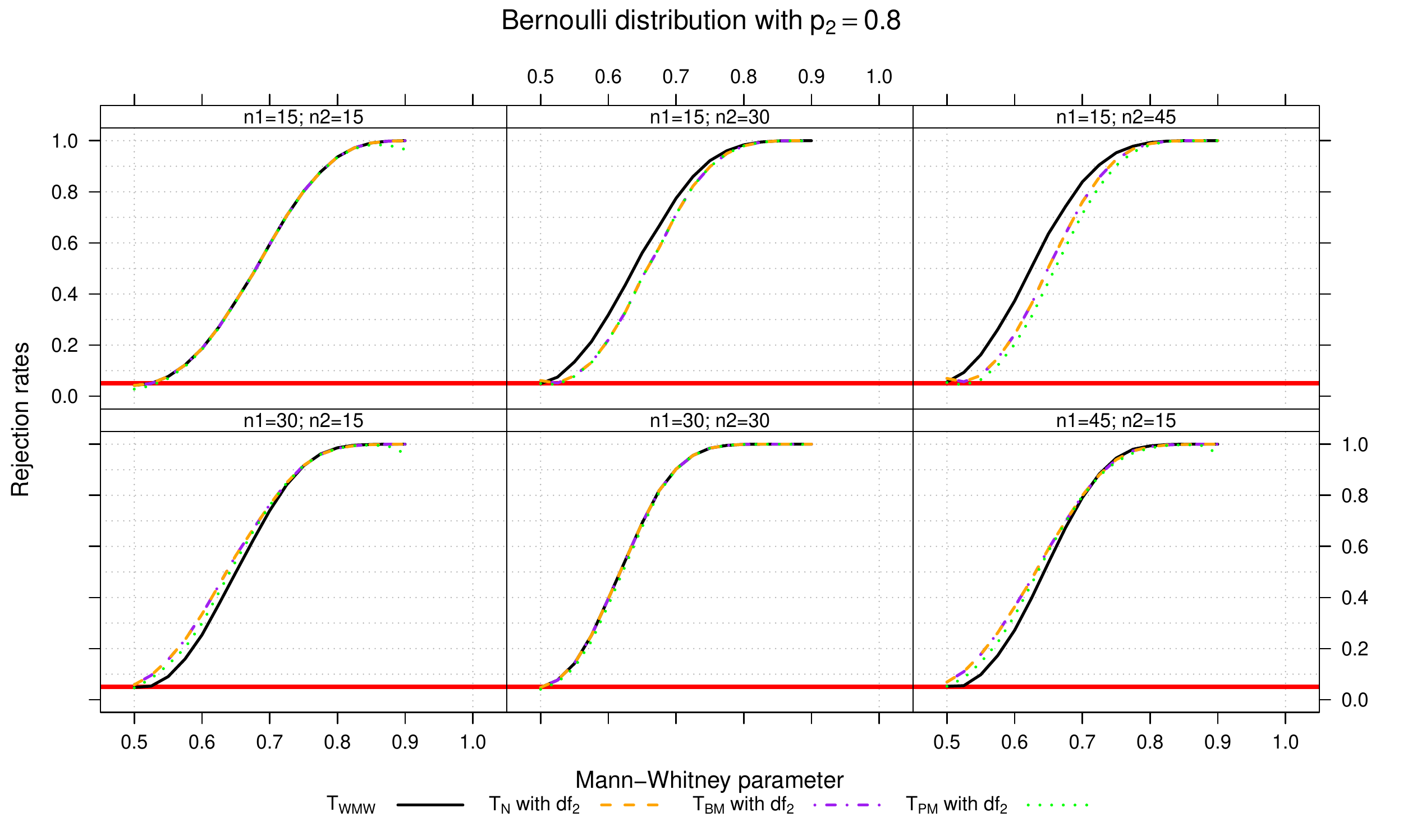}}
\caption{Power graphs for Bernoulli distributions based on 10\,000 simulation runs}{}
\label{fig:app2}
\end{figure}

Tables \ref{tab:app1} to \ref{tab:app2} show more detailed simulation results in terms of type I error rates for scenarios already treated in the main manuscript, while Table \ref{tab:app3} reports type I error rates for Pauly et al.'s studentised permutation approach [\citeyear{asen}] for exponential and binomial distributions not considered before. Figures \ref{fig:app1} to \ref{fig:app2} depict power curves for a range of distributions, whereas Tables \ref{tab:app4} to \ref{tab:app5} provide power results for Pauly et al.'s studentised permutation approach as regards a Mann-Whitney effect of $p=0.7$.

\begin{table}[ht!]
\small\sf\centering
\caption{Mean variance estimates as regards normal distributions $F_{1}=\mathcal{N}(0,\sigma_{1}^{2})$ and $F_{2}=\mathcal{N}(0,\sigma_{2}^{2})$ based on 100\,000 replications, with $\mathbb{V}(\widehat{p})$ and $sep$ denoting the true variance estimand and the relative frequency of the occurrence of completely separated samples respectively}
\begin{tabular}{rrrrcccccc}
  \hline
  $n_{1}$ & $n_{2}$ & $\sigma_{1}$ & $\sigma_{2}$  & $\mathbb{V}(\widehat{p})$ & $\widehat{\sigma}_{N}^{2}$ & $\widehat{\sigma}_{WMW}^{2}$ & $\widehat{\sigma}_{BM}^{2}$ & $\widehat{\sigma}_{PM}^{2}$ & sep \\ \hline
   7 &  7 & 1 & 1 & 0.02551020 & 0.02551208 & 0.02551020 & 0.02721286 & 0.02790682 & 0.00057 \\ 
  10 &  7 & 1 & 1 & 0.02142857 & 0.02142512 & 0.02142857 & 0.02261603 & 0.02321162 & 0.00009 \\ 
   7 & 10 & 1 & 1 & 0.02142857 & 0.02145043 & 0.02142857 & 0.02264110 & 0.02323616 & 0.00011 \\ 
  10 & 10 & 1 & 1 & 0.01750000 & 0.01749250 & 0.01750000 & 0.01832616 & 0.01881817 & 0.00000 \\ 
  15 & 15 & 1 & 1 & 0.01148148 & 0.01147906 & 0.01148148 & 0.01184935 & 0.01211924 & 0.00000 \\ 
  30 & 15 & 1 & 1 & 0.00851852 & 0.00851658 & 0.00851852 & 0.00870183 & 0.00884963 & 0.00000 \\ 
  15 & 30 & 1 & 1 & 0.00851852 & 0.00851616 & 0.00851852 & 0.00870137 & 0.00884913 & 0.00000 \\ 
  30 & 30 & 1 & 1 & 0.00564815 & 0.00564761 & 0.00564815 & 0.00574021 & 0.00582036 & 0.00000 \\ 
  15 & 45 & 1 & 1 & 0.00753086 & 0.00752970 & 0.00753086 & 0.00765317 & 0.00775451 & 0.00000 \\ 
  15 & 60 & 1 & 1 & 0.00703704 & 0.00703405 & 0.00703704 & 0.00712667 & 0.00720374 & 0.00000 \\ 
  15 & 75 & 1 & 1 & 0.00674074 & 0.00674097 & 0.00674074 & 0.00681506 & 0.00687722 & 0.00000 \\ 
  45 & 15 & 1 & 1 & 0.00753086 & 0.00752664 & 0.00753086 & 0.00765009 & 0.00775143 & 0.00000 \\ 
  60 & 15 & 1 & 1 & 0.00703704 & 0.00703678 & 0.00703704 & 0.00712939 & 0.00720646 & 0.00000 \\ 
  75 & 15 & 1 & 1 & 0.00674074 & 0.00674270 & 0.00674074 & 0.00681678 & 0.00687894 & 0.00000 \\ \hline
   7 &  7 & 1 & 3 & 0.02887661 & 0.02888177 & 0.02551020 & 0.03002283 & 0.03024767 & 0.00192 \\ 
  10 &  7 & 1 & 3 & 0.02785149 & 0.02784806 & 0.02142857 & 0.02864640 & 0.02885283 & 0.00100 \\ 
   7 & 10 & 1 & 3 & 0.02089686 & 0.02089052 & 0.02142857 & 0.02169008 & 0.02199699 & 0.00017 \\ 
  10 & 10 & 1 & 3 & 0.01997431 & 0.01997592 & 0.01750000 & 0.02053410 & 0.02078088 & 0.00010 \\ 
  15 & 15 & 1 & 3 & 0.01319212 & 0.01319536 & 0.01148148 & 0.01344344 & 0.01359980 & 0.00000 \\ 
  30 & 15 & 1 & 3 & 0.01253662 & 0.01253765 & 0.00851852 & 0.01266181 & 0.01274569 & 0.00000 \\ 
  15 & 30 & 1 & 3 & 0.00712746 & 0.00712714 & 0.00851852 & 0.00725117 & 0.00734695 & 0.00000 \\ 
  30 & 30 & 1 & 3 & 0.00653401 & 0.00653393 & 0.00564815 & 0.00659594 & 0.00664655 & 0.00000 \\ 
  15 & 45 & 1 & 3 & 0.00510591 & 0.00510523 & 0.00753086 & 0.00518793 & 0.00525571 & 0.00000 \\ 
  15 & 60 & 1 & 3 & 0.00409514 & 0.00409414 & 0.00703704 & 0.00415615 & 0.00420844 & 0.00000 \\ 
  15 & 75 & 1 & 3 & 0.00348867 & 0.00348763 & 0.00674074 & 0.00353725 & 0.00357981 & 0.00000 \\ 
  45 & 15 & 1 & 3 & 0.01231812 & 0.01231834 & 0.00753086 & 0.01240100 & 0.01245807 & 0.00000 \\ 
  60 & 15 & 1 & 3 & 0.01220887 & 0.01221170 & 0.00703704 & 0.01227375 & 0.01231707 & 0.00000 \\ 
  75 & 15 & 1 & 3 & 0.01214332 & 0.01214928 & 0.00674074 & 0.01219889 & 0.01223372 & 0.00000 \\ \hline
   7 &  7 & 1 & 5 & 0.03104145 & 0.03103983 & 0.02551020 & 0.03182096 & 0.03174387 & 0.00386 \\ 
  10 &  7 & 1 & 5 & 0.03054540 & 0.03053915 & 0.02142857 & 0.03108554 & 0.03106235 & 0.00300 \\ 
   7 & 10 & 1 & 5 & 0.02199142 & 0.02198072 & 0.02142857 & 0.02252815 & 0.02262856 & 0.00039 \\ 
  10 & 10 & 1 & 5 & 0.02156546 & 0.02156787 & 0.01750000 & 0.02194920 & 0.02203857 & 0.00026 \\ 
  15 & 15 & 1 & 5 & 0.01429217 & 0.01429766 & 0.01148148 & 0.01446706 & 0.01455033 & 0.00001 \\ 
  30 & 15 & 1 & 5 & 0.01400327 & 0.01400415 & 0.00851852 & 0.01408901 & 0.01413427 & 0.00000 \\ 
  15 & 30 & 1 & 5 & 0.00735018 & 0.00735051 & 0.00851852 & 0.00743522 & 0.00749513 & 0.00000 \\ 
  30 & 30 & 1 & 5 & 0.00710368 & 0.00710316 & 0.00564815 & 0.00714553 & 0.00717718 & 0.00000 \\ 
  15 & 45 & 1 & 5 & 0.00503618 & 0.00503644 & 0.00753086 & 0.00509293 & 0.00513693 & 0.00000 \\ 
  15 & 60 & 1 & 5 & 0.00387919 & 0.00387916 & 0.00703704 & 0.00392152 & 0.00395603 & 0.00000 \\ 
  15 & 75 & 1 & 5 & 0.00318499 & 0.00318451 & 0.00674074 & 0.00321842 & 0.00324679 & 0.00000 \\ 
  45 & 15 & 1 & 5 & 0.01390698 & 0.01390648 & 0.00753086 & 0.01396295 & 0.01399380 & 0.00000 \\ 
  60 & 15 & 1 & 5 & 0.01385883 & 0.01386002 & 0.00703704 & 0.01390245 & 0.01392595 & 0.00000 \\ 
  75 & 15 & 1 & 5 & 0.01382994 & 0.01383470 & 0.00674074 & 0.01386858 & 0.01388745 & 0.00000 \\ 
   \hline
\end{tabular}
\label{tab:app1}
\end{table}
\vspace*{\fill}
\newpage
\begin{table}[ht!]
\small\sf\centering
\caption{Type I error rates for normal distributions $F_{1}=\mathcal{N}(0,\sigma_{1}^{2})$ and $F_{2}=\mathcal{N}(0,\sigma_{2}^{2})$ based on 100\,000 replications at a two-sided nominal significance level of $\alpha=0.05$ as regards the test statistics $T_{WMW}$ and $T_{N}$ with $t$-approximation and different degrees of freedom}
\begin{tabular}{rrrrcccccc}
  \hline
  $n_{1}$ & $n_{2}$ & $\sigma_{1}$ & $\sigma_{2}$ & $T_{WMW} $ & $T_{N} \, (df)$ & $T_{N} \, (df_{1})$ & $T_{N} \, (df_{2})$ & $T_{N} \, (df_{3})$ & $T_{N} \, (df_{4})$ \\ 
  \hline
   7 &  7 & 1 & 1 & 0.05318 & 0.06381 & 0.05821 & 0.05527 & 0.05477 & 0.06041 \\ 
  10 &  7 & 1 & 1 & 0.04348 & 0.06358 & 0.06034 & 0.05428 & 0.05297 & 0.06242 \\ 
   7 & 10 & 1 & 1 & 0.04290 & 0.06265 & 0.05972 & 0.05399 & 0.05251 & 0.06152 \\ 
  10 & 10 & 1 & 1 & 0.05320 & 0.06310 & 0.06058 & 0.05696 & 0.05148 & 0.06199 \\ 
  15 & 15 & 1 & 1 & 0.05072 & 0.05809 & 0.05736 & 0.05651 & 0.04973 & 0.05786 \\ 
  30 & 15 & 1 & 1 & 0.04906 & 0.05595 & 0.05518 & 0.05417 & 0.05116 & 0.05587 \\ 
  15 & 30 & 1 & 1 & 0.04911 & 0.05588 & 0.05505 & 0.05431 & 0.05121 & 0.05585 \\ 
  30 & 30 & 1 & 1 & 0.04950 & 0.05336 & 0.05327 & 0.05306 & 0.04917 & 0.05332 \\ 
  15 & 45 & 1 & 1 & 0.04891 & 0.05535 & 0.05453 & 0.05340 & 0.05308 & 0.05533 \\ 
  15 & 60 & 1 & 1 & 0.04889 & 0.05401 & 0.05292 & 0.05192 & 0.05333 & 0.05401 \\ 
  15 & 75 & 1 & 1 & 0.04959 & 0.05562 & 0.05432 & 0.05292 & 0.05596 & 0.05555 \\ 
  45 & 15 & 1 & 1 & 0.04945 & 0.05556 & 0.05457 & 0.05329 & 0.05305 & 0.05551 \\ 
  60 & 15 & 1 & 1 & 0.04930 & 0.05509 & 0.05388 & 0.05262 & 0.05436 & 0.05502 \\ 
  75 & 15 & 1 & 1 & 0.04918 & 0.05434 & 0.05309 & 0.05164 & 0.05512 & 0.05426 \\ \hline
   7 &  7 & 1 & 3 & 0.07223 & 0.06006 & 0.04757 & 0.04572 & 0.05560 & 0.04757 \\ 
  10 &  7 & 1 & 3 & 0.08066 & 0.05910 & 0.05070 & 0.04509 & 0.06123 & 0.05169 \\ 
   7 & 10 & 1 & 3 & 0.04122 & 0.05436 & 0.05316 & 0.05091 & 0.04560 & 0.05369 \\ 
  10 & 10 & 1 & 3 & 0.07141 & 0.05720 & 0.05517 & 0.05172 & 0.05195 & 0.05541 \\ 
  15 & 15 & 1 & 3 & 0.06833 & 0.05484 & 0.05383 & 0.05235 & 0.05151 & 0.05380 \\ 
  30 & 15 & 1 & 3 & 0.10568 & 0.05408 & 0.05265 & 0.05111 & 0.05651 & 0.05341 \\ 
  15 & 30 & 1 & 3 & 0.03163 & 0.05351 & 0.05326 & 0.05299 & 0.04671 & 0.05313 \\ 
  30 & 30 & 1 & 3 & 0.07001 & 0.05367 & 0.05338 & 0.05308 & 0.05215 & 0.05338 \\ 
  15 & 45 & 1 & 3 & 0.01618 & 0.05240 & 0.05210 & 0.05185 & 0.04451 & 0.05226 \\ 
  15 & 60 & 1 & 3 & 0.00965 & 0.05170 & 0.05157 & 0.05130 & 0.04416 & 0.05175 \\ 
  15 & 75 & 1 & 3 & 0.00616 & 0.05321 & 0.05297 & 0.05263 & 0.04573 & 0.05328 \\ 
  45 & 15 & 1 & 3 & 0.12749 & 0.05448 & 0.05322 & 0.05153 & 0.05882 & 0.05419 \\ 
  60 & 15 & 1 & 3 & 0.14104 & 0.05398 & 0.05269 & 0.05116 & 0.05974 & 0.05373 \\ 
  75 & 15 & 1 & 3 & 0.14588 & 0.05344 & 0.05207 & 0.05068 & 0.05964 & 0.05307 \\ \hline
   7 &  7 & 1 & 5 & 0.08821 & 0.06395 & 0.03841 & 0.03694 & 0.06155 & 0.03766 \\ 
  10 &  7 & 1 & 5 & 0.09850 & 0.06031 & 0.04039 & 0.03648 & 0.06859 & 0.04089 \\ 
   7 & 10 & 1 & 5 & 0.04932 & 0.04751 & 0.04705 & 0.04573 & 0.04233 & 0.04723 \\ 
  10 & 10 & 1 & 5 & 0.08485 & 0.04977 & 0.04885 & 0.04618 & 0.04762 & 0.04876 \\ 
  15 & 15 & 1 & 5 & 0.08132 & 0.05361 & 0.05249 & 0.05108 & 0.05200 & 0.05246 \\ 
  30 & 15 & 1 & 5 & 0.12597 & 0.05277 & 0.05160 & 0.05022 & 0.05638 & 0.05224 \\ 
  15 & 30 & 1 & 5 & 0.03419 & 0.05273 & 0.05244 & 0.05207 & 0.04689 & 0.05203 \\ 
  30 & 30 & 1 & 5 & 0.08136 & 0.05328 & 0.05297 & 0.05257 & 0.05253 & 0.05292 \\ 
  15 & 45 & 1 & 5 & 0.01548 & 0.05063 & 0.05052 & 0.05039 & 0.04300 & 0.05038 \\ 
  15 & 60 & 1 & 5 & 0.00770 & 0.05121 & 0.05117 & 0.05111 & 0.04305 & 0.05110 \\ 
  15 & 75 & 1 & 5 & 0.00431 & 0.05140 & 0.05130 & 0.05116 & 0.04373 & 0.05134 \\ 
  45 & 15 & 1 & 5 & 0.15064 & 0.05355 & 0.05230 & 0.05092 & 0.05875 & 0.05322 \\ 
  60 & 15 & 1 & 5 & 0.16777 & 0.05333 & 0.05194 & 0.05023 & 0.05951 & 0.05302 \\ 
  75 & 15 & 1 & 5 & 0.17407 & 0.05311 & 0.05171 & 0.05004 & 0.05931 & 0.05282 \\  
   \hline
\end{tabular}
\end{table}
\vspace*{\fill}
\newpage
\begin{table}[ht!]
\small\sf\centering
\caption{Type I error rates for normal distributions $F_{1}=\mathcal{N}(0,\sigma_{1}^{2})$ and $F_{2}=\mathcal{N}(0,\sigma_{2}^{2})$ based on 100\,000 replications at a two-sided nominal significance level of $\alpha=0.05$ as regards the test statistics $T_{WMW}$ and $T_{BM}$ with $t$-approximation and different degrees of freedom}
\begin{tabular}{rrrrcccccc}
  \hline
  $n_{1}$ & $n_{2}$ & $\sigma_{1}$ & $\sigma_{2}$ & $T_{WMW} $ & $T_{BM} \, (df)$ & $T_{BM} \, (df_{1})$ & $T_{BM} \, (df_{2})$ & $T_{BM} \, (df_{3})$ & $T_{BM} \, (df_{4})$ \\ 
  \hline
   7 &  7 & 1 & 1 & 0.05318 & 0.05759 & 0.05527 & 0.04796 & 0.04444 & 0.05527 \\ 
  10 &  7 & 1 & 1 & 0.04348 & 0.05732 & 0.05453 & 0.05003 & 0.04652 & 0.05593 \\ 
   7 & 10 & 1 & 1 & 0.04290 & 0.05653 & 0.05416 & 0.04975 & 0.04611 & 0.05516 \\ 
  10 & 10 & 1 & 1 & 0.05320 & 0.05625 & 0.05454 & 0.05225 & 0.04642 & 0.05547 \\ 
  15 & 15 & 1 & 1 & 0.05072 & 0.05454 & 0.05381 & 0.05290 & 0.04624 & 0.05432 \\ 
  30 & 15 & 1 & 1 & 0.04906 & 0.05341 & 0.05266 & 0.05183 & 0.04871 & 0.05335 \\ 
  15 & 30 & 1 & 1 & 0.04911 & 0.05372 & 0.05303 & 0.05207 & 0.04904 & 0.05361 \\ 
  30 & 30 & 1 & 1 & 0.04950 & 0.05170 & 0.05160 & 0.05138 & 0.04745 & 0.05168 \\ 
  15 & 45 & 1 & 1 & 0.04891 & 0.05353 & 0.05275 & 0.05167 & 0.05116 & 0.05347 \\ 
  15 & 60 & 1 & 1 & 0.04889 & 0.05244 & 0.05159 & 0.05044 & 0.05190 & 0.05241 \\ 
  15 & 75 & 1 & 1 & 0.04959 & 0.05420 & 0.05290 & 0.05144 & 0.05464 & 0.05418 \\ 
  45 & 15 & 1 & 1 & 0.04945 & 0.05362 & 0.05264 & 0.05150 & 0.05125 & 0.05354 \\ 
  60 & 15 & 1 & 1 & 0.04930 & 0.05329 & 0.05239 & 0.05120 & 0.05261 & 0.05331 \\ 
  75 & 15 & 1 & 1 & 0.04918 & 0.05292 & 0.05174 & 0.05055 & 0.05384 & 0.05289 \\ \hline
   7 &  7 & 1 & 3 & 0.07223 & 0.05751 & 0.04572 & 0.04206 & 0.05082 & 0.04572 \\ 
  10 &  7 & 1 & 3 & 0.08066 & 0.05703 & 0.04741 & 0.04320 & 0.05842 & 0.04918 \\ 
   7 & 10 & 1 & 3 & 0.04122 & 0.05191 & 0.05114 & 0.04734 & 0.04171 & 0.04881 \\ 
  10 & 10 & 1 & 3 & 0.07141 & 0.05358 & 0.05201 & 0.04893 & 0.04961 & 0.05216 \\ 
  15 & 15 & 1 & 3 & 0.06833 & 0.05267 & 0.05154 & 0.05036 & 0.04949 & 0.05156 \\ 
  30 & 15 & 1 & 3 & 0.10568 & 0.05276 & 0.05137 & 0.04995 & 0.05525 & 0.05208 \\ 
  15 & 30 & 1 & 3 & 0.03163 & 0.05159 & 0.05137 & 0.05111 & 0.04474 & 0.05127 \\ 
  30 & 30 & 1 & 3 & 0.07001 & 0.05265 & 0.05244 & 0.05218 & 0.05117 & 0.05245 \\ 
  15 & 45 & 1 & 3 & 0.01618 & 0.05026 & 0.05011 & 0.04994 & 0.04294 & 0.05022 \\ 
  15 & 60 & 1 & 3 & 0.00965 & 0.05014 & 0.05002 & 0.04978 & 0.04290 & 0.05016 \\ 
  15 & 75 & 1 & 3 & 0.00616 & 0.05166 & 0.05146 & 0.05114 & 0.04444 & 0.05177 \\ 
  45 & 15 & 1 & 3 & 0.12749 & 0.05373 & 0.05237 & 0.05080 & 0.05804 & 0.05319 \\ 
  60 & 15 & 1 & 3 & 0.14104 & 0.05353 & 0.05210 & 0.05044 & 0.05906 & 0.05312 \\ 
  75 & 15 & 1 & 3 & 0.14588 & 0.05294 & 0.05153 & 0.05009 & 0.05903 & 0.05264 \\ \hline
   7 &  7 & 1 & 5 & 0.08821 & 0.06286 & 0.03694 & 0.03496 & 0.05923 & 0.03694 \\ 
  10 &  7 & 1 & 5 & 0.09850 & 0.05970 & 0.03836 & 0.03563 & 0.06734 & 0.03960 \\ 
   7 & 10 & 1 & 5 & 0.04932 & 0.04667 & 0.04620 & 0.04335 & 0.04022 & 0.04383 \\ 
  10 & 10 & 1 & 5 & 0.08485 & 0.04802 & 0.04721 & 0.04453 & 0.04680 & 0.04718 \\ 
  15 & 15 & 1 & 5 & 0.08132 & 0.05215 & 0.05087 & 0.04987 & 0.05054 & 0.05077 \\ 
  30 & 15 & 1 & 5 & 0.12597 & 0.05223 & 0.05097 & 0.04948 & 0.05565 & 0.05149 \\ 
  15 & 30 & 1 & 5 & 0.03419 & 0.05133 & 0.05112 & 0.05079 & 0.04559 & 0.05074 \\ 
  30 & 30 & 1 & 5 & 0.08136 & 0.05252 & 0.05217 & 0.05186 & 0.05181 & 0.05218 \\ 
  15 & 45 & 1 & 5 & 0.01548 & 0.04907 & 0.04899 & 0.04890 & 0.04202 & 0.04884 \\ 
  15 & 60 & 1 & 5 & 0.00770 & 0.04998 & 0.04992 & 0.04984 & 0.04198 & 0.04980 \\ 
  15 & 75 & 1 & 5 & 0.00431 & 0.05018 & 0.05010 & 0.05003 & 0.04260 & 0.05012 \\ 
  45 & 15 & 1 & 5 & 0.15064 & 0.05315 & 0.05184 & 0.05044 & 0.05825 & 0.05266 \\ 
  60 & 15 & 1 & 5 & 0.16777 & 0.05297 & 0.05155 & 0.04993 & 0.05919 & 0.05266 \\ 
  75 & 15 & 1 & 5 & 0.17407 & 0.05279 & 0.05132 & 0.04986 & 0.05906 & 0.05249 \\ 
   \hline
\end{tabular}
\end{table}
\vspace*{\fill}
\newpage
\begin{table}[ht!]
\small\sf\centering
\caption{Type I error rates for normal distributions $F_{1}=\mathcal{N}(0,\sigma_{1}^{2})$ and $F_{2}=\mathcal{N}(0,\sigma_{2}^{2})$ based on 100\,000 replications at a two-sided nominal significance level of $\alpha=0.05$ as regards the test statistics $T_{WMW}$ and $T_{PM}$ with $t$-approximation and different degrees of freedom}
\begin{tabular}{rrrrcccccc}
  \hline
  $n_{1}$ & $n_{2}$ & $\sigma_{1}$ & $\sigma_{2}$ & $T_{WMW} $ & $T_{PM} \, (df)$ & $T_{PM} \, (df_{1})$ & $T_{PM} \, (df_{2})$ & $T_{PM} \, (df_{3})$ & $T_{PM} \, (df_{4})$ \\ 
  \hline
   7 &  7 & 1 & 1 & 0.05318 & 0.05639 & 0.05139 & 0.04304 & 0.04444 & 0.05411 \\ 
  10 &  7 & 1 & 1 & 0.04348 & 0.05525 & 0.05165 & 0.04725 & 0.04510 & 0.05371 \\ 
   7 & 10 & 1 & 1 & 0.04290 & 0.05494 & 0.05134 & 0.04708 & 0.04464 & 0.05322 \\ 
  10 & 10 & 1 & 1 & 0.05320 & 0.05370 & 0.05181 & 0.04856 & 0.04380 & 0.05302 \\ 
  15 & 15 & 1 & 1 & 0.05072 & 0.05183 & 0.05109 & 0.05012 & 0.04405 & 0.05163 \\ 
  30 & 15 & 1 & 1 & 0.04906 & 0.05163 & 0.05095 & 0.05001 & 0.04691 & 0.05158 \\ 
  15 & 30 & 1 & 1 & 0.04911 & 0.05177 & 0.05110 & 0.05004 & 0.04717 & 0.05169 \\ 
  30 & 30 & 1 & 1 & 0.04950 & 0.05013 & 0.04998 & 0.04978 & 0.04604 & 0.05013 \\ 
  15 & 45 & 1 & 1 & 0.04891 & 0.05221 & 0.05140 & 0.05040 & 0.04985 & 0.05220 \\ 
  15 & 60 & 1 & 1 & 0.04889 & 0.05144 & 0.05046 & 0.04945 & 0.05056 & 0.05138 \\ 
  15 & 75 & 1 & 1 & 0.04959 & 0.05293 & 0.05174 & 0.05046 & 0.05350 & 0.05288 \\ 
  45 & 15 & 1 & 1 & 0.04945 & 0.05208 & 0.05122 & 0.05002 & 0.04985 & 0.05202 \\ 
  60 & 15 & 1 & 1 & 0.04930 & 0.05223 & 0.05121 & 0.04995 & 0.05119 & 0.05217 \\ 
  75 & 15 & 1 & 1 & 0.04918 & 0.05185 & 0.05075 & 0.04955 & 0.05288 & 0.05181 \\ \hline
   7 &  7 & 1 & 3 & 0.07223 & 0.05626 & 0.04437 & 0.03917 & 0.05082 & 0.04532 \\ 
  10 &  7 & 1 & 3 & 0.08066 & 0.05629 & 0.04612 & 0.04175 & 0.05760 & 0.05443 \\ 
   7 & 10 & 1 & 3 & 0.04122 & 0.05375 & 0.04814 & 0.04540 & 0.04084 & 0.04766 \\ 
  10 & 10 & 1 & 3 & 0.07141 & 0.05220 & 0.05021 & 0.04705 & 0.04767 & 0.05061 \\ 
  15 & 15 & 1 & 3 & 0.06833 & 0.05131 & 0.05036 & 0.04926 & 0.04825 & 0.05034 \\ 
  30 & 15 & 1 & 3 & 0.10568 & 0.05191 & 0.05047 & 0.04919 & 0.05437 & 0.05111 \\ 
  15 & 30 & 1 & 3 & 0.03163 & 0.05026 & 0.05003 & 0.04973 & 0.04355 & 0.04990 \\ 
  30 & 30 & 1 & 3 & 0.07001 & 0.05178 & 0.05157 & 0.05124 & 0.05034 & 0.05156 \\ 
  15 & 45 & 1 & 3 & 0.01618 & 0.04904 & 0.04881 & 0.04859 & 0.04187 & 0.04892 \\ 
  15 & 60 & 1 & 3 & 0.00965 & 0.04901 & 0.04879 & 0.04853 & 0.04174 & 0.04907 \\ 
  15 & 75 & 1 & 3 & 0.00616 & 0.05047 & 0.05023 & 0.04978 & 0.04333 & 0.05054 \\ 
  45 & 15 & 1 & 3 & 0.12749 & 0.05307 & 0.05171 & 0.05041 & 0.05746 & 0.05266 \\ 
  60 & 15 & 1 & 3 & 0.14104 & 0.05299 & 0.05155 & 0.05000 & 0.05860 & 0.05271 \\ 
  75 & 15 & 1 & 3 & 0.14588 & 0.05257 & 0.05124 & 0.04973 & 0.05871 & 0.05227 \\ \hline
   7 &  7 & 1 & 5 & 0.08821 & 0.06181 & 0.03626 & 0.03319 & 0.05923 & 0.03663 \\ 
  10 &  7 & 1 & 5 & 0.09850 & 0.05944 & 0.03790 & 0.03504 & 0.06700 & 0.05817 \\ 
   7 & 10 & 1 & 5 & 0.04932 & 0.05706 & 0.04410 & 0.04229 & 0.03960 & 0.04338 \\ 
  10 & 10 & 1 & 5 & 0.08485 & 0.04748 & 0.04596 & 0.04376 & 0.04546 & 0.04606 \\ 
  15 & 15 & 1 & 5 & 0.08132 & 0.05136 & 0.05019 & 0.04928 & 0.04981 & 0.05016 \\ 
  30 & 15 & 1 & 5 & 0.12597 & 0.05169 & 0.05037 & 0.04907 & 0.05529 & 0.05113 \\ 
  15 & 30 & 1 & 5 & 0.03419 & 0.05050 & 0.05021 & 0.04985 & 0.04479 & 0.04985 \\ 
  30 & 30 & 1 & 5 & 0.08136 & 0.05194 & 0.05166 & 0.05128 & 0.05128 & 0.05165 \\ 
  15 & 45 & 1 & 5 & 0.01548 & 0.04823 & 0.04810 & 0.04800 & 0.04138 & 0.04784 \\ 
  15 & 60 & 1 & 5 & 0.00770 & 0.04914 & 0.04908 & 0.04898 & 0.04093 & 0.04902 \\ 
  15 & 75 & 1 & 5 & 0.00431 & 0.04915 & 0.04910 & 0.04899 & 0.04195 & 0.04906 \\ 
  45 & 15 & 1 & 5 & 0.15064 & 0.05276 & 0.05164 & 0.05015 & 0.05778 & 0.05235 \\ 
  60 & 15 & 1 & 5 & 0.16777 & 0.05273 & 0.05118 & 0.04979 & 0.05890 & 0.05236 \\ 
  75 & 15 & 1 & 5 & 0.17407 & 0.05257 & 0.05114 & 0.04961 & 0.05894 & 0.05217 \\ 
   \hline
\end{tabular}
\end{table}
\vspace*{\fill}
\begin{table}[ht!]
\small\sf\centering
\caption{Mean variance estimates as regards the 5-point distributions with latent $F_{1}=\mathcal{B}(\alpha_{1},\beta_{1})$ and $F_{2}=\mathcal{B}(5,4)$ based on 100\,000 replications, with $\mathbb{V}(\widehat{p})$ and $sep$ denoting the true variance estimand and the relative frequency of the occurrence of completely separated samples respectively}
\begin{tabular}{rrrrccccccc}
  \hline
  $n_{1}$ & $n_{2}$ & $\alpha_{1}$ & $\beta_{1}$ & $\mathbb{V}(\widehat{p})$ & $\widehat{\sigma}_{N}^{2}$ & $\widehat{\sigma}_{WMW}^{2}$ & $\widehat{\sigma}_{BM}^{2}$ & $\widehat{\sigma}_{PM}^{2}$ & sep &\\
  \hline
   7 &  7 & 5      & 4 & 0.02135081 & 0.02134561 & 0.02136648 & 0.02178287 & 0.02333416 & 0.00009 \\ 
  10 &  7 & 5      & 4 & 0.01808278 & 0.01810689 & 0.01809247 & 0.01841288 & 0.01955837 & 0.00001 \\ 
   7 & 10 & 5      & 4 & 0.01808278 & 0.01808238 & 0.01809247 & 0.01838821 & 0.01953452 & 0.00001 \\ 
  10 & 10 & 5      & 4 & 0.01485400 & 0.01486281 & 0.01486064 & 0.01507701 & 0.01592065 & 0.00000 \\ 
  15 & 15 & 5      & 4 & 0.00985519 & 0.00985734 & 0.00985846 & 0.00995248 & 0.01035631 & 0.00000 \\ 
  30 & 15 & 5      & 4 & 0.00736765 & 0.00737328 & 0.00736927 & 0.00742084 & 0.00762987 & 0.00000 \\ 
  15 & 30 & 5      & 4 & 0.00736765 & 0.00737174 & 0.00736927 & 0.00741930 & 0.00762835 & 0.00000 \\ 
  30 & 30 & 5      & 4 & 0.00490385 & 0.00490477 & 0.00490462 & 0.00492854 & 0.00503659 & 0.00000 \\ 
  15 & 45 & 5      & 4 & 0.00653847 & 0.00654040 & 0.00653950 & 0.00657210 & 0.00671303 & 0.00000 \\ 
  15 & 60 & 5      & 4 & 0.00612388 & 0.00612619 & 0.00612402 & 0.00614996 & 0.00625626 & 0.00000 \\ 
  15 & 75 & 5      & 4 & 0.00587513 & 0.00587719 & 0.00587532 & 0.00589620 & 0.00598153 & 0.00000 \\ 
  45 & 15 & 5      & 4 & 0.00653847 & 0.00653712 & 0.00653950 & 0.00656879 & 0.00670973 & 0.00000 \\ 
  60 & 15 & 5      & 4 & 0.00612388 & 0.00612755 & 0.00612402 & 0.00615131 & 0.00625762 & 0.00000 \\ 
  75 & 15 & 5      & 4 & 0.00587513 & 0.00587810 & 0.00587532 & 0.00589710 & 0.00598242 & 0.00000 \\ \hline
   7 &  7 & 1.2071 & 1 & 0.02458929 & 0.02473559 & 0.02303112 & 0.02530736 & 0.02629003 & 0.00037 \\ 
  10 &  7 & 1.2071 & 1 & 0.01858647 & 0.01864873 & 0.01961235 & 0.01904997 & 0.01984263 & 0.00002 \\ 
   7 & 10 & 1.2071 & 1 & 0.02308031 & 0.02321632 & 0.01929195 & 0.02361703 & 0.02434381 & 0.00012 \\ 
  10 & 10 & 1.2071 & 1 & 0.01712232 & 0.01719060 & 0.01596299 & 0.01747059 & 0.01805212 & 0.00002 \\ 
  15 & 15 & 1.2071 & 1 & 0.01136812 & 0.01140776 & 0.01056297 & 0.01153218 & 0.01182431 & 0.00000 \\ 
  30 & 15 & 1.2071 & 1 & 0.00673455 & 0.00673987 & 0.00797797 & 0.00680209 & 0.00696066 & 0.00000 \\ 
  15 & 30 & 1.2071 & 1 & 0.01026745 & 0.01029732 & 0.00774917 & 0.01035954 & 0.01051018 & 0.00000 \\ 
  30 & 30 & 1.2071 & 1 & 0.00566068 & 0.00566779 & 0.00524136 & 0.00569890 & 0.00578042 & 0.00000 \\ 
  15 & 45 & 1.2071 & 1 & 0.00990409 & 0.00993534 & 0.00680155 & 0.00997680 & 0.01007819 & 0.00000 \\ 
  15 & 60 & 1.2071 & 1 & 0.00972377 & 0.00974447 & 0.00632434 & 0.00977558 & 0.00985204 & 0.00000 \\ 
  15 & 75 & 1.2071 & 1 & 0.00961620 & 0.00963242 & 0.00603741 & 0.00965731 & 0.00971867 & 0.00000 \\ 
  45 & 15 & 1.2071 & 1 & 0.00518592 & 0.00518722 & 0.00710263 & 0.00522868 & 0.00533715 & 0.00000 \\ 
  60 & 15 & 1.2071 & 1 & 0.00441069 & 0.00441226 & 0.00666124 & 0.00444337 & 0.00452578 & 0.00000 \\ 
  75 & 15 & 1.2071 & 1 & 0.00394525 & 0.00394590 & 0.00639503 & 0.00397079 & 0.00403718 & 0.00000 \\
   \hline
\end{tabular}
\end{table}
\vspace*{\fill}
\newpage
\begin{table}[ht!]
\small\sf\centering
\caption{Type I error rates for the 5-point distributions with latent $F_{1}=\mathcal{B}(\alpha_{1},\beta_{1})$ and $F_{2}=\mathcal{B}(5,4)$ based on 100\,000 replications at a two-sided nominal significance level of $\alpha=0.05$ as regards the test statistics $T_{WMW}$ and $T_{N}$ with $t$-approximation and different degrees of freedom}
\begin{tabular}{rrrrcccccc}
  \hline
  $n_{1}$ & $n_{2}$ & $\alpha_{1}$ & $\beta_{1}$ & $T_{WMW} $ & $T_{N} \, (df)$ & $T_{N} \, (df_{1})$ & $T_{N} \, (df_{2})$ & $T_{N} \, (df_{3})$ & $T_{N} \, (df_{4})$ \\ 
  \hline
   7 &  7 & 5      & 4 & 0.04611 & 0.06294 & 0.05821 & 0.05628 & 0.04818 & 0.06200 \\ 
  10 &  7 & 5      & 4 & 0.04761 & 0.06245 & 0.05867 & 0.05391 & 0.04932 & 0.06146 \\ 
   7 & 10 & 5      & 4 & 0.04769 & 0.06237 & 0.05892 & 0.05457 & 0.04928 & 0.06124 \\ 
  10 & 10 & 5      & 4 & 0.04832 & 0.06102 & 0.05821 & 0.05450 & 0.04865 & 0.06063 \\ 
  15 & 15 & 5      & 4 & 0.04875 & 0.05580 & 0.05517 & 0.05440 & 0.04825 & 0.05570 \\ 
  30 & 15 & 5      & 4 & 0.04814 & 0.05393 & 0.05308 & 0.05212 & 0.04889 & 0.05389 \\ 
  15 & 30 & 5      & 4 & 0.04902 & 0.05532 & 0.05466 & 0.05391 & 0.05067 & 0.05526 \\ 
  30 & 30 & 5      & 4 & 0.04857 & 0.05226 & 0.05202 & 0.05175 & 0.04802 & 0.05225 \\ 
  15 & 45 & 5      & 4 & 0.04923 & 0.05476 & 0.05375 & 0.05258 & 0.05255 & 0.05480 \\ 
  15 & 60 & 5      & 4 & 0.05026 & 0.05531 & 0.05411 & 0.05292 & 0.05466 & 0.05537 \\ 
  15 & 75 & 5      & 4 & 0.04959 & 0.05517 & 0.05403 & 0.05263 & 0.05590 & 0.05518 \\ 
  45 & 15 & 5      & 4 & 0.04898 & 0.05401 & 0.05294 & 0.05186 & 0.05199 & 0.05404 \\ 
  60 & 15 & 5      & 4 & 0.04868 & 0.05392 & 0.05296 & 0.05194 & 0.05360 & 0.05393 \\ 
  75 & 15 & 5      & 4 & 0.04856 & 0.05345 & 0.05208 & 0.05090 & 0.05408 & 0.05345 \\ \hline
   7 &  7 & 1.2071 & 1 & 0.05763 & 0.06234 & 0.05709 & 0.05126 & 0.05274 & 0.05870 \\ 
  10 &  7 & 1.2071 & 1 & 0.04264 & 0.05924 & 0.05621 & 0.05281 & 0.04510 & 0.05696 \\ 
   7 & 10 & 1.2071 & 1 & 0.07446 & 0.06349 & 0.05771 & 0.05213 & 0.05829 & 0.06074 \\ 
  10 & 10 & 1.2071 & 1 & 0.05798 & 0.05992 & 0.05740 & 0.05432 & 0.05162 & 0.05843 \\ 
  15 & 15 & 1.2071 & 1 & 0.05579 & 0.05340 & 0.05245 & 0.05146 & 0.04793 & 0.05281 \\ 
  30 & 15 & 1.2071 & 1 & 0.03218 & 0.05412 & 0.05383 & 0.05345 & 0.04716 & 0.05398 \\ 
  15 & 30 & 1.2071 & 1 & 0.08897 & 0.05436 & 0.05315 & 0.05188 & 0.05484 & 0.05390 \\ 
  30 & 30 & 1.2071 & 1 & 0.05827 & 0.05173 & 0.05149 & 0.05120 & 0.04915 & 0.05156 \\ 
  15 & 45 & 1.2071 & 1 & 0.10304 & 0.05312 & 0.05175 & 0.05027 & 0.05639 & 0.05277 \\ 
  15 & 60 & 1.2071 & 1 & 0.11529 & 0.05395 & 0.05294 & 0.05152 & 0.05833 & 0.05380 \\ 
  15 & 75 & 1.2071 & 1 & 0.12079 & 0.05404 & 0.05257 & 0.05096 & 0.05932 & 0.05384 \\ 
  45 & 15 & 1.2071 & 1 & 0.02028 & 0.05380 & 0.05342 & 0.05296 & 0.04658 & 0.05386 \\ 
  60 & 15 & 1.2071 & 1 & 0.01493 & 0.05258 & 0.05207 & 0.05145 & 0.04643 & 0.05276 \\ 
  75 & 15 & 1.2071 & 1 & 0.01190 & 0.05324 & 0.05252 & 0.05180 & 0.04777 & 0.05340 \\ 
   \hline
\end{tabular}
\end{table}
\vspace*{\fill}
\newpage
\begin{table}[ht!]
\small\sf\centering
\caption{Type I error rates for the 5-point distributions with latent $F_{1}=\mathcal{B}(\alpha_{1},\beta_{1})$ and $F_{2}=\mathcal{B}(5,4)$ based on 100\,000 replications at a two-sided nominal significance level of $\alpha=0.05$ as regards the test statistics $T_{WMW}$ and $T_{BM}$ with $t$-approximation and different degrees of freedom}
\begin{tabular}{rrrrcccccc}
  \hline
  $n_{1}$ & $n_{2}$ & $\alpha_{1}$ & $\beta_{1}$ & $T_{WMW} $ & $T_{BM} \, (df)$ & $T_{BM} \, (df_{1})$ & $T_{BM} \, (df_{2})$ & $T_{BM} \, (df_{3})$ & $T_{BM} \, (df_{4})$ \\ 
  \hline
   7 &  7 & 5      & 4 & 0.04611 & 0.06189 & 0.05808 & 0.05045 & 0.04776 & 0.05866 \\ 
  10 &  7 & 5      & 4 & 0.04761 & 0.06048 & 0.05711 & 0.05298 & 0.04742 & 0.05976 \\ 
   7 & 10 & 5      & 4 & 0.04769 & 0.06040 & 0.05723 & 0.05350 & 0.04756 & 0.05953 \\ 
  10 & 10 & 5      & 4 & 0.04832 & 0.05881 & 0.05668 & 0.05316 & 0.04747 & 0.05814 \\ 
  15 & 15 & 5      & 4 & 0.04875 & 0.05496 & 0.05421 & 0.05315 & 0.04744 & 0.05485 \\ 
  30 & 15 & 5      & 4 & 0.04814 & 0.05318 & 0.05232 & 0.05140 & 0.04836 & 0.05308 \\ 
  15 & 30 & 5      & 4 & 0.04902 & 0.05477 & 0.05409 & 0.05315 & 0.04977 & 0.05469 \\ 
  30 & 30 & 5      & 4 & 0.04857 & 0.05159 & 0.05142 & 0.05115 & 0.04748 & 0.05154 \\ 
  15 & 45 & 5      & 4 & 0.04923 & 0.05417 & 0.05315 & 0.05201 & 0.05199 & 0.05416 \\ 
  15 & 60 & 5      & 4 & 0.05026 & 0.05486 & 0.05361 & 0.05248 & 0.05422 & 0.05490 \\ 
  15 & 75 & 5      & 4 & 0.04959 & 0.05480 & 0.05362 & 0.05228 & 0.05545 & 0.05481 \\ 
  45 & 15 & 5      & 4 & 0.04898 & 0.05342 & 0.05239 & 0.05144 & 0.05127 & 0.05339 \\ 
  60 & 15 & 5      & 4 & 0.04868 & 0.05351 & 0.05263 & 0.05142 & 0.05318 & 0.05348 \\ 
  75 & 15 & 5      & 4 & 0.04856 & 0.05303 & 0.05181 & 0.05054 & 0.05369 & 0.05308 \\ \hline
   7 &  7 & 1.2071 & 1 & 0.05763 & 0.05988 & 0.05598 & 0.04755 & 0.05076 & 0.05678 \\ 
  10 &  7 & 1.2071 & 1 & 0.04264 & 0.05560 & 0.05366 & 0.05086 & 0.04357 & 0.05462 \\ 
   7 & 10 & 1.2071 & 1 & 0.07446 & 0.06077 & 0.05624 & 0.05046 & 0.05599 & 0.05909 \\ 
  10 & 10 & 1.2071 & 1 & 0.05798 & 0.05816 & 0.05533 & 0.05293 & 0.04954 & 0.05635 \\ 
  15 & 15 & 1.2071 & 1 & 0.05579 & 0.05221 & 0.05141 & 0.05031 & 0.04682 & 0.05171 \\ 
  30 & 15 & 1.2071 & 1 & 0.03218 & 0.05309 & 0.05288 & 0.05251 & 0.04634 & 0.05305 \\ 
  15 & 30 & 1.2071 & 1 & 0.08897 & 0.05357 & 0.05250 & 0.05131 & 0.05417 & 0.05325 \\ 
  30 & 30 & 1.2071 & 1 & 0.05827 & 0.05115 & 0.05092 & 0.05065 & 0.04840 & 0.05100 \\ 
  15 & 45 & 1.2071 & 1 & 0.10304 & 0.05249 & 0.05135 & 0.04976 & 0.05595 & 0.05218 \\ 
  15 & 60 & 1.2071 & 1 & 0.11529 & 0.05373 & 0.05260 & 0.05116 & 0.05793 & 0.05350 \\ 
  15 & 75 & 1.2071 & 1 & 0.12079 & 0.05372 & 0.05224 & 0.05064 & 0.05903 & 0.05348 \\ 
  45 & 15 & 1.2071 & 1 & 0.02028 & 0.05292 & 0.05242 & 0.05196 & 0.04578 & 0.05306 \\ 
  60 & 15 & 1.2071 & 1 & 0.01493 & 0.05177 & 0.05129 & 0.05075 & 0.04573 & 0.05192 \\ 
  75 & 15 & 1.2071 & 1 & 0.01190 & 0.05247 & 0.05186 & 0.05111 & 0.04696 & 0.05266 \\ 
   \hline
\end{tabular}
\end{table}
\vspace*{\fill}
\newpage
\begin{table}[ht!]
\small\sf\centering
\caption{Type I error rates for the 5-point distributions with latent $F_{1}=\mathcal{B}(\alpha_{1},\beta_{1})$ and $F_{2}=\mathcal{B}(5,4)$ based on 100\,000 replications at a two-sided nominal significance level of $\alpha=0.05$ as regards the test statistics $T_{WMW}$ and $T_{PM}$ with $t$-approximation and different degrees of freedom}
\begin{tabular}{rrrrcccccc}
  \hline
  $n_{1}$ & $n_{2}$ & $\alpha_{1}$ & $\beta_{1}$ & $T_{WMW} $ & $T_{PM} \, (df)$ & $T_{PM} \, (df_{1})$ & $T_{PM} \, (df_{2})$ & $T_{PM} \, (df_{3})$ & $T_{PM} \, (df_{4})$ \\ 
  \hline
   7 &  7 & 5      & 4 & 0.04611 & 0.05712 & 0.05114 & 0.04129 & 0.04050 & 0.05519 \\ 
  10 &  7 & 5      & 4 & 0.04761 & 0.05410 & 0.05124 & 0.04431 & 0.04190 & 0.05298 \\ 
   7 & 10 & 5      & 4 & 0.04769 & 0.05465 & 0.05167 & 0.04426 & 0.04192 & 0.05363 \\ 
  10 & 10 & 5      & 4 & 0.04832 & 0.05175 & 0.05097 & 0.04798 & 0.04230 & 0.05153 \\ 
  15 & 15 & 5      & 4 & 0.04875 & 0.05055 & 0.05006 & 0.04910 & 0.04307 & 0.05048 \\ 
  30 & 15 & 5      & 4 & 0.04814 & 0.04998 & 0.04915 & 0.04825 & 0.04543 & 0.04992 \\ 
  15 & 30 & 5      & 4 & 0.04902 & 0.05166 & 0.05076 & 0.04990 & 0.04677 & 0.05164 \\ 
  30 & 30 & 5      & 4 & 0.04857 & 0.04915 & 0.04894 & 0.04875 & 0.04518 & 0.04914 \\ 
  15 & 45 & 5      & 4 & 0.04923 & 0.05158 & 0.05071 & 0.04961 & 0.04925 & 0.05168 \\ 
  15 & 60 & 5      & 4 & 0.05026 & 0.05260 & 0.05161 & 0.05041 & 0.05220 & 0.05276 \\ 
  15 & 75 & 5      & 4 & 0.04959 & 0.05300 & 0.05186 & 0.05055 & 0.05373 & 0.05302 \\ 
  45 & 15 & 5      & 4 & 0.04898 & 0.05093 & 0.05008 & 0.04909 & 0.04871 & 0.05101 \\ 
  60 & 15 & 5      & 4 & 0.04868 & 0.05173 & 0.05076 & 0.04940 & 0.05106 & 0.05176 \\ 
  75 & 15 & 5      & 4 & 0.04856 & 0.05124 & 0.05019 & 0.04906 & 0.05204 & 0.05126 \\ \hline
   7 &  7 & 1.2071 & 1 & 0.05763 & 0.05605 & 0.05002 & 0.04387 & 0.04589 & 0.05261 \\ 
  10 &  7 & 1.2071 & 1 & 0.04264 & 0.05278 & 0.04991 & 0.04506 & 0.04032 & 0.05047 \\ 
   7 & 10 & 1.2071 & 1 & 0.07446 & 0.05799 & 0.05296 & 0.04612 & 0.05315 & 0.05534 \\ 
  10 & 10 & 1.2071 & 1 & 0.05798 & 0.05378 & 0.05169 & 0.04872 & 0.04558 & 0.05254 \\ 
  15 & 15 & 1.2071 & 1 & 0.05579 & 0.04953 & 0.04858 & 0.04752 & 0.04430 & 0.04898 \\ 
  30 & 15 & 1.2071 & 1 & 0.03218 & 0.05090 & 0.05065 & 0.05029 & 0.04391 & 0.05086 \\ 
  15 & 30 & 1.2071 & 1 & 0.08897 & 0.05202 & 0.05091 & 0.04977 & 0.05278 & 0.05161 \\ 
  30 & 30 & 1.2071 & 1 & 0.05827 & 0.04972 & 0.04946 & 0.04917 & 0.04702 & 0.04956 \\ 
  15 & 45 & 1.2071 & 1 & 0.10304 & 0.05132 & 0.05009 & 0.04876 & 0.05475 & 0.05104 \\ 
  15 & 60 & 1.2071 & 1 & 0.11529 & 0.05284 & 0.05160 & 0.05029 & 0.05679 & 0.05268 \\ 
  15 & 75 & 1.2071 & 1 & 0.12079 & 0.05288 & 0.05140 & 0.05008 & 0.05821 & 0.05266 \\ 
  45 & 15 & 1.2071 & 1 & 0.02028 & 0.05040 & 0.05004 & 0.04966 & 0.04394 & 0.05048 \\ 
  60 & 15 & 1.2071 & 1 & 0.01493 & 0.04991 & 0.04935 & 0.04874 & 0.04385 & 0.05004 \\ 
  75 & 15 & 1.2071 & 1 & 0.01190 & 0.05064 & 0.05003 & 0.04917 & 0.04494 & 0.05078 \\ 
   \hline
\end{tabular}
\label{tab:app2}
\end{table}
\vspace*{\fill}
\newpage
\begin{table}[ht!]
\small\sf\centering
\caption{Type I error rates for exponential and binomial distributions at a two-sided nominal significance level of $\alpha=0.05$ for the studentised permutation tests based on 10\,000 random permutations for each of the 10\,000 replications}
\begin{tabular}{rrcccccccc}
  \hline
  $n_{1}$ & $n_{2}$ & $F_{1}$ & $F_{2}$ & $\widetilde{T}_{N}$ & $\widetilde{T}_{BM}$ & $\widetilde{T}_{PM}$ & $\widetilde{T}_{N}^{Logit}$ & $\widetilde{T}_{BM}^{Logit}$ & $\widetilde{T}_{PM}^{Logit}$ \\ 
  \hline
   7 &  7 & $\mathcal{E}(1)$ & $\mathcal{E}(1)$ & 0.0488 & 0.0506 & 0.0502 & 0.0493 & 0.0484 & 0.0479 \\ 
   7 & 10 & $\mathcal{E}(1)$ & $\mathcal{E}(1)$ & 0.0484 & 0.0495 & 0.0496 & 0.0477 & 0.0485 & 0.0487 \\ 
  10 &  7 & $\mathcal{E}(1)$ & $\mathcal{E}(1)$ & 0.0460 & 0.0473 & 0.0473 & 0.0465 & 0.0475 & 0.0473 \\ 
  10 & 10 & $\mathcal{E}(1)$ & $\mathcal{E}(1)$ & 0.0503 & 0.0501 & 0.0504 & 0.0509 & 0.0508 & 0.0509 \\ 
  15 & 15 & $\mathcal{E}(1)$ & $\mathcal{E}(1)$ & 0.0503 & 0.0501 & 0.0502 & 0.0503 & 0.0504 & 0.0502 \\ 
  15 & 30 & $\mathcal{E}(1)$ & $\mathcal{E}(1)$ & 0.0478 & 0.0481 & 0.0484 & 0.0484 & 0.0481 & 0.0480 \\ 
  30 & 15 & $\mathcal{E}(1)$ & $\mathcal{E}(1)$ & 0.0533 & 0.0531 & 0.0531 & 0.0531 & 0.0530 & 0.0531 \\ 
  30 & 30 & $\mathcal{E}(1)$ & $\mathcal{E}(1)$ & 0.0508 & 0.0508 & 0.0507 & 0.0506 & 0.0506 & 0.0507 \\ 
  15 & 45 & $\mathcal{E}(1)$ & $\mathcal{E}(1)$ & 0.0498 & 0.0497 & 0.0498 & 0.0503 & 0.0501 & 0.0502 \\ 
  45 & 15 & $\mathcal{E}(1)$ & $\mathcal{E}(1)$ & 0.0499 & 0.0499 & 0.0499 & 0.0506 & 0.0504 & 0.0501 \\ \hline
   7 &  7 & $\mathcal{B}(5,0.6)$ & $\mathcal{B}(5,0.6)$ & 0.0319 & 0.0321 & 0.0325 & 0.0308 & 0.0309 & 0.0311 \\ 
   7 & 10 & $\mathcal{B}(5,0.6)$ & $\mathcal{B}(5,0.6)$ & 0.0341 & 0.0345 & 0.0350 & 0.0346 & 0.0348 & 0.0350 \\ 
  10 &  7 & $\mathcal{B}(5,0.6)$ & $\mathcal{B}(5,0.6)$ & 0.0353 & 0.0355 & 0.0356 & 0.0347 & 0.0349 & 0.0353 \\ 
  10 & 10 & $\mathcal{B}(5,0.6)$ & $\mathcal{B}(5,0.6)$ & 0.0420 & 0.0415 & 0.0415 & 0.0424 & 0.0425 & 0.0422 \\ 
  15 & 15 & $\mathcal{B}(5,0.6)$ & $\mathcal{B}(5,0.6)$ & 0.0453 & 0.0454 & 0.0456 & 0.0459 & 0.0461 & 0.0458 \\ 
  15 & 30 & $\mathcal{B}(5,0.6)$ & $\mathcal{B}(5,0.6)$ & 0.0508 & 0.0507 & 0.0510 & 0.0505 & 0.0505 & 0.0507 \\ 
  30 & 15 & $\mathcal{B}(5,0.6)$ & $\mathcal{B}(5,0.6)$ & 0.0469 & 0.0470 & 0.0470 & 0.0470 & 0.0469 & 0.0469 \\ 
  30 & 30 & $\mathcal{B}(5,0.6)$ & $\mathcal{B}(5,0.6)$ & 0.0498 & 0.0498 & 0.0497 & 0.0496 & 0.0496 & 0.0496 \\ 
  15 & 45 & $\mathcal{B}(5,0.6)$ & $\mathcal{B}(5,0.6)$ & 0.0491 & 0.0490 & 0.0493 & 0.0489 & 0.0489 & 0.0490 \\ 
  45 & 15 & $\mathcal{B}(5,0.6)$ & $\mathcal{B}(5,0.6)$ & 0.0514 & 0.0514 & 0.0515 & 0.0515 & 0.0516 & 0.0516 \\ 
   \hline
\end{tabular}
\label{tab:app3}
\end{table}
\vspace*{\fill}
\newpage
\begin{figure}[ht!]
\centerline{\includegraphics[scale=0.65]{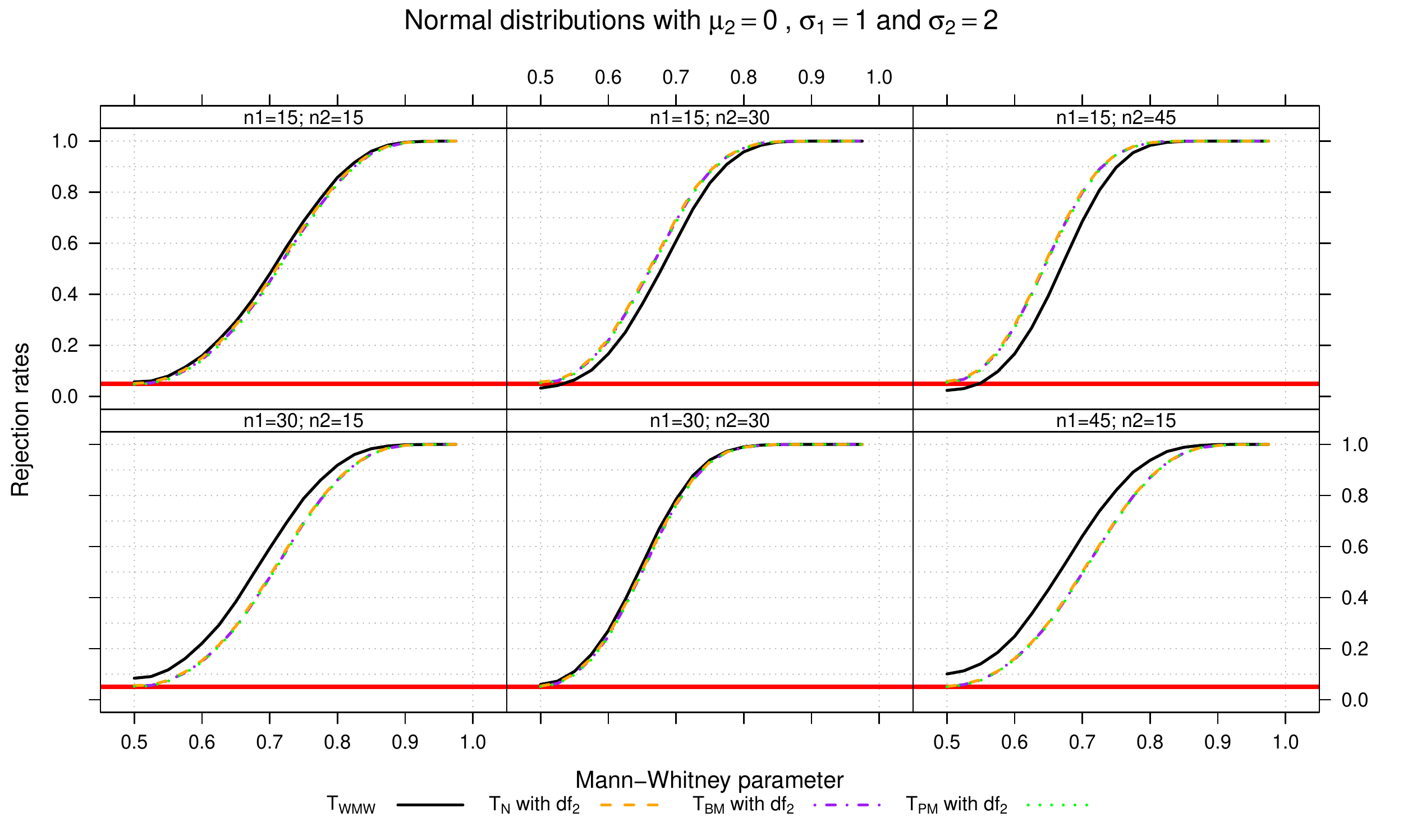}}
\caption{Power graphs for normal distributions based on 10\,000 simulation runs}{}
\end{figure}
\vspace*{\fill}
\newpage
\begin{figure}[ht!]
\centerline{\includegraphics[scale=0.65]{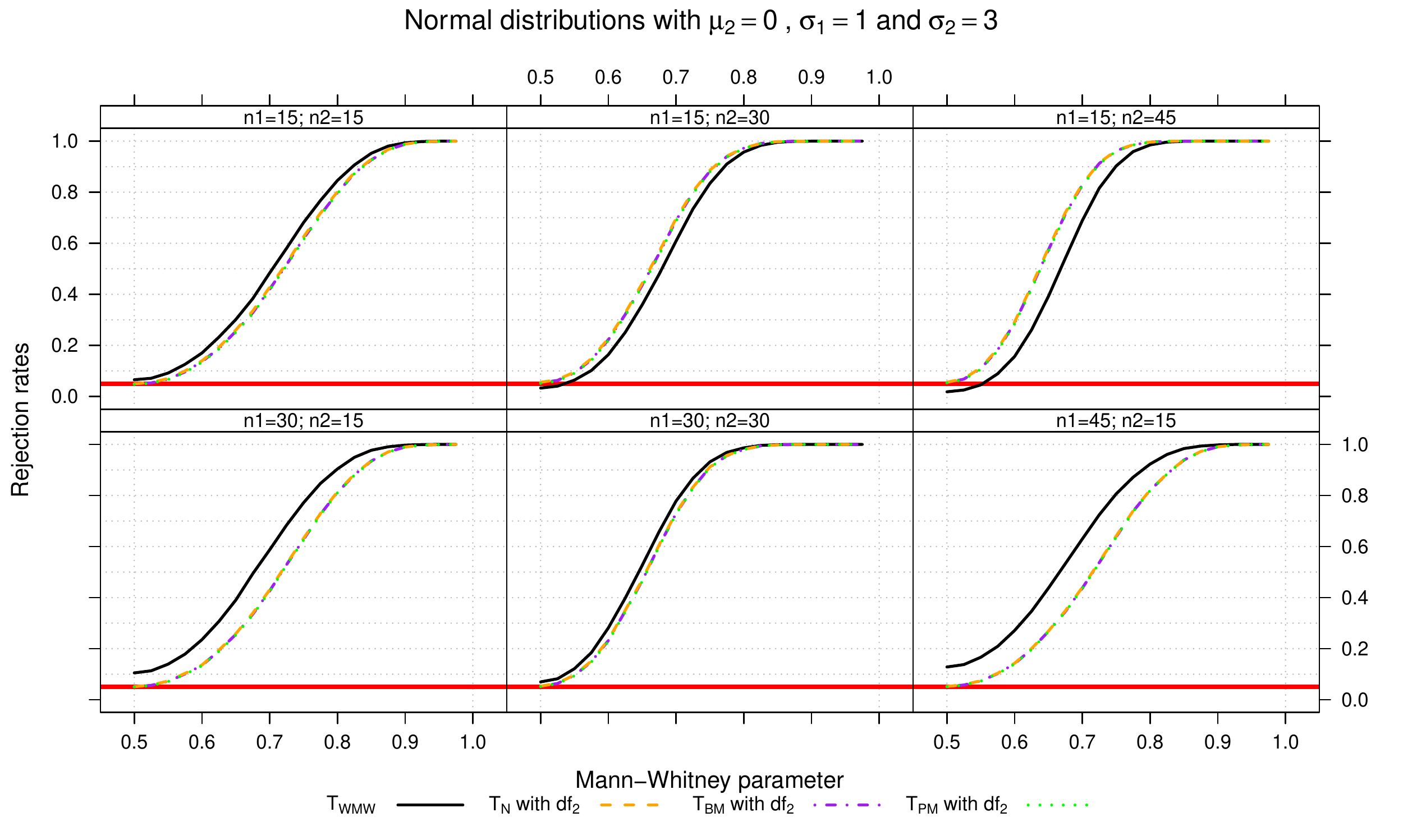}}
\caption{Power graphs for normal distributions based on 10\,000 simulation runs}{}
\end{figure}
\begin{figure}[ht!]
\centerline{\includegraphics[scale=0.65]{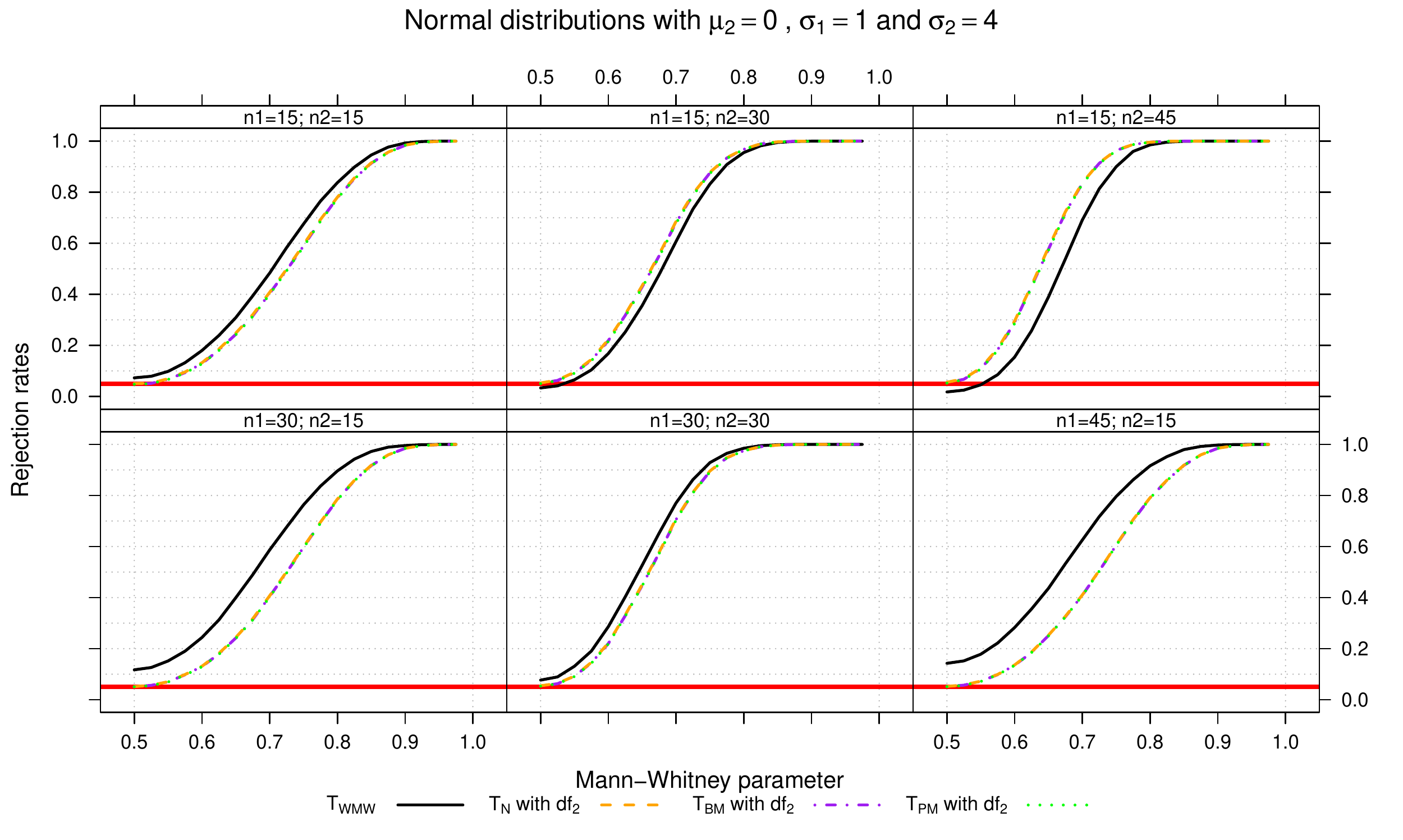}}
\caption{Power graphs for normal distributions based on 10\,000 simulation runs}{}
\end{figure}
\vspace*{\fill}
\newpage
\begin{figure}[ht!]
\centerline{\includegraphics[scale=0.65]{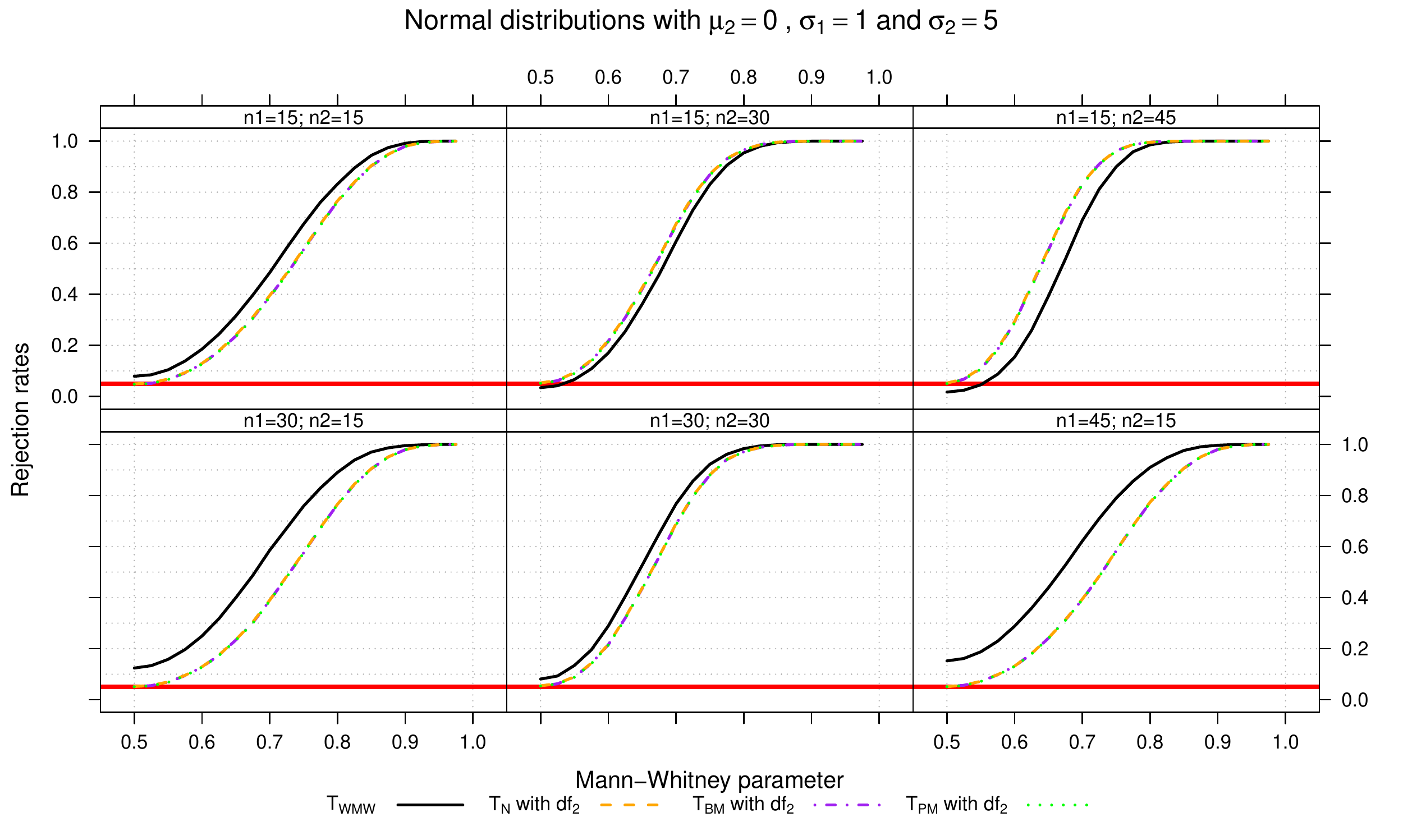}}
\caption{Power graphs for normal distributions based on 10\,000 simulation runs}{}
\end{figure}
%
\begin{figure}[ht!]
\centerline{\includegraphics[scale=0.65]{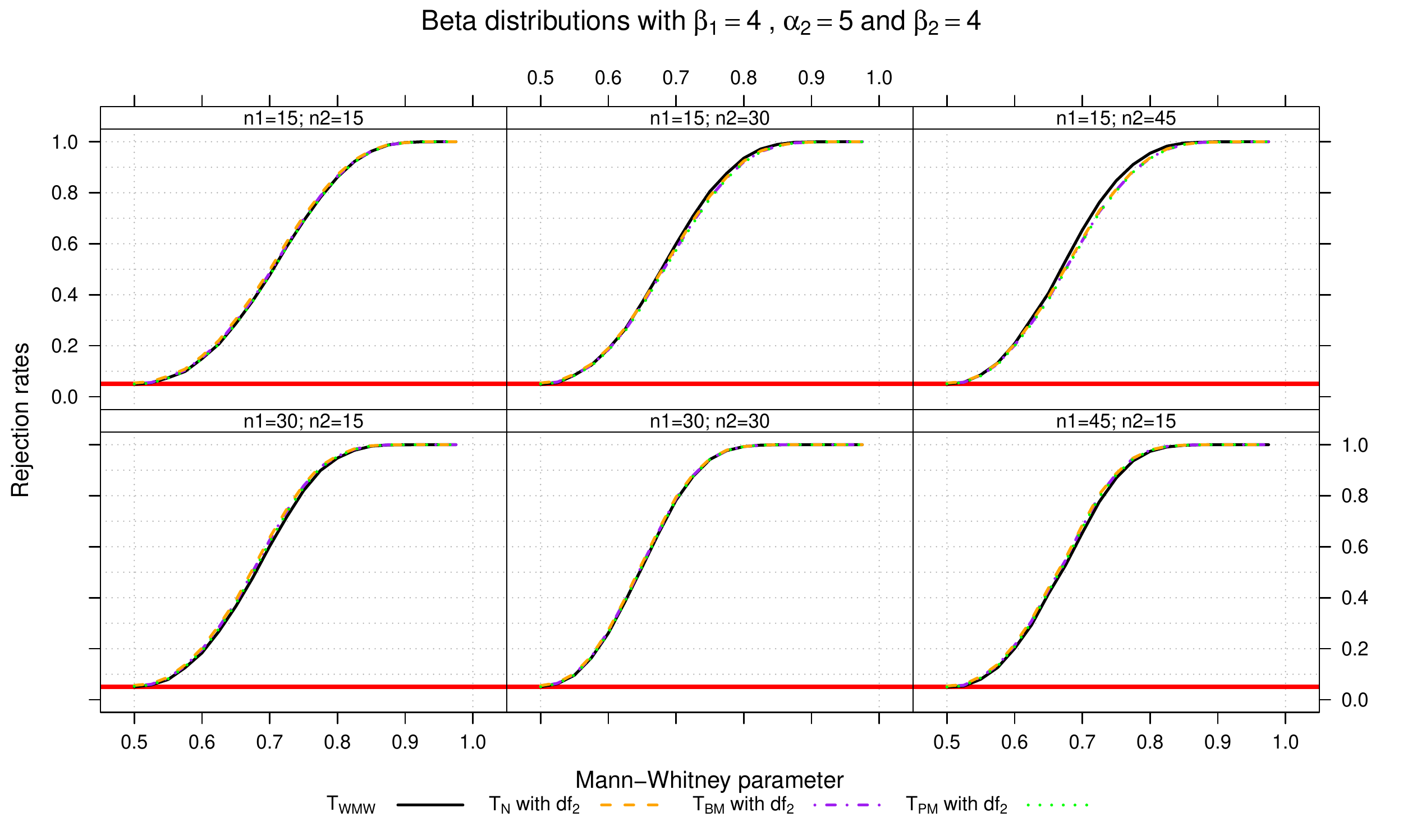}}
\caption{Power graphs for Beta distributions based on 10\,000 simulation runs}{}
\end{figure}
\vspace*{\fill}
\newpage
\begin{figure}[ht!]
\centerline{\includegraphics[scale=0.65]{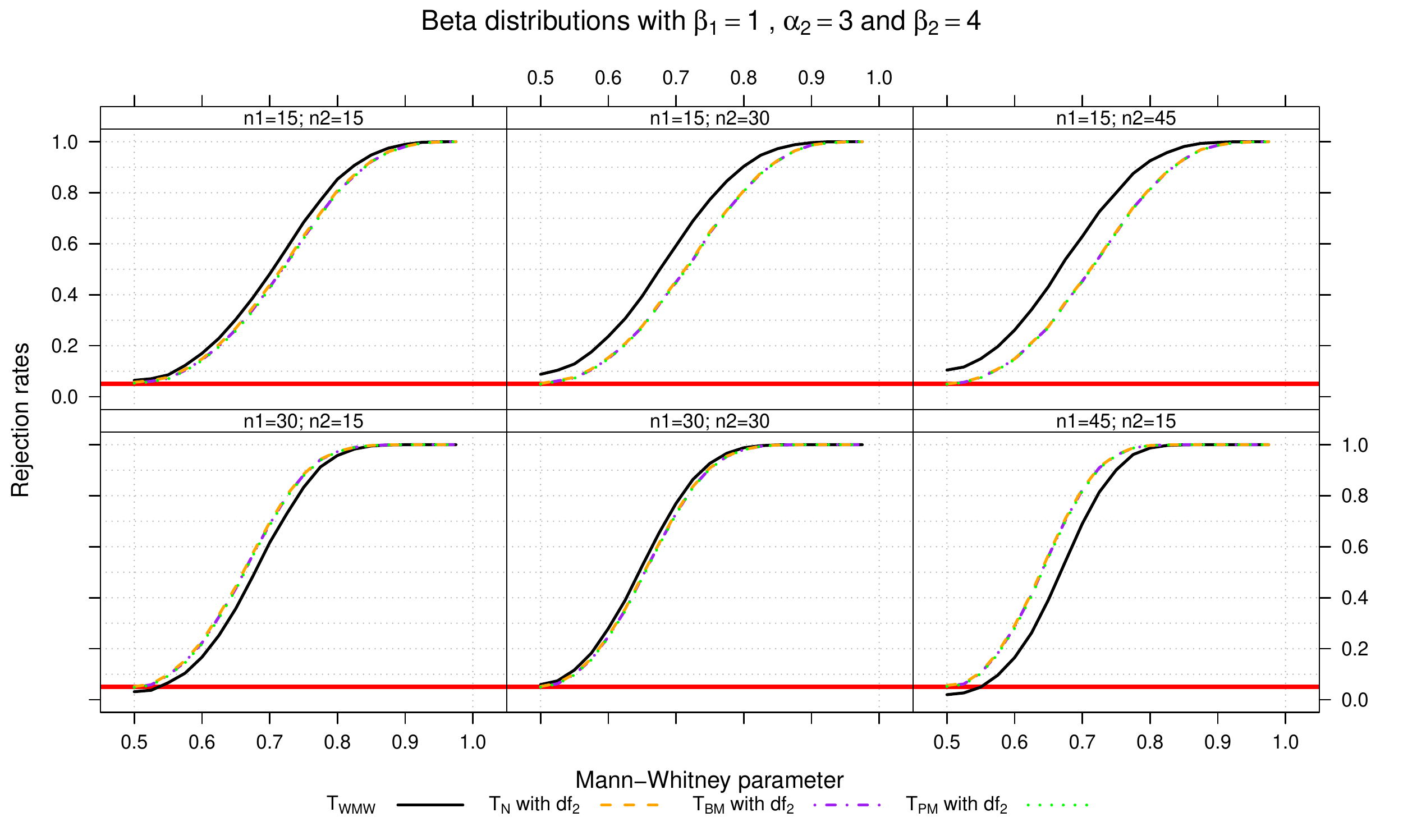}}
\caption{Power graphs for Beta distributions based on 10\,000 simulation runs}{}
\end{figure}
\begin{figure}[ht!]
\centerline{\includegraphics[scale=0.65]{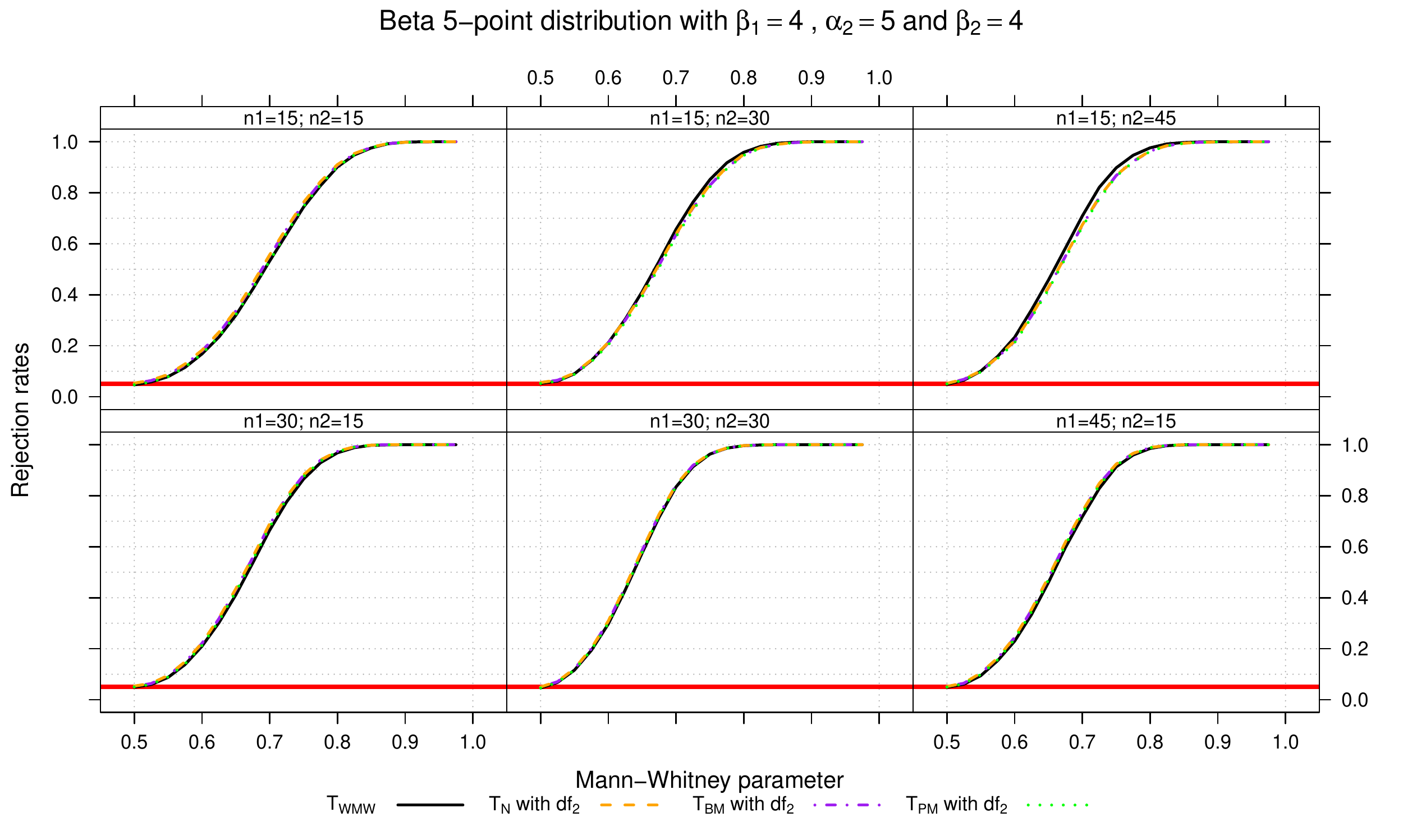}}
\caption{Power graphs for Beta 5-point distributions based on 10\,000 simulation runs}{}
\end{figure}
\vspace*{\fill}
%
\newpage
\begin{figure}[ht!]
\centerline{\includegraphics[scale=0.65]{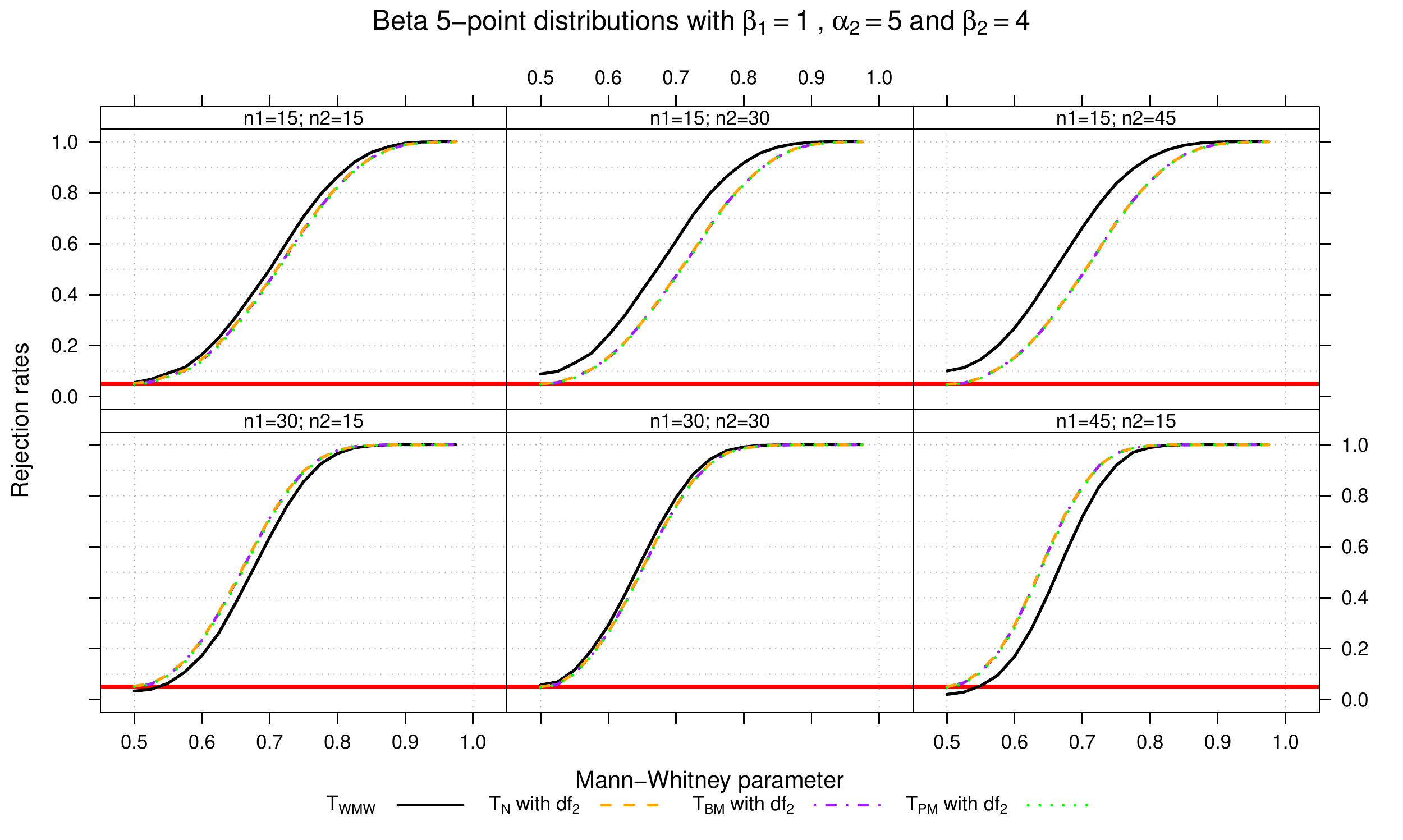}}
\caption{Power graphs for Beta 5-point distributions based on 10\,000 simulation runs}{}
\end{figure}
\begin{figure}[ht!]
\centerline{\includegraphics[scale=0.65]{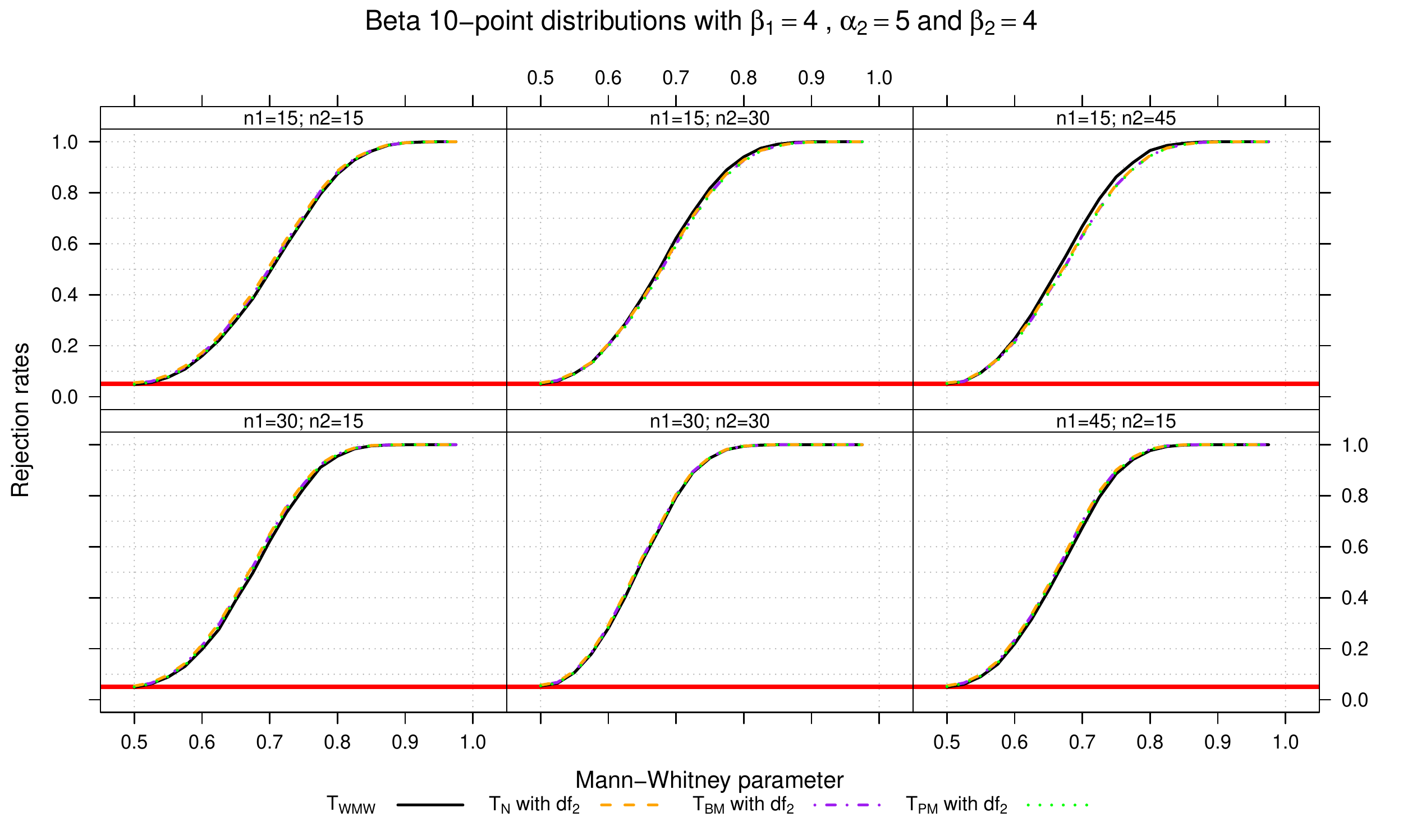}}
\caption{Power graphs for Beta 10-point distributions based on 10\,000 simulation runs}{}
\end{figure}
\vspace*{\fill}
%
\newpage
\begin{figure}[ht!]
\centerline{\includegraphics[scale=0.65]{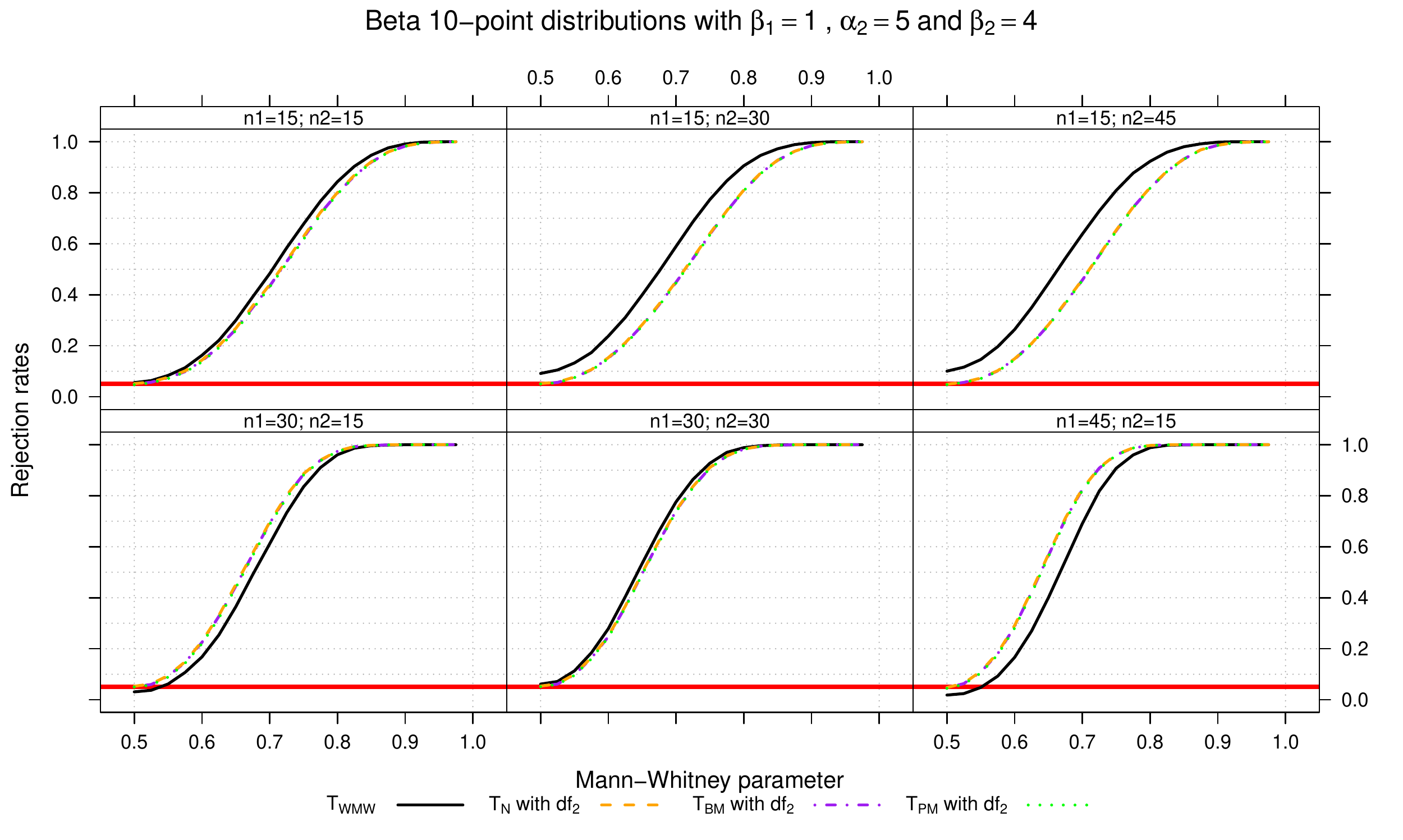}}
\caption{Power graphs for Beta 10-point distributions based on 10\,000 simulation runs}{}
\end{figure}
\begin{figure}[ht!]
\centerline{\includegraphics[scale=0.65]{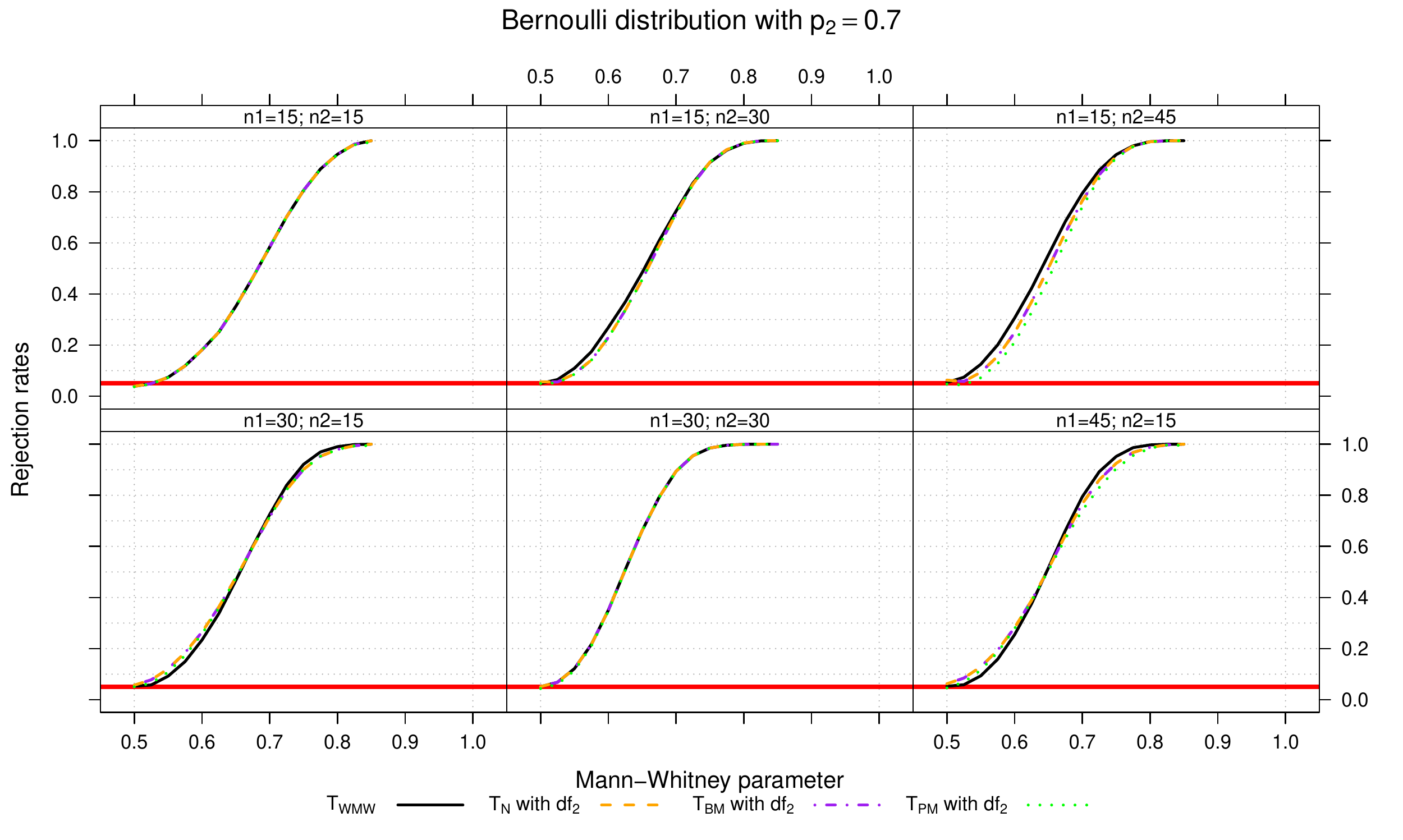}}
\caption{Power graphs for Bernoulli distributions based on 10\,000 simulation runs}{}
\end{figure}
\vspace*{\fill}
\newpage
\newpage
\begin{table}[ht!]
\small\sf\centering
\caption{Power results for Mann-Whitney parameter $p=0.7$ as regards normal distributions $F_{1} = \mathcal{N}(\mu_{1},\sigma_{1}^{2})$ and $F_{2} = \mathcal{N}(\mu_{2},\sigma_{2}^{2})$ at a two-sided nominal significance level of $\alpha = 0.05$ for the studentised permutation tests based on 10 000 random permutations for each of the 10 000 replications}
\begin{tabular}{rrcccccccccc}
  \hline
  $n_{1}$ & $n_{2}$ & $\mu_{1}$ & $\mu_{2}$ & $\sigma_{1}$ & $\sigma_{2}$ & $\widetilde{T}_{N}$ & $\widetilde{T}_{BM}$ & $\widetilde{T}_{PM}$ & $\widetilde{T}_{N}^{Logit}$ & $\widetilde{T}_{BM}^{Logit}$ & $\widetilde{T}_{PM}^{Logit}$ \\ 
  \hline
   7 &  7 & -0.7416 & 0 & 1 & 1 & 0.2264 & 0.2294 & 0.2293 & 0.2253 & 0.2224 & 0.2248 \\ 
   7 & 10 & -0.7416 & 0 & 1 & 1 & 0.2719 & 0.2742 & 0.2746 & 0.2676 & 0.2695 & 0.2706 \\ 
  10 &  7 & -0.7416 & 0 & 1 & 1 & 0.2734 & 0.2760 & 0.2757 & 0.2695 & 0.2715 & 0.2708 \\ 
  10 & 10 & -0.7416 & 0 & 1 & 1 & 0.3365 & 0.3364 & 0.3358 & 0.3374 & 0.3373 & 0.3367 \\ 
  15 & 15 & -0.7416 & 0 & 1 & 1 & 0.4883 & 0.4879 & 0.4887 & 0.4882 & 0.4882 & 0.4879 \\ 
  15 & 30 & -0.7416 & 0 & 1 & 1 & 0.6116 & 0.6116 & 0.6116 & 0.6076 & 0.6079 & 0.6088 \\ 
  30 & 15 & -0.7416 & 0 & 1 & 1 & 0.6046 & 0.6048 & 0.6045 & 0.6032 & 0.6037 & 0.6034 \\ 
  30 & 30 & -0.7416 & 0 & 1 & 1 & 0.7839 & 0.7841 & 0.7843 & 0.7848 & 0.7843 & 0.7839 \\ 
  15 & 45 & -0.7416 & 0 & 1 & 1 & 0.6531 & 0.6529 & 0.6533 & 0.6497 & 0.6502 & 0.6502 \\ 
  45 & 15 & -0.7416 & 0 & 1 & 1 & 0.6442 & 0.6441 & 0.6445 & 0.6395 & 0.6401 & 0.6398 \\ \hline
   7 &  7 & -1.6583 & 0 & 1 & 3 & 0.2110 & 0.2335 & 0.2353 & 0.1712 & 0.1740 & 0.1831 \\ 
   7 & 10 & -1.6583 & 0 & 1 & 3 & 0.2677 & 0.2748 & 0.2841 & 0.2604 & 0.2733 & 0.2781 \\ 
  10 &  7 & -1.6583 & 0 & 1 & 3 & 0.2449 & 0.2555 & 0.2598 & 0.1789 & 0.1827 & 0.1843 \\ 
  10 & 10 & -1.6583 & 0 & 1 & 3 & 0.3052 & 0.3147 & 0.3219 & 0.2793 & 0.2870 & 0.2920 \\ 
  15 & 15 & -1.6583 & 0 & 1 & 3 & 0.4314 & 0.4358 & 0.4398 & 0.4140 & 0.4199 & 0.4259 \\ 
  15 & 30 & -1.6583 & 0 & 1 & 3 & 0.6650 & 0.6667 & 0.6686 & 0.6747 & 0.6783 & 0.6803 \\ 
  30 & 15 & -1.6583 & 0 & 1 & 3 & 0.4582 & 0.4623 & 0.4669 & 0.4184 & 0.4241 & 0.4282 \\ 
  30 & 30 & -1.6583 & 0 & 1 & 3 & 0.7209 & 0.7237 & 0.7253 & 0.7134 & 0.7161 & 0.7183 \\ 
  15 & 45 & -1.6583 & 0 & 1 & 3 & 0.8098 & 0.8109 & 0.8111 & 0.8216 & 0.8221 & 0.8224 \\ 
  45 & 15 & -1.6583 & 0 & 1 & 3 & 0.4560 & 0.4598 & 0.4625 & 0.4079 & 0.4133 & 0.4188 \\ 
   \hline
\end{tabular}
\label{tab:app4}
\end{table}

\begin{table}[ht!]
\small\sf\centering
\caption{Power results for Mann-Whitney parameter $p=0.7$ as regards 5-point Beta distributions with latent $F_{1} = \mathcal{B}(\alpha_{1},\beta_{1})$ and $F_{2} = \mathcal{B}(5,4)$ at a two-sided nominal significance level of $\alpha = 0.05$ for the studentised permutation tests based on 10 000 random permutations for each of the 10 000 replications}
\begin{tabular}{rrcccccccccc}
  \hline
  $n_{1}$ & $n_{2}$ & $\alpha_{1}$ & $\beta_{1}$ & $\widetilde{T}_{N}$ & $\widetilde{T}_{BM}$ & $\widetilde{T}_{PM}$ & $\widetilde{T}_{N}^{Logit}$ & $\widetilde{T}_{BM}^{Logit}$ & $\widetilde{T}_{PM}^{Logit}$ \\ 
  \hline
   7 &  7 & 2.86332 & 4 & 0.1948 & 0.1944 & 0.1936 & 0.1915 & 0.1905 & 0.1899 \\ 
   7 & 10 & 2.86332 & 4 & 0.2476 & 0.2477 & 0.2480 & 0.2458 & 0.2445 & 0.2455 \\ 
  10 &  7 & 2.86332 & 4 & 0.2555 & 0.2559 & 0.2551 & 0.2546 & 0.2553 & 0.2564 \\ 
  10 & 10 & 2.86332 & 4 & 0.3263 & 0.3260 & 0.3267 & 0.3251 & 0.3251 & 0.3261 \\ 
  15 & 15 & 2.86332 & 4 & 0.5060 & 0.5059 & 0.5054 & 0.5058 & 0.5058 & 0.5059 \\ 
  15 & 30 & 2.86332 & 4 & 0.6268 & 0.6270 & 0.6281 & 0.6228 & 0.6230 & 0.6236 \\ 
  30 & 15 & 2.86332 & 4 & 0.6586 & 0.6584 & 0.6584 & 0.6598 & 0.6596 & 0.6598 \\ 
  30 & 30 & 2.86332 & 4 & 0.8314 & 0.8314 & 0.8313 & 0.8313 & 0.8312 & 0.8314 \\ 
  15 & 45 & 2.86332 & 4 & 0.6697 & 0.6697 & 0.6706 & 0.6628 & 0.6631 & 0.6646 \\ 
  45 & 15 & 2.86332 & 4 & 0.7183 & 0.7180 & 0.7183 & 0.7192 & 0.7189 & 0.7191 \\ \hline
   7 &  7 & 0.57606 & 1 & 0.2158 & 0.2168 & 0.2238 & 0.1907 & 0.1914 & 0.1956 \\ 
   7 & 10 & 0.57606 & 1 & 0.2437 & 0.2464 & 0.2531 & 0.1997 & 0.2004 & 0.2032 \\ 
  10 &  7 & 0.57606 & 1 & 0.2710 & 0.2745 & 0.2787 & 0.2722 & 0.2761 & 0.2839 \\ 
  10 & 10 & 0.57606 & 1 & 0.3126 & 0.3154 & 0.3200 & 0.2924 & 0.2953 & 0.2998 \\ 
  15 & 15 & 0.57606 & 1 & 0.4630 & 0.4647 & 0.4686 & 0.4500 & 0.4516 & 0.4569 \\ 
  15 & 30 & 0.57606 & 1 & 0.5012 & 0.5027 & 0.5065 & 0.4704 & 0.4726 & 0.4773 \\ 
  30 & 15 & 0.57606 & 1 & 0.6977 & 0.6983 & 0.6991 & 0.7069 & 0.7079 & 0.7090 \\ 
  30 & 30 & 0.57606 & 1 & 0.7631 & 0.7634 & 0.7653 & 0.7578 & 0.7588 & 0.7603 \\ 
  15 & 45 & 0.57606 & 1 & 0.4971 & 0.4985 & 0.5020 & 0.4576 & 0.4593 & 0.4649 \\ 
  45 & 15 & 0.57606 & 1 & 0.8100 & 0.8103 & 0.8097 & 0.8237 & 0.8241 & 0.8239 \\ 
   \hline
\end{tabular}
\end{table}
\vspace*{\fill}
\newpage
\begin{table}[ht!]
\small\sf\centering
\caption{Power results for Mann-Whitney parameter $p=0.7$ as regards exponential and binomial distributions at a two-sided nominal significance level of $\alpha=0.05$ for the studentised permutation tests based on 10\,000 random permutations for each of the 10\,000 replications}
\begin{tabular}{rrcccccccc}
  \hline
  $n_{1}$ & $n_{2}$ & $F_{1}$ & $F_{2}$ & $\widetilde{T}_{N}$ & $\widetilde{T}_{BM}$ & $\widetilde{T}_{PM}$ & $\widetilde{T}_{N}^{Logit}$ & $\widetilde{T}_{BM}^{Logit}$ & $\widetilde{T}_{PM}^{Logit}$ \\ 
  \hline
   7 &  7 & $\mathcal{E}(2.33333)$ & $\mathcal{E}(1)$ & 0.2302 & 0.2336 & 0.2328 & 0.2274 & 0.2256 & 0.2281 \\ 
   7 & 10 & $\mathcal{E}(2.33333)$ & $\mathcal{E}(1)$ & 0.2784 & 0.2809 & 0.2810 & 0.2817 & 0.2831 & 0.2839 \\ 
  10 &  7 & $\mathcal{E}(2.33333)$ & $\mathcal{E}(1)$ & 0.2765 & 0.2797 & 0.2810 & 0.2638 & 0.2658 & 0.2659 \\ 
  10 & 10 & $\mathcal{E}(2.33333)$ & $\mathcal{E}(1)$ & 0.3417 & 0.3420 & 0.3424 & 0.3412 & 0.3419 & 0.3418 \\ 
  15 & 15 & $\mathcal{E}(2.33333)$ & $\mathcal{E}(1)$ & 0.4797 & 0.4789 & 0.4796 & 0.4799 & 0.4801 & 0.4796 \\ 
  15 & 30 & $\mathcal{E}(2.33333)$ & $\mathcal{E}(1)$ & 0.6429 & 0.6425 & 0.6423 & 0.6478 & 0.6471 & 0.6464 \\ 
  30 & 15 & $\mathcal{E}(2.33333)$ & $\mathcal{E}(1)$ & 0.5646 & 0.5655 & 0.5661 & 0.5562 & 0.5570 & 0.5572 \\ 
  30 & 30 & $\mathcal{E}(2.33333)$ & $\mathcal{E}(1)$ & 0.7921 & 0.7921 & 0.7917 & 0.7922 & 0.7923 & 0.7922 \\ 
  15 & 45 & $\mathcal{E}(2.33333)$ & $\mathcal{E}(1)$ & 0.7002 & 0.6993 & 0.6992 & 0.7049 & 0.7040 & 0.7038 \\ 
  45 & 15 & $\mathcal{E}(2.33333)$ & $\mathcal{E}(1)$ & 0.5947 & 0.5956 & 0.5974 & 0.5808 & 0.5823 & 0.5832 \\ \hline
   7 &  7 & $\mathcal{B}(5,0.43129)$ & $\mathcal{B}(5,0.6)$ & 0.2072 & 0.2090 & 0.2088 & 0.2044 & 0.2041 & 0.2049 \\ 
   7 & 10 & $\mathcal{B}(5,0.43129)$ & $\mathcal{B}(5,0.6)$ & 0.2602 & 0.2616 & 0.2620 & 0.2559 & 0.2566 & 0.2582 \\ 
  10 &  7 & $\mathcal{B}(5,0.43129)$ & $\mathcal{B}(5,0.6)$ & 0.2679 & 0.2674 & 0.2676 & 0.2653 & 0.2653 & 0.2667 \\ 
  10 & 10 & $\mathcal{B}(5,0.43129)$ & $\mathcal{B}(5,0.6)$ & 0.3323 & 0.3324 & 0.3325 & 0.3323 & 0.3321 & 0.3327 \\ 
  15 & 15 & $\mathcal{B}(5,0.43129)$ & $\mathcal{B}(5,0.6)$ & 0.5140 & 0.5138 & 0.5131 & 0.5150 & 0.5152 & 0.5148 \\ 
  15 & 30 & $\mathcal{B}(5,0.43129)$ & $\mathcal{B}(5,0.6)$ & 0.6426 & 0.6426 & 0.6430 & 0.6393 & 0.6395 & 0.6410 \\ 
  30 & 15 & $\mathcal{B}(5,0.43129)$ & $\mathcal{B}(5,0.6)$ & 0.6339 & 0.6341 & 0.6342 & 0.6327 & 0.6322 & 0.6323 \\ 
  30 & 30 & $\mathcal{B}(5,0.43129)$ & $\mathcal{B}(5,0.6)$ & 0.8161 & 0.8162 & 0.8160 & 0.8165 & 0.8165 & 0.8162 \\ 
  15 & 45 & $\mathcal{B}(5,0.43129)$ & $\mathcal{B}(5,0.6)$ & 0.6803 & 0.6805 & 0.6812 & 0.6738 & 0.6738 & 0.6746 \\ 
  45 & 15 & $\mathcal{B}(5,0.43129)$ & $\mathcal{B}(5,0.6)$ & 0.6834 & 0.6835 & 0.6834 & 0.6825 & 0.6824 & 0.6824 \\ 
   \hline
\end{tabular}
\label{tab:app5}
\end{table}
\vspace*{\fill}

\end{document}